\documentclass[preprint,prd,aps,showpacs,showkeys,nofootinbib]{revtex4}
\usepackage{graphicx}
\usepackage{dcolumn}
\usepackage{bm}
\begin{document}

\preprint{SNUTP 05-016}
\preprint{hep-ph/0602056}

\title{Mixing among the neutral Higgs bosons and rare $B$
decays in the CP violating MSSM}

\author{Tai-Fu Feng$^{a,b,c,d}$, Xue-Qian Li$^{c,e}$, Jukka Maalampi$^b$}
\affiliation{$^{a}$ Center for Theoretical Physics, Seoul National University,
Seoul, 151-742, Korea}
\affiliation{$^b$ Department of Physics, 40014 University of Jyv\"askyl\"a,Finland}

\affiliation{$^c$ CCAST (World Laboratory), P. O. Box 8730, Beijing 100080, China}

\affiliation{$^d$ Department of Physics, Dalian University of Technology, Dalian 116024, China}

\affiliation{$^e$ Department of Physics, Nankai University, Tianjin 300071, China}

\date{\today}

\begin{abstract}
Considering corrections from two-loop Feynman diagrams which involve
gluino at large $\tan\beta$, we analyze the effects of possible CP
phases on the rare $B$ decays: $\bar{ B}_{_{s}} \rightarrow
l^+l^-$ and $\bar{B}\rightarrow Kl^+l^-$ in the CP violating minimal
supersymmetric extension of the standard model.
It is shown that the results of exact two loop
calculations obviously differ from that including one-loop
contributions plus threshold radiative corrections. The numerical
analysis indicates that the possibly large CP phases strongly affect
the theoretical estimation of the branching ratios, and this results
coincide with the conclusion of some other works appearing in recent
literature.
\end{abstract}


\maketitle

\section{Introduction}

\indent\indent The rare $B$ decays serve as a good probe for the
new physics beyond the standard model (SM) since they do not
seriously suffer from the uncertainties caused by the long
distance effects. The forthcoming experiments of the $B$ factories
will make more precise measurements on the rare processes and it
is believed that those measurements should set even stricter
constraints on the parameter space of new physics. Among those
plausible new physics scenarios beyond the SM, the simplest and
most favorable extension is the Minimal Supersymmetric Model
(MSSM) \cite{Gunion}. In the MSSM, there are five physical scalars
(Higgs) compared to the SM where there is only one. The
contribution of those neutral Higgs bosons to the rare $B$
processes have been extensively discussed in literature. The main
conclusion of the analysis is that the branching ratios of the
processes such as $\bar{B}_{_s}\rightarrow
l^+l^-,\;\bar{B}_{_s}\rightarrow Kl^+l^-$ {\it etc.} are enhanced
for larger values of $\tan\beta$, which is the ratio of the two
vacuum expectation values (VEVs) of the neutral Higgs fields, even
within the minimal flavor violating (MFV)\footnote{That is, the
Cabibbo-Kobayashi-Maskawa (CKM) matrix is  assumed to be the only
source of flavor mixing.} supersymmetry scenario.

In addition to the SM CP phase which exists in the CKM matrix,
there are three more possible sources for the CP violation phases
in the MSSM Lagrangian. The first one is the $\mu$ parameter in
the superpotential which is complex and the second source is the
corresponding soft breaking parameters, namely, the complex masses
of the $SU(3)\times SU(2)\times U(1)$ gauginos in the soft
breaking terms induce three CP phases. As the third one, there are
several CP phases emerging from the scalar soft mass matrices
${\bf m}_{_{Q,U,D,L,R}}^2$ and the soft trilinear coupling
matrices ${\bf A}_{_{U,D,E}}$. Generally, only the off-diagonal
elements of the soft mass matrices can be complex due to the
hermiticity of these matrices. By contrast, the trilinear coupling
matrices ${\bf A}_{_{U,D,E}}$ can have complex diagonal
elements\cite{Brhlik}. Actually, the CP phases must be constrained
by the present experimental results. The most rigorous constraints
on those CP phases come from the experimental bounds of the
electron and neutron electric dipole moments (EDMs), which are
$d_{_e}< 4.3\times 10^{-27}e\cdot cm$ \cite{Commins} and $d_{_n}<
6.5\times 10^{-26}e\cdot cm$\cite{Harris}, respectively. The bound
of the EDM of $H_{_g}^{199}$: $d_{_{H_{g}^{199}}} < 9.\times
10^{-28}e\cdot cm$\cite{Lamoreaux} is also measured with high
accuracy. In order to make the theoretical prediction consistent
with the experimental data, three approaches are adopted in
literature. One possibility is to make the CP phases sufficiently
small, i.e. smaller than $10^{-2}$ \cite{cp1}. Alternatively one
can also assume a mass suppression by making the supersymmetry
spectra heavy enough, i.e. in a range of several TeV\cite{cp2}, or
invoke a cancellation among different contributions to the fermion
EDMs \cite{cp3}.

In the second scenario, a series of
works\cite{Pilaftsis1,Pilaftsis2,Pilaftsis3, Pilaftsis4} analyzes
the mixing among the neutral Higgs bosons in the CP violating
MSSM. Considering the constraints from the experimental upper
bounds of the electron and neutron electric dipole moments (EDMs),
the soft trilinear coupling for the third generation scalar quarks
can have large CP phases. Due to the large Yukawa couplings for
the third generation quarks, the radiative corrections can lead to
a large mixing among the would-be CP-even and CP-odd neutral Higgs
bosons. This mixing causes drastic changes for the couplings
between the neutral Higgs and  quarks, the neutral Higgs and gauge
bosons, and the self couplings among the Higgs fields, thus the
consequence is that the lower bound of the mass of the lightest
neutral Higgs is relaxed to $60\;{\rm GeV}$.

Presently, the calculations of the effective hamiltonian for the
transition $b\rightarrow sl^+l^-$ are presented by the authors of
Ref.\cite{Buras1,He,Skiba,Grossman,Guetta,Logan} within the MSSM
and THDM models. However, those works are all focusing on the
CP-conserving processes. As pointed out in Ref.\cite{Ibrahim}, the
CP violation phases which induce mixing among the neutral Higgs
bosons affect the effective Hamiltonian of $b\rightarrow sl^+l^-$
significantly, especially for larger $\tan\beta$ values. However,
their analysis \cite{Ibrahim} only included the leading terms in
$\tan^3\beta$ which originate from the counter diagrams.

In this work, we analyze the rare $B$-decays: $B\rightarrow
X_{_s}\gamma$, $\bar{B}_{_{s}}^0 \rightarrow l^+l^-$ and
$\bar{B}\rightarrow Kl^+l^-$ ($l=\mu,\;\tau$) in the CP violating
MSSM. In order to reduce the number of free parameters, we assume
no additional sources of flavor violation other than the CKM
matrix elements. In our calculation, we consider all  possible
contributions from the one-loop box, $\gamma-$, $Z-$ and
Higgs-penguin diagrams, as well as the threshold radiative
corrections to the effective Hamiltonian for $b\rightarrow
sl^+l^-$ at first. The threshold radiative corrections indicate a
sum of the SUSY QCD and SUSY electroweak corrections to the
effective vertex $H_u^*D^cQ$ at larger $\tan\beta$
values\cite{Babu,Pilaftsis5,Carena1,Pierce}. Although threshold
radiative corrections approximate the exact two-loop results
adequately when the supersymmetry energy scale is sufficiently
high, the authors of Ref. \cite{Borzumati,Feng1} pointed out that
the difference between threshold radiative approximation and exact
two loop calculation is obvious in some regions of the parameter
space of the MSSM. As a comparison and complement to the results
of threshold radiative corrections which were already derived and
evaluated in literature, we present calculations of the exact
two-loop corrections from gluino at large $\tan\beta$ here. Those
two-loop results, which include the gluino corrections to the
penguin vertices $\bar{s}bZ$, $\bar{s}bH$, and the gluino
corrections to the four fermion interaction $\bar{s}b\bar{l}l$,
are explicitly presented in this work. The gluino corrections to
the penguin vertex $\bar{s}b\gamma$ can be found in our previous
analysis\cite{Feng1}.

In the second section, we will present the modified couplings
involving the neutral Higgs bosons in the CP violating MSSM. The
effective Lagrangian for rare $B$ decay modes is given in the
third section. We also show the newly derived theoretical
formulations of those decay branching ratios {\it etc.}  in this
section. Section {\bf IV} is devoted to the numerical analysis and
discussion, our conclusion is made in the last section. Some
tedious formulae are collected in the appendices.

\section{Mixing among neutral Higgs bosons and relevant couplings}

\indent\indent In this section, we present the modified Yukawa
couplings of the neutral Higgs to the squarks and sleptons where
the effects induced by the CP violation phases are taken
into account.
The most general form of the superpotential which has the gauge
invariance and retains all the conservation laws of the SM is
written as
\begin{eqnarray}
{\cal W}=\mu_{_H}\epsilon_{ij}\hat{H}^1_i\hat{H}_j^2+\epsilon_{ij}
h_{_{IJ}}^l\hat{H}_i^1\hat{L}_j^I\hat{R}^J+\epsilon_{ij}
h_{_{IJ}}^d\hat{H}_i^1\hat{Q}_j^I\hat{D}^J+\epsilon_{ij}
h_{_{IJ}}^u\hat{H}_i^2\hat{Q}_j^I\hat{U}^J, \label{sp}
\end{eqnarray}
where $\hat{H}^1,\;\hat{H}^2$ are the Higgs superfields;
$\hat{Q}^{I}$ and $\hat{L}^{I}$ are quark and lepton superfields
in doublets of the weak SU(2), where I=1, 2, 3 are the indices of
generations; the rest superfields $\hat{U}^{I}$, $\hat{D}^{I}$ and
$\hat{R}^{I}$ are the scalar quark superfields of u- and d-types
and charged leptons in singlets of the weak SU(2) respectively.
Indices i, j are contracted for  SU(2), and $h^{l}$, $h^{u,d}$ are
the Yukawa couplings. To explicitly break supersymmetry, the
soft-supersymmetry breaking terms are introduced as
\begin{eqnarray}
&&{\cal L}_{soft}=-m_{_{H^1}}^2H_i^{1*}H_i^1-m_{_{H^2}}^2H_i^{2*}H_i^2
-m_{_{L^{I}}}^2\tilde{L}_i^{I*}\tilde{L}_i^{I}-m_{_{R^{I}}}^2
\tilde{R}^{I*}\tilde{R}^{I}
-m_{_{Q^{I}}}^2\tilde{Q}_i^{I*}\tilde{Q}_i^{I}
-m_{_{U^{I}}}^2\tilde{U}^{I*}\tilde{U}^{I}\nonumber \\
&&\hspace{1.4cm}-m_{_{D^{I}}}^2\tilde{D}^{I*}\tilde{D}^{I}
+(m_1\lambda_B\lambda_1+m_2\lambda_A^i\lambda_A^i
+m_3\lambda_G^a\lambda_G^a+h.c.) +\Big[\mu
B\epsilon_{ij}H_i^1H_j^2
+\epsilon_{ij}A^l_{_{I}}H_{i}^{1}\tilde{L}^{I}_{j}\tilde{R}^{I}
\nonumber \\
&&\hspace{1.4cm}
+\epsilon_{ij}A^d_{_{I}}H_{i}^{1}\tilde{Q}^{I}_{j}\tilde{D}^{I}
+\epsilon_{ij}A^u_{_{I}}H_{i}^{2}\tilde{Q}^{I}_{j}
\tilde{U}^{I}+h.c.\Big], \label{soft}
\end{eqnarray}
where
$m_{_{H^1}}^2,\;m_{_{H^2}}^2,\;m_{_{L^{I}}}^2,\;m_{_{R^{I}}}^2,
\;m_{_{Q^{I}}}^2,\;m_{_{U^{I}}}^2$ and $m_{_{D^{I}}}^2$ are the
squared masses of the superparticles, $m_3,\;m_2,\;m_1$ denote the
masses of $\lambda_G^a\;(a=1,\;2,\;
\cdots\;8),\;\lambda_A^i\;(i=1,\;2,\;3)$ and $\lambda_B$, which
are the $SU(3)\times SU(2)\times U(1)$ gauginos. $B$ is a free
parameter in unit of mass. With the soft-supersymmetry breaking
terms in Eq.(\ref{soft}), we can study the phenomenology in the
framework of the MSSM.

By the effective potential which accounts for the two-loop Yukawa
and QCD corrections via the renormalization group equation (RGE),
the squared mass matrix of the neutral Higgs bosons is written as:
\begin{eqnarray}
&&\hspace{-0.5cm}{\bf \hat{m}}_{_{H^0}}^2=\nonumber\\&&\nonumber\\
&&\hspace{-0.5cm}\left({\tiny
\begin{array}{llc}\left(\begin{array}{l}
m_a^2s_{_\beta}^2-{8m_{_{\rm w}}^2s_{_{\rm w}}^2\over e^2}
\Big[\lambda_1c_{_\beta}^2\\
+{\bf Re}(\lambda_5)s_{_\beta}^2+
{\bf Re}(\lambda_6)s_{_\beta}c_{_\beta}\Big]
\end{array}\right)
&\left(\begin{array}{l}-m_a^2s_{_\beta}c_{_\beta}
-{8m_{_{\rm w}}^2s_{_{\rm w}}^2\over e^2}\Big[\Big(\lambda_3+\lambda_4
\Big)s_{_\beta}c_{_\beta}\\+{\bf Re}(\lambda_6)c_{_\beta}^2
+{\bf Re}(\lambda_7)s_{_\beta}^2\Big]
\end{array}\right)
&\left(\begin{array}{l}{\bf Im}(\lambda_5)
s_{_\beta}\\+{\bf Im}(\lambda_6)c_{_\beta}\end{array}\right)\\\\
\left(\begin{array}{l}-m_a^2s_{_\beta}c_{_\beta}
-{8m_{_{\rm w}}^2s_{_{\rm w}}^2\over e^2}\Big[\Big(\lambda_3+\lambda_4
\Big)s_{_\beta}c_{_\beta}\\+{\bf Re}(\lambda_6)c_{_\beta}^2
+{\bf Re}(\lambda_7)s_{_\beta}^2\end{array}\right)&
\left(\begin{array}{l}m_a^2c_{_\beta}^2
-{8m_{_{\rm w}}^2s_{_{\rm w}}^2\over e^2}
\Big[\lambda_2s_{_\beta}^2\\+{\bf Re}(\lambda_5)c_{_\beta}^2+
{\bf Re}(\lambda_7)s_{_\beta}c_{_\beta}\Big]\end{array}\right)
&\left(\begin{array}{l}{\bf Im}(\lambda_5)s_{_\beta}\\
+{\bf Im}(\lambda_6)c_{_\beta}\end{array}\right)\\\\
\hspace{0.3cm}{\bf Im}(\lambda_5)s_{_\beta}+{\bf Im}(\lambda_6)c_{_\beta}&
\hspace{0.3cm}{\bf Im}(\lambda_5)c_{_\beta}+{\bf Im}(\lambda_7)s_{_\beta}&
m_a^2
\end{array}}\right)
\label{nhmass}
\end{eqnarray}
with the squared mass $m_a^2$ being
\begin{equation}
m_a^2=m_{_{H^\pm}}^2-{4m_{_{\rm w}}^2s_{_{\rm w}}^2\over e^2}\Big(
{1\over 2}\lambda_4-{\bf Re}(\lambda_5)\Big)\;, \label{ma}
\end{equation}
where the parameters $m_{_{H^\pm}}$ represent the  masses of the
physical charged Higgs-bosons. The concrete expressions of the
parameters $\lambda_i \;(i=1,2,\cdots,7)$ can be found in Ref.
\cite{Pilaftsis2}. Since the squared mass matrix of neutral Higgs
$m_{_{H^0}}^2$ is symmetric, we can find an orthogonal matrix
${\cal Z}_{_H}$ to diagonalize it:
\begin{equation}
{\cal Z}_{_H}^T{\bf \hat{m}}_{_{H^0}}^2{\cal Z}_{_H}=diag(m^2_{_{H_1^0}},
m^2_{_{H_2^0}},m^2_{_{H_3^0}})\;.
\end{equation}

Correspondingly, the modified interactions related to our
calculation are listed below:
\begin{eqnarray}
&&{\cal L}_{_{H_k^0\bar\chi_{_\beta}^+\chi_{_\alpha}^+}}
={e\over \sqrt{2}s_{_{\rm w}}}H_k^0\bar\chi_{_\beta}^+
\bigg\{\Big(\kappa_{_{H^{^k}}}^1\Big)_{_{\beta\alpha}}\omega_-+
\Big(\kappa_{_{H^{^k}}}^2\Big)_{_{\beta\alpha}}\omega_+
\bigg\}\chi_{_\alpha}^+
\;,\nonumber\\
&&{\cal L}_{_{H_k^0\tilde{t}_{_\beta}^*\tilde{t}_{_\alpha}}}
={em_{_{\rm w}}\over s_{_{\rm w}}c_{_{\rm w}}^2}\Big(\zeta_{_{tH}}^k\Big)_{_{ji}}
H_k^0\tilde{t}_i\tilde{t}_j^*\;, \nonumber\\
&&{\cal L}_{_{H_k^0\tilde{D}_{_\beta}^*\tilde{D}_{_\alpha}}}={em_{_{\rm
w}}\over s_{_{\rm w}}c_{_{\rm w}}^2}\Big(\zeta_{_{sH}}^k\Big)_{_{ji}}
H_k^0\tilde{s}_i\tilde{s}_j^*\;,\nonumber\\
&&{\cal L}_{_{H_i^0H^{^\pm}W^{^\mp}}}={e\over 2s_{_{\rm w}}}\Big\{
\Big[s_{_\beta}{\cal Z}_{_H}^{2i}+c_{_\beta}{\cal
Z}_{_H}^{3i}+i{\cal
Z}_{_H}^{1i}\Big]W_\mu^+\Big(H_i^0(i{\partial}^\mu
H^-)-(i{\partial}^\mu H_i^0)H^-\Big)\nonumber\\
&&\hspace{2.2cm}-\Big[s_{_\beta}{\cal Z}_{_H}^{2i}+c_{_\beta}{\cal
Z}_{_H}^{3i}-i{\cal
Z}_{_H}^{1i}\Big]W_\mu^-\Big(H_i^0(i{\partial}^\mu
H^+)-(i{\partial}^\mu H_i^0)H^+\Big)\Big\}\;,\nonumber\\
&&{\cal L}_{_{H_i^0G^{^\pm}W^{^\mp}}}={e\over 2s_{_{\rm w}}}\Big\{
\Big[-c_{_\beta}{\cal Z}_{_H}^{2i}+s_{_\beta}{\cal
Z}_{_H}^{3i}\Big]W_\mu^+\Big(H_i^0(i{\partial}^\mu
G^-)-(i{\partial}^\mu H_i^0)G^-\Big)\nonumber\\
&&\hspace{2.2cm}-\Big[-c_{_\beta}{\cal
Z}_{_H}^{2i}+s_{_\beta}{\cal
Z}_{_H}^{3i}\Big]W_\mu^-\Big(H_i^0(i{\partial}^\mu
G^+)-(i{\partial}^\mu H_i^0)G^+\Big)\Big\}\;,\nonumber\\
&&{\cal L}_{_{H_i^0W^{^\pm}W^{^\mp}}}={em_{_{\rm w}}\over s_{_{\rm
w}}}\Big(c_{_\beta}{\cal Z}_{_H}^{2i}+s_{_\beta}{\cal
Z}_{_H}^{3i}\Big)H_i^0W_\mu^+W^{-\mu}\;.
\label{coupling1}
\end{eqnarray}
The couplings
$\kappa_{_{H^{^\rho}}}^{1,2},\;\zeta_{_{(t,s)H}}^\rho$ are
presented in appendix \ref{2b}. Since the SUSY-QCD modifies the
Yukawa coupling $\bar{t}bH^+$ remarkably\cite{Carena1} for larger
$\tan\beta$ values,  we should consider the SUSY-QCD effects on
the effective lagrangian which only includes one loop
contributions. Re-summing the dominant supersymmetric corrections
for larger $\tan\beta$ to all orders, as well as the leading and
next-to-leading logarithms of the standard QCD corrections, one
can write the interaction as\cite{Babu, Pilaftsis5}
\begin{eqnarray}
&&{\cal L}_{_{H\bar{q}q}}={em_{_b}(Q)\over2m_{_{\rm w}}s_{_{\rm
w}}s_{_\beta}c_{_\beta}}\bigg\{V_{_{tb}}^*\chi_{_{FC}}\Big[
c_{_\beta}{\cal Z}_{_H}^{3k}-s_{_\beta}{\cal Z}_{_H}^{2k}-i{\cal
Z}_{_H}^{1k}\Big]\Big[V_{_{td}}\bar{b}\omega_-d+V_{_{ts}}\bar{b}
\omega_-s\Big]H_k^0\nonumber\\&&\hspace{1.3cm}-s_{_\beta}
\chi_{_{B}}\Big[{\cal Z}_{_H}^{2k} +\Delta {\cal Z
}_{_H}^{3k}+i\Big(s_{_\beta}-c_{_\beta} \Delta\Big){\cal Z
}_{_H}^{1k}\Big]\bar{b}\omega_-bH_k^0+h.c.\bigg\}
\nonumber\\&&\hspace{1.3cm}+{e\over\sqrt{2}s_{_{\rm
w}}}\bigg\{V_{_{tb}}^*\Big[{m_{_t}(Q)\over m_{_{\rm w}}\tan\beta
}\bar{b}\omega_+t+\xi_{_H}^b{m_{_b}(Q)\over m_{_{\rm w}}
}\tan\beta\bar{b}\omega_-t\Big]H^-+V_{_{tb}}^*\Big[{m_{_t}(Q)\over
m_{_{\rm w}}}\bar{b}\omega_+t\nonumber\\&&\hspace{1.3cm}
-\xi_{_G}^b{m_{_b}(Q)\over m_{_{\rm
w}}}\bar{b}\omega_-t\Big]G^-+\sum\limits
\xi_{_{CKM}}^{ib*}V_{_{ib}}^*\Big[{m_{_{u_i}}(Q)\over m_{_{\rm
w}}\tan\beta }\bar{b}\omega_+u_i+{m_{_b}(Q)\over m_{_{\rm w}}
}\tan\beta\bar{b}\omega_-u_i\Big]H^-\nonumber\\&&\hspace{1.3cm}+\sum\limits
\xi_{_{CKM}}^{ib*}V_{_{ib}}^*\Big[{m_{_{u_i}}(Q)\over m_{_{\rm w}}
}\bar{b}\omega_+u_i-{m_{_b}(Q)\over m_{_{\rm w}}
}\bar{b}\omega_-u_i\Big]G^-+\sum\limits
\xi_{_{CKM}}^{ti*}V_{_{ti}}^*\Big[{m_{_{t}}(Q)\over m_{_{\rm
w}}\tan\beta }\bar{d}_i\omega_+t\nonumber\\&&\hspace{1.3cm}
+{m_{_{d_i}}(Q)\over m_{_{\rm w}}
}\tan\beta\bar{d}_i\omega_-t\Big]H^-+\sum\limits
\xi_{_{CKM}}^{ti*}V_{_{ti}}^*\Big[{m_{_{t}}(Q)\over m_{_{\rm w}}
}\bar{d}_i\omega_+t-{m_{_{d_i}}(Q)\over m_{_{\rm w}}
}\bar{d}_i\omega_-t\Big]G^-+h.c.\bigg\}\;, \label{modhtb}
\end{eqnarray}
where $Q$ is the characteristic energy scale of the process. The
$\tan\beta$-enhancing radiative corrections are
\begin{eqnarray}
&&\chi_{_{B}}={1\over 1+\tan\beta\Delta}\;,\nonumber\\
&&\chi_{_{FC}}=-\chi_{_{B}}{em_{_t}(Q)\tan\beta\Delta
^{^{EW}}\over\sqrt{2}m_{_{\rm w}}s_{_{\rm
w}}s_{_\beta}[1+\tan\beta\Delta^{^{S}}]}\;,\nonumber\\
&&\xi_{_H}^b=\chi_{_{B}}\Big(1+\Delta^{^{S}}-h_{_t}\cot\beta\Delta
^{^{EW}}\Big)\;,\nonumber\\
&&\xi_{_G}^b=\chi_{_{B}}\Big(1+\Delta\Big)\;,\nonumber\\
&&\xi_{_{CKM}}^{ub}=\xi_{_{CKM}}^{cb}=\xi_{_{CKM}}^{td}
=\xi_{_{CKM}}^{ts}={1+\tan\beta\Delta\over1+\tan\beta\Delta^{^{S}}}
\;,\nonumber\\
&&\xi_{_{CKM}}^{tb}=\xi_{_{CKM}}^{ud}=\xi_{_{CKM}}^{cs}=1\;,\nonumber\\
&&\Delta=\Delta^{^{S}}+h_{_t}\Delta^{^{EW}}\;,\nonumber\\
&&\Delta^{^{S}}={2\alpha_{_s}\over 3\pi}m_{_{\tilde g}}^*\mu_{_H}^*
I(m_{_{\tilde{b}_L}},m_{_{\tilde{b}_R}},|m_{_{\tilde
g}}|)\;,\nonumber\\
&&\Delta^{^{EW}}={1\over 64\pi^2}\mu_{_H}^*A_{_t}
I(m_{_{\tilde{t}_L}},m_{_{\tilde{t}_R}},|\mu_{_H}|)\;, \label{radi}
\end{eqnarray}
with the vertex function\cite{Pierce}
\begin{eqnarray}
&&I(a,b,c)={1\over
(a^2-b^2)(b^2-c^2)(a^2-c^2)}\Big[a^2b^2\ln\Big({a^2\over b^2
}\Big)+b^2c^2\ln\Big({b^2\over c^2}\Big)+c^2a^2\ln\Big({c^2\over
a^2}\Big)\Big]\;. \label{vertex}
\end{eqnarray}
While deriving Eq. (\ref{modhtb}) and Eq. (\ref{radi}), we
consider the fact that
$m_{_t}\gg(m_{_c},\;m_{_u});\;m_{_b}\gg(m_{_s},\;m_{_d})$ and
$|V_{_{tb}}|\gg
(|V_{_{ts}}|,\;|V_{_{td}}|,\;|V_{_{ub}}|,\;|V_{_{cb}}|)$.
Computations without these approximations were given in
Ref.\cite{Pilaftsis5}. The running quark masses are evaluated by
\begin{eqnarray}
&&m_{_t}(Q)=U_{_6}(Q, m_{_t})\cdot
m_{_t}(m_{_t})\;,\nonumber\\
&&m_{_b}(Q)=U_{_6}(Q, m_{_t})\cdot U_{_5}(m_{_t},m_{_b})\cdot
m_{_b}(m_{_b}),\label{runmass}
\end{eqnarray}
where we have assume that there are no other colored particles
with masses between $Q$ and $m_{_t}$. The evolution factor
$U_{_f}$ reads
\begin{eqnarray}
&&U_{_f}(Q_2,Q_1)=\Big({\alpha_{_s}(Q_2)\over
\alpha_{_s}(Q_1)}\Big)^{d_{_f}}\Big[1+{\alpha_{_s}(Q_1)-\alpha_{_s}(Q_2)
\over 4\pi}J_{_f}\Big]\;,\nonumber\\
&&d_{_f}={12\over 33-2f}\;,\nonumber\\
&&J_{_f}=-{8982-504f+40f^2\over 3(33-f)^2}\;, \label{evaluation}
\end{eqnarray}
where, $f$ is the number of active quark flavors. With those
preparations given above, we can discuss the rare $B$ decays in the
CP violating MSSM.

\section{The effective Hamilton and decay width for rare $B$ decays}

\subsection{The effective Hamiltonian}

\indent\indent The processes  which we are interested in, are
$\bar{B}_{_s}\rightarrow l^+l^-$ and $\bar{B}\rightarrow Kl^+l^-$,
both of them originate from the  transition $b\rightarrow s$.
Integrating out the heavy degrees of freedom  in the full theory,
an effective Hamiltonian is obtained \cite{Buras2}:
\begin{eqnarray}
&&H_{_{eff}}=-{4G_{_F}\over\sqrt{2}}V_{_{ts}}^{*}V_{_{tb}}
\bigg\{\sum\limits_{i=1}^{10}C_{i}(\mu){\cal
O}_{i}(\mu)+\sum\limits_{i=9}^{10}C_i^\prime{\cal O
}_i^\prime+C_{_S}{\cal O}_{_S}+C_{_P}{\cal O}_{_P}
+C_{_S}^\prime{\cal O}_{_S}^\prime+C_{_P}^\prime{\cal O
}_{_P}^\prime\bigg\}\;,\label{Ham}
\end{eqnarray}
with $C_2(m_{_{\rm w}})=\Big(\xi_{_{\rm
CKM}}^{ts}\Big)^*\xi_{_{\rm CKM}}^{tb}$, $V_{_{ij}}$ represent the
physical CKM entries, and $\xi_{_{\rm CKM}}^{ti}$ is the
$\tan\beta-$enhanced radiative corrections to the effective CKM
entries. The operators in the effective Hamiltonian are:
\begin{eqnarray}
&&{\cal O}_1=\Big(\bar{s}_\alpha\gamma_\mu\omega_-c_\beta\Big)
\Big(\bar{c}_\beta\gamma^\mu\omega_-b_\alpha\Big)\;,\nonumber\\
&&{\cal O}_2=\Big(\bar{s}_\alpha\gamma_\mu\omega_-c_\alpha\Big)
\Big(\bar{c}_\beta\gamma^\mu\omega_-b_\beta\Big)\;,\nonumber\\
&&{\cal
O}_3=\Big(\bar{s}_\alpha\gamma_\mu\omega_-c_\alpha\Big)\sum\limits_{q=u,d,s,c,b}
\Big(\bar{q}_\beta\gamma^\mu\omega_-q_\beta\Big)\;,\nonumber\\
&&{\cal
O}_4=\Big(\bar{s}_\alpha\gamma_\mu\omega_-c_\beta\Big)\sum\limits_{q=u,d,s,c,b}
\Big(\bar{q}_\beta\gamma^\mu\omega_-q_\alpha\Big)\;,\nonumber\\
&&{\cal
O}_5=\Big(\bar{s}_\alpha\gamma_\mu\omega_-c_\alpha\Big)\sum\limits_{q=u,d,s,c,b}
\Big(\bar{q}_\beta\gamma^\mu\omega_+q_\beta\Big)\;,\nonumber\\
&&{\cal
O}_6=\Big(\bar{s}_\alpha\gamma_\mu\omega_-c_\beta\Big)\sum\limits_{q=u,d,s,c,b}
\Big(\bar{q}_\beta\gamma^\mu\omega_+q_\alpha\Big)\;,\nonumber\\
&&{\cal O}_7={em_{_b}\over (4\pi)^2}\bar{s}_\alpha F\cdot
\sigma\omega_+b_\alpha\;,
\nonumber\\
&&{\cal O}_8={em_{_b}\over (4\pi)^2}\bar{s}_\alpha
T_{\alpha\beta}^aG^a\cdot
\sigma\omega_+b_\beta\;,\nonumber\\
&&{\cal O}_9={e^2\over(4\pi)^2}\Big(\bar{s}_\alpha\gamma_\mu
\omega_-b_\alpha\Big)\Big(\bar l\gamma^\mu l\Big)\;,\nonumber\\
&&{\cal O}_{10}={e^2\over(4\pi)^2}\Big(\bar{s}_\alpha\gamma_\mu
\omega_-b_\alpha\Big)\Big(\bar l\gamma^\mu\gamma_5 l\Big)\;,\nonumber\\
&&{\cal O}_{9}^\prime={e^2\over(4\pi)^2}\Big(\bar{s}_\alpha\gamma_\mu
\omega_+b_\alpha\Big)\Big(\bar l\gamma^\mu l\Big)\;,\nonumber\\
&&{\cal O}_{10}^\prime={e^2\over(4\pi)^2}\Big(\bar{s}_\alpha\gamma_\mu
\omega_+b_\alpha\Big)\Big(\bar l\gamma^\mu\gamma_5 l\Big)\;,\nonumber\\
&&{\cal O}_{_S}={e^2\over(4\pi)^2}\Big(\bar{s}_\alpha
\omega_+b_\alpha\Big)\Big(\bar l l\Big)\;,\nonumber\\
&&{\cal O}_{_P}={e^2\over(4\pi)^2}\Big(\bar{s}_\alpha
\omega_+b_\alpha\Big)\Big(\bar l\gamma_5 l\Big)\;,\nonumber\\
&&{\cal O}_{_S}^\prime={e^2\over(4\pi)^2}\Big(\bar{s}_\alpha
\omega_-b_\alpha\Big)\Big(\bar l l\Big)\;,\nonumber\\
&&{\cal O}_{_P}^\prime={e^2\over(4\pi)^2}\Big(\bar{s}_\alpha
\omega_-b_\alpha\Big)\Big(\bar l\gamma_5 l\Big)\;.
\nonumber\\
\label{operator}
\end{eqnarray}

\begin{figure}
\setlength{\unitlength}{1mm}
\begin{center}
\begin{picture}(100,150)(55,90)
\put(20,80){\includegraphics{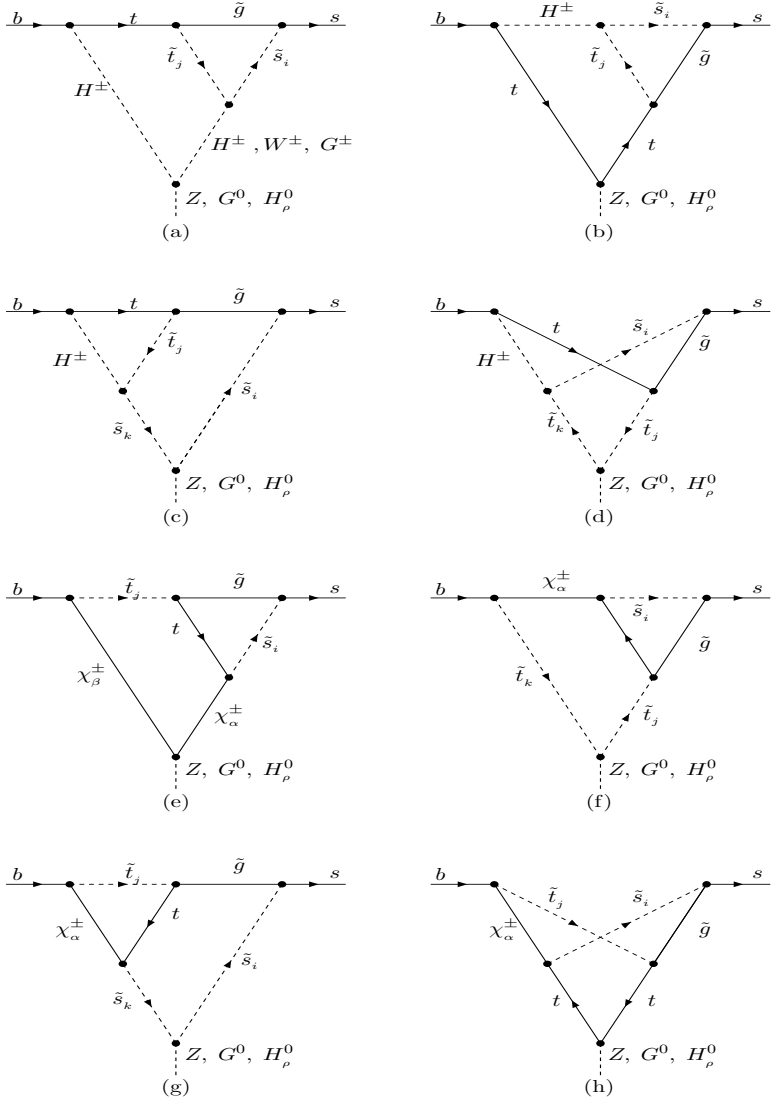}}
\end{picture}
\caption[]{The corrections of two-loop diagrams involving gluino
(later we abbreviate them as "the two loop gluino corrections") to
the penguin vertices $\bar{s}bZ,\;\bar{s}bH$ at large
$\tan\beta$.} \label{fig1}
\end{center}
\end{figure}

In our calculations, we adopt the Feynman rules in the 't
Hooft-Feynman gauge with $\xi=1$. Within the framework of the MFV
CP violating MSSM, the one loop Feynman diagrams that contribute
to the effective Hamiltonian (Eq.\ref{Ham}) can be found in the
literature.  For example, the one-loop $\gamma-,\;Z-$ penguin
diagrams are presented in Ref.\cite{Fengt} and the Higgs penguin
and box diagrams are given in Ref.\cite{Bobeth}. At the
electro-weak scale, the one loop Wilson coefficients in
Eq.\ref{Ham} are divided into several pieces:
\begin{eqnarray}
&&C_i(\mu_{_{\rm W}})=C_i^{\gamma}(\mu_{_{\rm W}})+C_i^{Z}(\mu_{_{\rm W}})
+C_i^{box}(\mu_{_{\rm W}}),\;\;(i=9,10)\nonumber\\
&&C_i^\prime(\mu_{_{\rm W}})=C_i^{{\prime\gamma}}(\mu_{_{\rm W}})
+C_i^{{\prime Z}}(\mu_{_{\rm W}})
+C_i^{{\prime box}}(\mu_{_{\rm W}}),\;\;(i=9,10)\nonumber\\
&&C_i(\mu_{_{\rm W}})=C_i^{H_i^0}(\mu_{_{\rm W}})+C_i^{count}(\mu_{_{\rm W}})
+C_i^{box}(\mu_{_{\rm W}})+C_i^{{resum}}(\mu_{_{\rm W}}),
\;\;(i=S,P)\nonumber\\
&&C_i^\prime(\mu_{_{\rm W}})=C_i^{{\prime H_i^0}}(\mu_{_{\rm W}})
+C_i^{{\prime count}}(\mu_{_{\rm W}})
+C_i^{{\prime box}}(\mu_{_{\rm W}}),\;\;(i=S,P).
\label{wil1}
\end{eqnarray}
Here, we collect the Wilson coefficients corresponding to the one
loop $\gamma-,\;Z-,\;H_{k}^0-$ penguin and box contributions in
the appendix \ref{1loop}. In order to include the threshold
radiative corrections of gluinos, we have replaced the tree-level
vertices by the corresponding interactions which are presented in
Eq. (\ref{modhtb}). Similarly, we can obtain the contributions
which originate from a resummation of high order threshold effects
on the Wilson coefficients of the operators involving down-type
quarks:
\begin{eqnarray}
&&C_{_S}^{^{\rm resum}}(\mu_{_{\rm W
}})={\sqrt{x_{_b}x_{_l}}\over2
s_{_\beta}c_{_\beta}^2x_{_{H_k^0}}}\chi_{_{FC}}^*{\cal
Z}_{_H}^{2k}\Big(c_{_\beta}{\cal Z}_{_H}^{3k}-s_{_\beta}{\cal
Z}_{_H}^{2k}+i{\cal Z}_{_H}^{1k}\Big)\nonumber\\
&&C_{_P}^{^{\rm resum}}(\mu_{_{\rm W
}})=-i{\sqrt{x_{_b}x_{_l}}\over2
c_{_\beta}^2x_{_{H_k^0}}}\chi_{_{FC}}^*{\cal
Z}_{_H}^{1k}\Big(c_{_\beta}{\cal Z}_{_H}^{3k}-s_{_\beta}{\cal
Z}_{_H}^{2k}+i{\cal Z}_{_H}^{1k}\Big)\;,
\end{eqnarray}
with $c_{_\beta}=\cos\beta,\;s_{_\beta}=\sin\beta,$ and $x_i={m_i^2
\over m_{_{\rm W}}^2}$.

As we mentioned in the introduction, there are obvious differences
between the results from exact two-loop calculations and that from
threshold radiative approximation which is derived in terms of the
heavy mass expanding method. In the large $\tan\beta$ scenario,
corrections from the two-loop Feynman diagrams including  gluino
to the penguin vertices $\bar{s}bZ,\;\bar{s}bH$ are drawn in
Fig.\ref{fig1}. Correspondingly, the corrections to the Wilson
coefficients from the two-loop penguin diagrams are written as
\begin{eqnarray}
&&\delta\Big(C_{_9}\Big)_{_{2P}}={\alpha_{_s}\over8\pi
s_{_{\rm w}}^2}{m_{_b}t_{_\beta}\over m_{_{\rm w}}}
\Big(1-4s_{_{\rm w}}^2\Big){\cal P}_{_Z}\;,\nonumber\\
&&\delta\Big(C_{_{10}}\Big)_{_{2P}}=-{\alpha_{_s}\over8\pi
s_{_{\rm w}}^2}{m_{_b}t_{_\beta}\over m_{_{\rm w}}}{\cal P}_{_Z}\;,\nonumber\\
&&\delta\Big(C_{_{S}}\Big)_{_{2P}}={\alpha_{_s}\over4\pi
s_{_{\rm w}}^2}{m_{_b}m_{_{l^I}}t_{_\beta}^2\over m_{_{\rm w}}^2}
\sum\limits_{\rho=1}^3\Big({\cal Z}_{_H}\Big)_{_{2\rho}}
{1\over s_{_\beta}x_{_{H^\rho}}}{\cal P}_{_{H^{^\rho}}}
\;,\nonumber\\
&&\delta\Big(C_{_{P}}\Big)_{_{2P}}={\alpha_{_s}\over4\pi
s_{_{\rm w}}^2c_{_{\rm w}}^2}{m_{_b}m_{_{l^I}}t_{_\beta}\over m_{_{\rm w}}^2}
{\cal P}_{_G}-i{\alpha_{_s}\over4\pi
s_{_{\rm w}}^2}{m_{_b}m_{_{l^I}}t_{_\beta}^2\over m_{_{\rm w}}^2}
\sum\limits_{\rho=1}^3\Big({\cal Z}_{_H}\Big)_{_{1\rho}}
{1\over x_{_{H^\rho}}}{\cal P}_{_{H^{^\rho}}}\;,\nonumber\\
\label{2l-peng}
\end{eqnarray}
with ${\cal P}_{_{Z,G}}=\sum\limits_{i=1}^8{\cal P}_{_{Z,G}}^{(i)}$,
${\cal P}_{_{H^{^\rho}}}=\sum\limits_{i=1}^{10}{\cal
P}_{_{H^{^\rho}}}^{(i)}$. We list the expression of those nonzero
form factors ${\cal P}_{_{Z,G}}^{(i)}, \;{\cal
P}_{_{H^{^\rho}}}^{(i)}$ in the appendix \ref{2t} in order to
shorten the length of text.

\begin{figure}
\setlength{\unitlength}{1mm}
\begin{center}
\begin{picture}(100,150)(55,90)
\put(20,80){\includegraphics{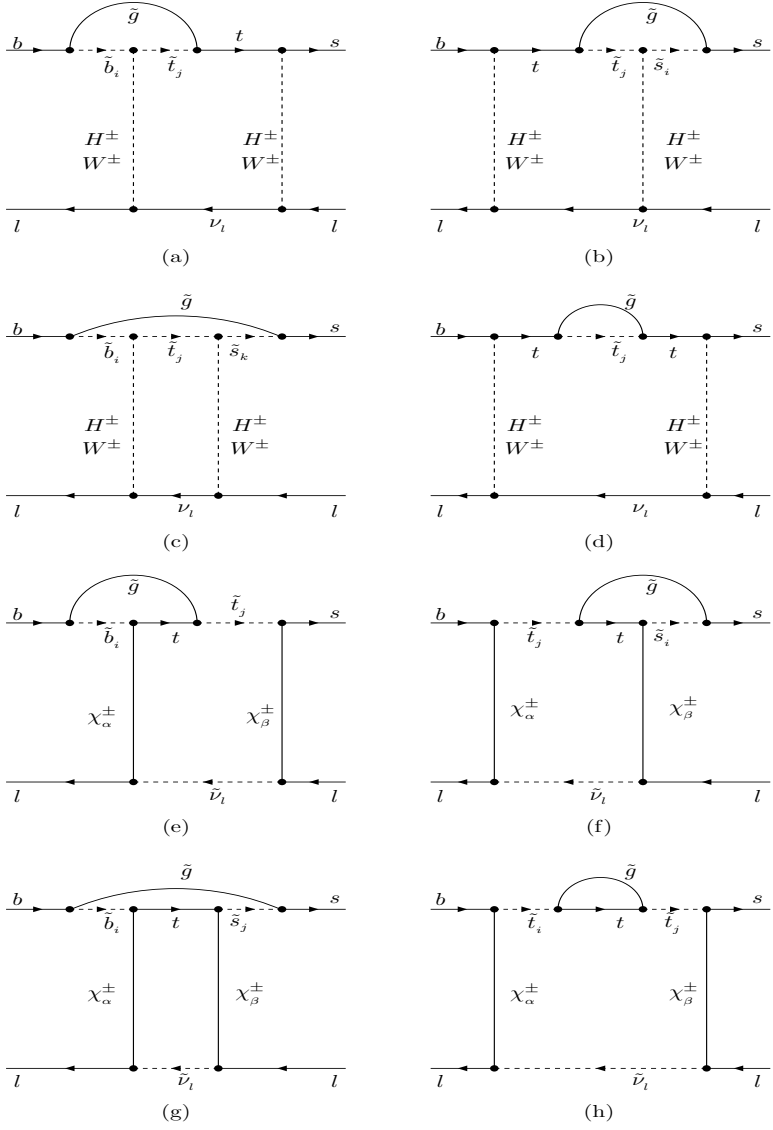}}
\end{picture}
\caption[]{The two loop gluino corrections to the four fermion interactions
$\bar{s}b\bar{l}l$ at large $\tan\beta$.}
\label{fig2}
\end{center}
\end{figure}

We should also include the two loop gluino correction to the four
fermion interactions $\bar{s}b\bar{l}l$ for completeness. The
Feynman diagrams are plotted in Fig.\ref{fig2}. Correspondingly,
the corrections to the Wilson coefficients from those two loop box
diagrams are formulated as
\begin{eqnarray}
&&\delta\Big(C_{_9}\Big)_{_{2B}}=-{2\alpha_{_s}\over3\pi
s_{_{\rm w}}^2}t_{_\beta}{\cal B}_{_V}\;,\nonumber\\
&&\delta\Big(C_{_{10}}\Big)_{_{2B}}=-{2\alpha_{_s}\over3\pi
s_{_{\rm w}}^2}t_{_\beta}{\cal B}_{_A}\;,\nonumber\\
&&\delta\Big(C_{_9}^\prime\Big)_{_{2B}}=-{2\alpha_{_s}\over3\pi
s_{_{\rm w}}^2}t_{_\beta}{\cal B}^\prime_{_V}\;,\nonumber\\
&&\delta\Big(C_{_{10}}^\prime\Big)_{_{2B}}=-{2\alpha_{_s}\over3\pi
s_{_{\rm w}}^2}t_{_\beta}{\cal B}^\prime_{_A}\;,\nonumber\\
&&\delta\Big(C_{_{S}}\Big)_{_{2B}}=-{2\alpha_{_s}\over3\pi
s_{_{\rm w}}^2}t_{_\beta}{\cal B}_{_{S}}
\;,\nonumber\\
&&\delta\Big(C_{_{P}}\Big)_{_{2B}}=-{2\alpha_{_s}\over3\pi
s_{_{\rm w}}^2}t_{_\beta}{\cal B}_{_{P}}\;,\nonumber\\
&&\delta\Big(C_{_{S}}^\prime\Big)_{_{2B}}=-{2\alpha_{_s}\over3\pi
s_{_{\rm w}}^2}t_{_\beta}{\cal B}^\prime_{_{S}}
\;,\nonumber\\
&&\delta\Big(C_{_{P}}^\prime\Big)_{_{2B}}=-{2\alpha_{_s}\over3\pi
s_{_{\rm w}}^2}t_{_\beta}{\cal B}^\prime_{_{P}}\;,
\label{2l-box}
\end{eqnarray}
with
\begin{eqnarray}
&&{\cal B}^\prime_{_{V,A}}={\cal B}^{\prime(1)}_{_{V,A}}
+{\cal B}^{\prime(2)}_{_{V,A}}\;,\nonumber\\
&&{\cal B}_{_{S,P}}=\sum\limits_{i=1}^9{\cal B}_{_{S,P}}^{(i)}
\;,\nonumber\\
&&{\cal B}^\prime_{_{S,P}}=\sum\limits_{i=1}^6{\cal B}_{_{S,P}}^{\prime(i)}\;.
\label{2l-bform}
\end{eqnarray}
The concrete expressions of those nonzero form factors ${\cal
B}^{\prime(i)} _{_{V,A}},\;{\cal B}_{_{S,P}}^{(i)},\;{\cal
B}_{_{S,P}}^{\prime(i)}$ can be found in the appendix \ref{2b}. In
Eq. \ref{2l-peng} and Eq. \ref{2l-box}, we adopt the
$\overline{MS}$ scheme to remove the UV-divergences which are
caused by the divergent sub-diagrams \cite{Feng3}.

When the effective Lagrangian is applied to the hadronic processes
whose characteristic energy scale is about $m_b$, we should evolve
those Wilson coefficients from the weak scale down to the hadronic
scale. The running depends on the anomalous dimension matrix of the
concerned operators\cite{Buras3}. The Wilson coefficients obtained
at the weak scale are regarded as the initial conditions for the
renormalization group equations (RGEs). Up to the leading order
(LO), the Wilson coefficients at hadronic scale are given
as\cite{Grinstein,Buras4,Grigjanis}
\begin{eqnarray}
&&C_7(m_{_b})=\eta^{-{16\over23}}\Big[C_7(m_{_{\rm
W}})+{8\over3}\Big( C_8(m_{_{\rm W}})-{2021\over468}C_2(m_{_{\rm
W}})\Big)\Big(\eta^{2\over23}-1\Big) \nonumber\\
&&\hspace{1.8cm}+{389\over540}C_2(m_{_{\rm
W}})\Big(\eta^{10\over23}-1\Big)+{107\over702}C_2(m_{_{\rm W}})
\Big(\eta^{28\over23}-1\Big)\Big]\;, \nonumber\\
&&C_7^\prime(m_{_b})=\eta^{-{16\over23}}\Big[C_7^\prime(m_{_{\rm
W}})+{8\over3}C_8^\prime(m_{_{\rm W}})\Big(\eta^{2\over23}-1\Big)
\Big]\;, \nonumber\\
&&C_9(m_{_b})=C_9(m_{_{\rm W}})+{4\pi\over\alpha_{_s}(m_{_{\rm W
}})}\Big[-{4\over33}(1-\eta^{11\over23})+{8\over87}(1-\eta^{-{29\over23
}})\Big]C_2(m_{_{\rm W}})\;,\nonumber\\
&&C_{10}(m_{_b})=C_{10}(m_{_{\rm W}})\;, \label{hadwil}
\end{eqnarray}
with $\eta={\alpha_{_s}(m_{_b})\over\alpha_{_s}(m_{_{\rm W}})}$. The
most stringent constraint on the supersymmetry parameter space comes
from the rare decay of $B-$meson: $B\rightarrow X_{_s}\gamma$
\cite{CLEO}. The theoretical prediction on the branching ratio of
the inclusive process $B\rightarrow X_{_s}\gamma$ is given as
\cite{Grigjanis,Kagan}
\begin{eqnarray}
&&BR(B\rightarrow X_{_s}\gamma)={\Gamma(B\rightarrow
X_{_s}\gamma)\over\Gamma(B\rightarrow
X_{_c}e\bar{\nu}_{_e})}BR(B\rightarrow
X_{_c}e\bar{\nu}_{_e})\nonumber\\
&&\hspace{2.6cm}={|V_{_{ts}}^{*}V_{_{tb}}|^2\over|V_{_{cb}}|^2}{6\alpha_{_{em}}
|C_{_7}(m_{_b})|^2\over\pi\rho({m_{_c}\over
m_{_b}})\Big(1-{2\alpha_{_s}(m_{_b})\over3\pi}f({m_{_c}\over
m_{_b}})\Big)}BR(B\rightarrow X_{_c}e\bar{\nu}_{_e})\;,
\label{brbsg}
\end{eqnarray}
where $\alpha_{_{em}}$ is the $QED$ fine structure constant and the
phase factor is $$\rho({m_{_c}\over m_{_b}})=1-8\Big({m_{_c}\over
m_{_b}}\Big)^2+8\Big({m_{_c}\over m_{_b}}\Big)^6+\Big({m_{_c}\over
m_{_b}}\Big)^8-24\Big({m_{_c}\over m_{_b}}\Big)^4\ln {m_{_c}\over
m_{_b}}\;,$$ and the one-loop $QCD$ correction to the semileptonic
decay gives $f({m_{_c}\over m_{_b}})\simeq2.4$ \cite{Cabibbo}. When
we calculate the branching ratios of other rare processes, the
branching ratio of $B\rightarrow X_{_s}\gamma$ which is
experimentally measured with relatively high accuracy, must be
considered as a prior constraint.

In the SM, the CP asymmetry of $B\rightarrow X_{_s}\gamma$ process
\begin{eqnarray}
&&A_{_{CP}}(B\rightarrow X_{_s}\gamma)={\Gamma(\bar{B}\rightarrow
X_{_{\bar s}}\gamma)-\Gamma(B\rightarrow X_{_s}\gamma)\over
\Gamma(\bar{B}\rightarrow X_{_{\bar s}}\gamma)
+\Gamma(B\rightarrow X_{_s}\gamma)}
\label{eq9}
\end{eqnarray}
is calculated to be rather small: $A_{_{CP}}\sim 0.5\%$
\cite{Kagan}. By the recent experimental measurement \cite{Coan}
of the CP asymmetry, we have
\begin{eqnarray}
&&-0.30\le A_{_{CP}}(B\rightarrow X_{_s}\gamma)\le 0.14
\label{eq10}
\end{eqnarray}
at 95\% C.L. In other word, the studies of the direct CP asymmetry
in $B\rightarrow X_{_s}\gamma$ may uncover new sources of CP
violation which lie outside the SM. Applying Eq. \ref{brbsg}, the
CP asymmetry can be written as
\begin{eqnarray}
&&A_{_{CP}}(B\rightarrow X_{_s}\gamma)={\alpha_{_s}(\mu_{_b})\over
|C_{_7}(\mu_{_b})|^2}\bigg\{\bigg[{40\over81}-{8z\over9}\Big(\upsilon(z)
+b(z,\delta)\Big)\Big(1+{V_{_{us}}^*V_{_{ub}}\over
V_{_{ts}}^*V_{_{tb}}}\Big)\bigg]{\bf Im}[C_{_2}(\mu_{_b})C_{_7}^*(\mu_{_b})]
\nonumber\\&&\hspace{3.2cm}
-{4\over9}{\bf Im}[C_{_8}(\mu_{_b})C_{_7}^*(\mu_{_b})]
+{8z\over27}b(z,\delta){\bf Im}\Big[\Big(1+{V_{_{us}}^*V_{_{ub}}\over
V_{_{ts}}^*V_{_{tb}}}\Big)C_{_2}(\mu_{_b})C_{_8}^*(\mu_{_b})\Big]\bigg\}\;,
\label{eq12}
\end{eqnarray}
where $z=(m_{_c}/m_{_b})^2$, $\upsilon(z)$ and $b(z,\delta)$
can be found in \cite{Kagan1}.

\subsection{The decay width for two rare $B$ processes}

\subsubsection{$\bar{B}_{_s}\rightarrow l^+l^-\;(l=\mu,\;\tau)$}

\indent\indent The decay constant of the pseudoscalar meson
$\bar{B}_{_s}$ is defined as \cite{Skiba}:
\begin{eqnarray}
\langle 0|\bar{s}\gamma_\mu\gamma_5b|\bar{B}_{_s}(P)\rangle=ip_\mu
f_{_{B_s}}\;. \label{hm1}
\end{eqnarray}
With the equation of motion for quark fields, Eq. (\ref{hm1}), one
can  write
\begin{eqnarray}
&&\langle0|\bar{s}\gamma_5b|\bar{B}_{_s}(P)\rangle=if_{_{B_s}}{m_{_{B_s}}^2\over
m_{_b}+m_{_s}}\;.\label{hm2}
\end{eqnarray}
Correspondingly, we derive the branching ratio as
\begin{eqnarray}
&&BR(\bar{B}_{_s}\rightarrow l^+l^-)={\alpha_{_{em}}^2G_{_F}^2
|V_{_{ts}}^{*}V_{_{tb}}|^2\over16\pi^3}m_{_{B_s}}\tau_{_{B_s}}
\sqrt{1-4m_{_l}^2/m_{_{B_s}}^2}\bigg\{\Big(1-{4m_{_l}^2\over
m_{_{B_s}}^2}\Big)|F_{_S}|^2
\nonumber\\
&&\hspace{3.5cm} +|F_{_P}+2m_{_l}F_{_A}|^2\bigg\}\;, \label{br1}
\end{eqnarray}
with $m_{_{B_s}}$ and $\tau_{_{B_s}}$ denote mass and life time
of the meson $B_{_s}$ respectively, and
\begin{eqnarray}
&&F_{_S}=-{i\over2}m_{_{B_s}}^2f_{_{B_s}}\bigg\{{C_{_S}m_{_b}-C_{_S}^\prime
m_{_s}\over m_{_b}+m_{_s}}\bigg\}\;,\nonumber\\
&&F_{_P}=-{i\over2}m_{_{B_s}}^2f_{_{B_s}}\bigg\{{C_{_P}m_{_b}-C_{_P}^\prime
m_{_s}\over m_{_b}+m_{_s}}\bigg\}\;,\nonumber\\
&&F_{_A}=-{i\over2}f_{_{B_s}}\Big(C_{10}-C_{10}^\prime\Big)\;.
\label{factor2}
\end{eqnarray}
A point should be noted  that there a  CP asymmetry is observable
in this process
\begin{eqnarray}
&&A_{_{CP}}={\Gamma(B_{_s}\rightarrow \bar ll)
-\Gamma(\bar{B}_{_s}\rightarrow \bar ll)\over
\Gamma(B_{_s}\rightarrow \bar ll) +\Gamma(\bar{B}_{_s}\rightarrow
\bar ll)}
\nonumber\\
&&\hspace{1.5cm} ={2{\bf
Im}(\xi_{_{CP}})X_{_s}\over(1+|\xi_{_{CP}}|^2)(1+X_{_s}^2)}
\;,\label{cpbsdl}
\end{eqnarray}
and as indicated in the formula, it is induced by the mixing of
$B_{_s}$ and $\bar{B_{_s}}$ \cite{Huangcs}. Here,
\begin{eqnarray}
&&X_{_s}=\Delta m_{_{B_s}}/\Gamma_{_{B_s}}\;,\nonumber\\
&&\xi_{_{CP}}={V_{_{ts}}^*V_{_{tb}}(C_{_S}\sqrt{1-4m_{_l}^2/m_{_{B_s}}^2}
+C_{_P}+2m_{_l}C_{_{10}}/m_{_{B_s}})\over V_{_{ts}}V_{_{tb}}^*
(C_{_S}^*\sqrt{1-4m_{_l}^2/m_{_{B_s}}^2}-C_{_P}^*-2m_{_l}C_{_{10}}^*/m_{_{B_s}})}
\;,\label{cpbsdl1}
\end{eqnarray}
and $\Delta m_{_{B_s}}$ is the mass difference in $\bar{B}_{_s}
-B_{_s}$ mixing, $\Gamma_{_{B_s}}$ denotes decay width of the meson
$B_{_s}$.

\subsubsection{$\bar{B}\rightarrow Kl^+l^-\;(l=\mu,\;\tau)$}

\indent\indent According to Ref.\cite{Wirbel}, the nonzero
hadronic matrix elements for the exclusive decay
$\bar{B}\rightarrow Kl^+l^-$ are written as
\begin{eqnarray}
&&\langle K(k)|\bar{s}\gamma_\mu
b|\bar{B}(p)\rangle=f_+(q^2)(2p-q)_\mu
+\Big(f_0(q^2)-f_+(q^2)\Big){m_{_B}^2-m_{_K}^2\over
q^2}q_\mu\;,\nonumber\\
&&\langle K(k)|\bar{s}i\sigma_{\mu\nu}q^\nu
b|\bar{B}(p)\rangle=-{f_{_T}(q^2)\over
m_{_B}+m_{_K}}\Big(q^2(2p-q)_\mu-(m_{_B}^2-m_{_K}^2)q_\mu\Big)
\;,\nonumber\\
&&\langle K(k)|\bar{s}b|\bar{B}(p)\rangle={m_{_B}^2-m_{_K}^2\over
m_{_b}-m_{_s}}f_0(q^2)\;,
\label{hm3}
\end{eqnarray}
where $q^\mu=(p-k)^\mu$ is the four-momentum transferred to the
dilepton system. The resulting form factors are parameterized as
\begin{eqnarray}
&&f_0(s)=f_0(0)\exp\Big[c_1^0{s\over m_{_{B}}^2}+c_2^0({s\over
m_{_{B}}^2})^2 +c_3^0({s\over m_{_{B}}^2})^3\Big]\;,\nonumber\\
&&f_+(s)=f_+(0)\exp\Big[c_1^+{s\over m_{_{B}}^2}+c_2^+({s\over
m_{_{B}}^2})^2 +c_3^+({s\over m_{_{B}}^2})^3\Big]\;,\nonumber\\
&&f_{_T}(s)=f_{_T}(0)\exp\Big[c_1^T{s\over m_{_{B}}^2}+c_2^T({s\over
m_{_{B}}^2})^2 +c_3^T({s\over m_{_{B}}^2})^3\Big]\;.
\label{para}
\end{eqnarray}
With the effective Lagrangian Eq. (\ref{Ham}) and the hadronic
matrix elements Eq. (\ref{hm3}), we can write the transition
matrix elements for $\bar{B}\rightarrow Kl^+l^-$ as following
\begin{eqnarray}
&&{\cal M}=F_{_S}\bar{l}l+F_{_P}\bar{l}\gamma_5l+F_{_V}p^\mu\bar{l}\gamma_\mu
l+F_{_A}p^\mu\bar{l}\gamma_\mu\gamma_5l\;,
\label{rme}
\end{eqnarray}
where $p_\mu$ denotes the four-momentum of the initial $B$ meson,
and the form factors $F_i$ are Lorentz-invariant.  Following Ref.
\cite{Bobeth}, we define  $\theta$ as the angle between the
three-momenta ${\bf p}_{_l}$ and ${\bf p}_{_s}$ in the
center-of-mass frame of the dilepton. The energy-angular
distribution of the decay products is
\begin{eqnarray}
&&{d\Gamma(\bar{B}\rightarrow Kl^+l^-)\over
ds\;d\cos\theta}={\alpha_{_em}^2G_{_F}^2
|V_{_{ts}}^{*}V_{_{tb}}|^2\over2^{9}\pi^5m_{_B}^3}
\lambda^{1\over2}(m_{_B}^2,m_{_K}^2,s)\beta_{_l}\Big\{s\Big[\beta_{_l}^2
|F_{_S}|^2+|F_{_P}|^2\Big]
\nonumber\\
&&\hspace{3.5cm}
+{1\over4}\lambda(m_{_B}^2,m_{_K}^2,s)\Big[1-\beta_{_l}^2\cos^2\theta\Big]
\Big[|F_{_V}|^2+|F_{_A}|^2\Big]
\nonumber\\
&&\hspace{3.5cm}
+4m_{_l}^2m_{_B}^2|F_{_A}|^2
+2m_{_l}\Big[\lambda^{1\over2}(m_{_B}^2,m_{_K}^2,s)\beta_{_l}
{\bf Re}(F_{_S}F_{_V}^*)\cos\theta
\nonumber\\
&&\hspace{3.5cm}
+(m_{_B}^2-m_{_K}^2+s){\bf Re}(F_{_P}F_{_A}^*)\Big]\Big\}\;,
\label{dw1}
\end{eqnarray}
with
$s=(p_{_{l^+}}+p_{_{l^-}})^2\;,\;\;\beta_{_l}=\sqrt{1-{4m_{_l}^2\over
s}}$ and $\lambda(a,b,c)=a^2+b^2+c^2-2(ab+bc+ca)$. The kinematic
quantities $s,\;\cos\theta$ have natural bounds
\begin{equation}
4m_{_l}^2\le s\le(m_{_B}-m_{_K})^2,\;\;-1\le\cos\theta\le1\;.
\end{equation}
A particularly interesting quantity is the forward-backward
asymmetry
\begin{eqnarray}
&&A_{_{FB}}(s)=\frac{\int_{_0}^{^1}d\cos\theta{d\Gamma\over ds\;d\cos\theta}
-\int_{_{-1}}^{^0}d\cos\theta{d\Gamma\over ds\;d\cos\theta}}{\int_{_0}^{^1}d\cos
\theta{d\Gamma\over ds\;d\cos\theta}+\int_{_{-1}}^{^0}d\cos\theta{d\Gamma
\over ds\;d\cos\theta}}
\nonumber\\
&&\hspace{1.3cm}
={\alpha_{_em}^2G_{_F}^2|V_{_{ts}}^{*}V_{_{tb}}|^2\over2^{8}\pi^5m_{_B}^3}
m_{_l}\lambda(m_{_B}^2,m_{_K}^2,s)\beta_{_l}^2
{\bf Re}(F_{_S}F_{_V}^*)}/{{d\Gamma\over ds}\;,
\label{afb1}
\end{eqnarray}
where the dilepton invariant mass spectrum ${d\Gamma\over ds}$
is given as
\begin{eqnarray}
&&{d\Gamma(\bar{B}\rightarrow Kl^+l^-)\over
ds}={\alpha_{_em}^2G_{_F}^2
|V_{_{ts}}^{*}V_{_{tb}}|^2\over2^{8}\pi^5m_{_B}^3}
\lambda^{1\over2}(m_{_B}^2,m_{_K}^2,s)\beta_{_l}
\Big\{s\Big[\beta_{_l}^2|F_{_S}|^2+|F_{_P}|^2\Big]
\nonumber\\
&&\hspace{3.5cm}
+{1\over6}\lambda(m_{_B}^2,m_{_K}^2,s)\Big[1+{2m_{_l}^2\over
s}\Big]\Big[|F_{_V}|^2+|F_{_A}|^2\Big]
\nonumber\\
&&\hspace{3.5cm}
+4m_{_l}^2m_{_B}^2|F_{_A}|^2
+2m_{_l}(m_{_B}^2-m_{_K}^2+s){\bf Re}(F_{_P}F_{_A}^*)\Big\}\;.
\label{dw2}
\end{eqnarray}

The form factors are formulated as
\begin{eqnarray}
&&F_{_S}={1\over2}(m_{_{B}}^2-m_{_{K}}^2)f_0(s)\bigg\{{C_{_S}m_{_b}
+C_{_S}^\prime m_{_s}\over m_{_b}-m_{_s}}\bigg\}\;,\nonumber\\
&&F_{_P}=-m_{_l}C_{10}\bigg\{f_+(s)-{m_{_{B}}^2-m_{_{K}}^2\over s}
\Big[f_0(s)-f_+(s)\Big]\bigg\}
\nonumber\\
&&\hspace{1.0cm}
+{1\over2}(m_{_{B}}^2-m_{_{K}}^2)f_0(s)
\bigg\{{C_{_S}m_{_b}+C_{_S}^\prime m_{_s}\over
m_{_b}-m_{_s}}\bigg\}
\;,\nonumber\\
&&F_{_A}=C_{10}f_+(s)\;,\nonumber\\
&&F_{_V}=\bigg\{C_{9}f_+(s)+2C_{_7}m_{_b}{f_{_T}(s)\over m_{_B}
+m_{_K}}\bigg\}\;.
\end{eqnarray}
Our main interest is in the average forward-backward asymmetry
$<A_{_{FB}}>$, which can be achieved from Eq.\ref{afb1} by
integrating out the numerator and denominator separately over the
dilepton invariant mass-square $s$. Basing on the preparations
made above, we can numerically analyze the effects of various CP
phases on the rare $B$ decays in  next section.

\section{Numerical results}

\indent\indent In this section, we present our numerical results for
the rates of the rare $B$ decays in the MSSM. As mentioned above,
the most stringent constraint on the parameter space of 'new
physics' beyond the SM is the experimental bound on the branching
ratio of $B\rightarrow X_{_s} \gamma$:
$$2.\times 10^{-4}<BR(B\rightarrow X_{_s}\gamma)
<4.5\times 10^{-4}.$$ In our numerical analysis, we  take this as
a constraint for the parameter space of the CP violating MSSM. The
inputs of the SM sector are \cite{Data} $\alpha_{_{\rm
EW}}=1./128.8,\;m_{_{\rm W}}=80.23{\rm GeV}, \;m_{_{\rm
Z}}=91.18{\rm GeV},\;\alpha_{_s}(m_{_{\rm Z}})=0.117,\;
m_{_t}^{pole}=173.8{\rm GeV}.$ The on-shell running masses of top
and bottom quarks are related to the corresponding pole masses
through \cite{Pierce}

\begin{figure}
\setlength{\unitlength}{1mm}
\begin{center}
\begin{picture}(100,50)(55,90)
\put(50,50){\includegraphics{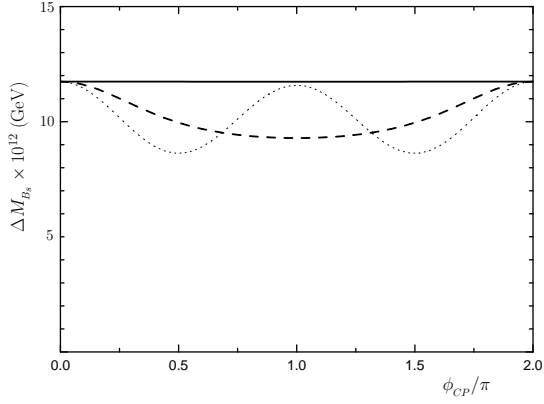}}
\end{picture}
\caption[]{Taking $\tan\beta=20$, and $\mu_{_H}=300\;({\rm GeV})$, $\Delta M_{_{B_s}}$ varies
with $\phi_{_{CP}}$ where (a)solid line stands for $\phi_{_{CP}}=\theta_{_2}$
as well as $\theta_{_3}=\theta_{_t}=0$, (b)dash line stands for $\phi_{_{CP}}=\theta_{_3}$
as well as $\theta_{_2}=\theta_{_t}=0$, and (c)dot line stands for $\phi_{_{CP}}=\theta_{_t}$
as well as $\theta_{_2}=\theta_{_3}=0$, respectively.}
\label{fig3}
\end{center}
\end{figure}

\begin{figure}
\setlength{\unitlength}{1mm}
\begin{center}
\begin{picture}(100,50)(55,90)
\put(50,50){\includegraphics{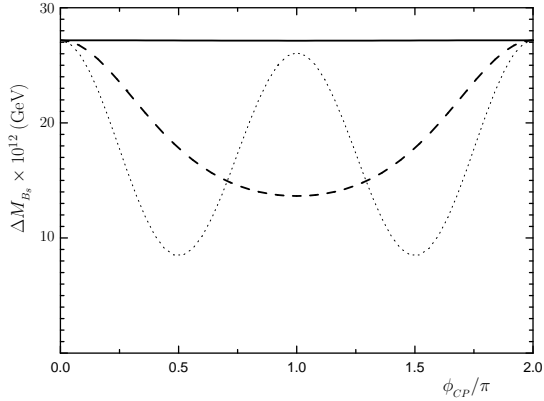}}
\end{picture}
\caption[]{Taking $\tan\beta=50$, and $\mu_{_H}=100\;({\rm GeV})$, $\Delta M_{_{B_s}}$ varies
with $\phi_{_{CP}}$ where (a)solid line stands for $\phi_{_{CP}}=\theta_{_2}$
as well as $\theta_{_3}=\theta_{_t}=0$, (b)dash line stands for $\phi_{_{CP}}=\theta_{_3}$
as well as $\theta_{_2}=\theta_{_t}=0$, and (c)dot line stands for $\phi_{_{CP}}=\theta_{_t}$
as well as $\theta_{_2}=\theta_{_3}=0$, respectively.}
\label{fig4}
\end{center}
\end{figure}

\begin{eqnarray}
&&m_{_t}(m_{_t}^{pole})={m_{_t}^{pole}\over1+5/(3\pi)\alpha_{_s}
(m_{_t}^{pole})}\;,\nonumber\\
&&m_{_b}(m_{_b}^{pole})={m_{_b}^{pole}\over1+5/(3\pi)\alpha_{_s}
(m_{_b}^{pole})}\;.
\label{shell-run}
\end{eqnarray}
For the physical CKM matrix elements, we adopt the Wolfenstein
parametrization and the corresponding parameters are set as
$A=0.85,\;\lambda=0.22,\; \rho=0.21,\;\eta=0.34$ which stand for
the central values permitted by present experiments \cite{Data}.
In the hadronic sector, $m_{_{B_s}}=5.37{\rm GeV},\;
m_{_{B}}=5.28{\rm GeV},\;m_{_{K}}=0.50{\rm GeV},\;$ and
$\tau_{_{B_s}}=1.46\times 10^{-12}s,\; \tau_{_{B}}=1.54\times
10^{-12}s,$ and the decay constants are
$f_{_{B}}=f_{_{B_s}}=0.21{\rm GeV}$.

\begin{figure}
\setlength{\unitlength}{1mm}
\begin{center}
\begin{picture}(80,120)(55,60)
\put(30,50){\includegraphics{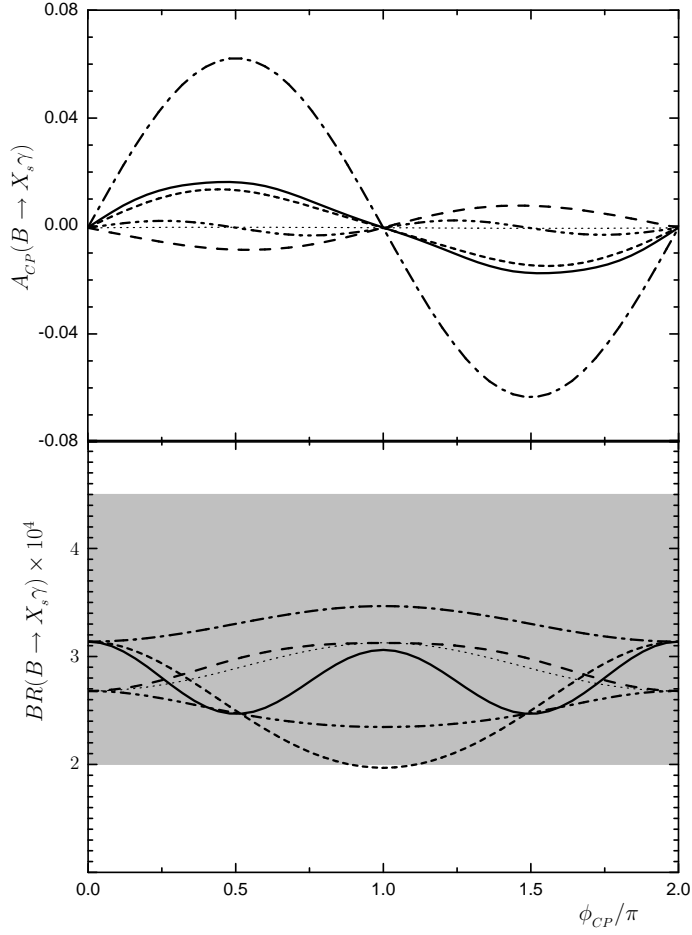}}
\end{picture}
\caption[]{$BR(B\rightarrow X_{_s}\gamma)$ and
$A_{_{CP}}(B\rightarrow X_{_s}\gamma)$ when $m_{_b}^{pole}=4.8{\rm
GeV},\;\tan\beta=20$, as well as
$\mu_{_H}=300\;({\rm GeV})$. In the figure, (a)solid line
represents rigorous two loop analysis at
$\phi_{_{CP}}=\theta_{_2}$ and $\theta_{_3}=\theta_{_t}=0$,
(b)dash line represents one loop results plus threshold radiative
corrections at $\phi_{_{CP}}=\theta_{_2}$ and
$\theta_{_3}=\theta_{_t}=0$, (c)dash-dot line represents rigorous
two loop analysis at $\phi_{_{CP}}=\theta_{_3}$ and
$\theta_{_2}=\theta_{_t}=0$, (d)dot line represents one loop
results plus threshold radiative corrections at
$\phi_{_{CP}}=\theta_{_3}$ and $\theta_{_2}=\theta_{_t}=0$,
(e)short-dash line represents rigorous two loop analysis at
$\phi_{_{CP}}=\theta_{_t}$ and $\theta_{_2}=\theta_{_3}=0$,
(f)dash-dot-dot line represents one loop results plus threshold
radiative corrections at $\phi_{_{CP}}=\theta_{_t}$ and
$\theta_{_2}=\theta_{_3}=0$.} \label{fig5}
\end{center}
\end{figure}

Beside the constraint from the branching ratio of $BR(B\rightarrow
X_{_s} \gamma)$, other strong constraints on the new CP violating
phases originate from the  $\bar{B}_{_s}- B_{_s}$ mixing and
resultant mass difference:
$$\Delta M_{_{B_{_s}}}>9.48\times10^{-12}\;({\rm GeV})$$ as well as
the mass difference in $\bar{B}_{_d}-B_{_d}$ mixing: $$\Delta
M_{_{B_{_d}}} =(3.304\pm0.046)\times10^{-13}\;({\rm GeV}).$$ With
$m_{_{H^\pm}}=300\;({\rm GeV})$ and $\tan\beta\ge10$, the
theoretical estimation of the contributions to $\Delta
M_{_{B_{_d}}}$ from the SM and two Higgs-doublet sectors fits the
experimental bound very well. For the SUSY sector, the sfermion
which contributes to $\Delta M_{_{B_{_s}}}$, may belong to either
the first generation or the third one. All the physical quantities
which we are interested in, do not depend on the first generation
parameters. Namely we are free to make our choice, no matter what
parameters we set for the third generation of sfermions, we can
assume that the new CP phases come from the first generation.
Adding the SUSY contributions to that from SM and charged Higgs
bosons, we still can make the total result satisfying the present
experimental bound. Moreover, for the theoretical prediction on
$\Delta M_{_{B_{_s}}}$, we only consider the contributions of
one-loop box diagrams and the threshold radiative corrections at
large $\tan\beta$ \cite{Pilaftsis5}. Beside those parameters in
the SM and two-Higgs-doublet sectors, the following supersymmetric
parameters should be involved in our calculations
$\mu_{_H},\;m_{_{2,3}},\;
m_{_{\tilde{Q}_{2,3}}},\;m_{_{\tilde{L}_{2,3}}},\;
m_{_{\tilde{t}}},\;m_{_{\tilde{b}}},\;m_{_{\tilde{s}}}$, as well
as the trilinear Yukawa couplings $A_{_{t}},\;
A_{_{b}},\;A_{_{s}}$.
So far, there are no model-independent constraints on the masses
of supersymmetric particle within the framework of CP conservative
MSSM. Based on the following assumptions
\begin{itemize}
\item $\tilde{\chi}_1^0\;(\tilde{\gamma})$ is the lightest
supersymmetric particle, \item except $\tilde{t}$ and $\tilde{b}$,
all scalar quarks are assumed to be degenerate in mass, i.e.
$m_{_{\tilde U}}=m_{_{\tilde D}} =m_{_{\tilde Q}}$,
\end{itemize}
with the Tevatron data, loose bounds on the Stop and Sbottom
masses are found \cite{Ratoff}
\begin{eqnarray}
&&m_{_{\tilde t}}>140\;{\rm GeV}\;{\rm or}\;<64\;{\rm GeV}\;,\nonumber\\
&&m_{_{\tilde b}}>210\;{\rm GeV}\;{\rm or}\;<32\;{\rm GeV}\;.
\label{sq-bound}
\end{eqnarray}
When we take into account the effect of the supersymmetric CP
phases, those bounds would be further relaxed. In spite of this
fact, we still take the bounds in Eq. (\ref{sq-bound}) seriously
in our later numerical computations.
Without losing too much generality, we fix
the supersymmetric parameters as: $|m_{_2}|=|m_{_3}|=300\;({\rm
GeV})$, $m_{_{\tilde{Q}_{2}}} =m_{_{\tilde{L}_{2}}}=10\;({\rm
TeV})$, $m_{_{\tilde{s}}}=1\;({\rm TeV})$, $A_{_{s}}=0\;({\rm
GeV})$, $m_{_{\tilde{Q}_{3}}}=m_{_{\tilde{L}_{3}}} =400\;({\rm
GeV})$, $m_{_{\tilde{b}}}=500\;({\rm GeV})$, and
$m_{_{\tilde{t}}}=|A_{_{t}}| =|A_{_{b}}|=200\;({\rm
GeV})$ unless otherwise noted. Now, our numerical results should be affected by the
following CP violating phases: $\theta_{_\mu}=\arg(\mu_{_H}),\;
\theta_{_2}=\arg(m_{_2}),\;\theta_{_3}=\arg(m_{_3}),
\;\theta_{_t}=\arg(A_{_{t}}),\;\theta_{_b}=\arg(A_{_{b}})$. In
order to reduce degrees of freedom further, we set $\theta_{_\mu}=
\theta_{_b}=0$. One of the reasons why we take this assumption on
the parameter space is that the loop calculation for diagrams
inducing the lepton and neutron's EDMs  restricts the argument to
be $\theta_{_\mu}\le \pi/(5\tan\beta)$ when those scalar fermions
of the first generation are heavy enough. Additionally, we find
that our numerical results depend on the argument $\theta_{_b}$
rather moderately. Moreover, for the model we employ here, the
mass of the lightest Higgs boson sets a strong constraint on the
parameter space of the new physics. As indicated in the literature
\cite{Pilaftsis1,Pilaftsis2,Pilaftsis3,Pilaftsis4,Pilaftsis5}, the
CP violation would cause changes to the neutral-Higgs-quark
coupling, neutral Higgs-gauge-boson coupling and self-coupling of
Higgs boson. The present experimental lower bound for the mass of
the lightest Higgs boson is relaxed to 60 GeV. In our numerical
analysis we take this constraint for the parameter space into
account. Now, we present our numerical results item by item. Since
the present experimental result of $\Delta M_{_{B_{_s}}}$
constrains the 'new' CP phases in our calculation strongly, we
discuss the mass difference in $\bar{B}_{_s}-B_{_s}$ mixing
firstly.

\begin{figure}
\setlength{\unitlength}{1mm}
\begin{center}
\begin{picture}(80,120)(55,60)
\put(30,50){\includegraphics{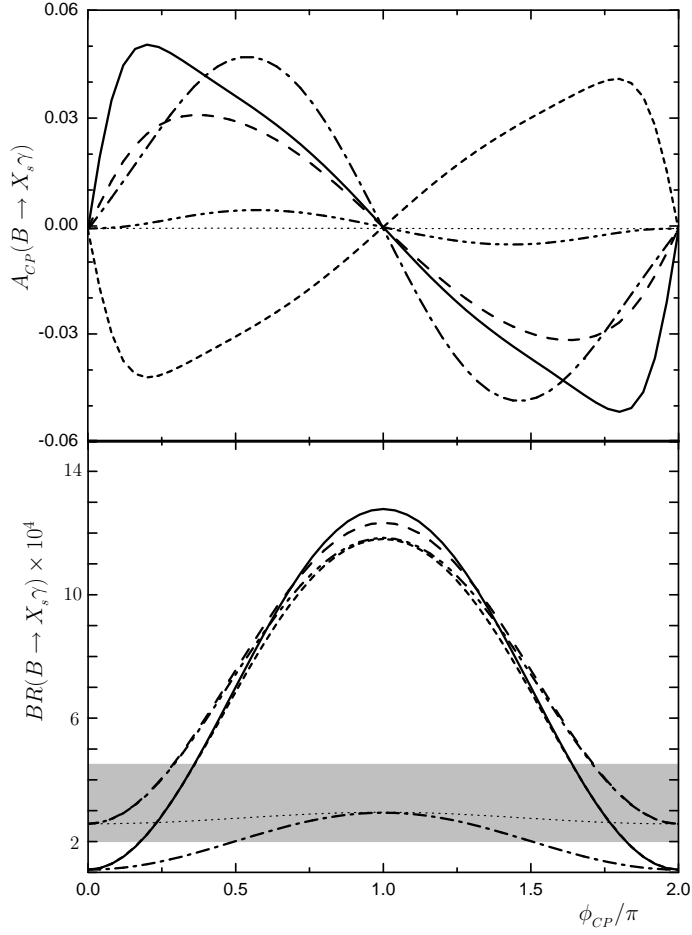}}
\end{picture}
\caption[]{$BR(B\rightarrow X_{_s}\gamma)$ and
$A_{_{CP}}(B\rightarrow X_{_s}\gamma)$ when $m_{_b}^{pole}=4.8{\rm
GeV},\;\tan\beta=50$, as well as
$\mu_{_H}=100\;({\rm GeV})$. In this figure, (a)solid line
represents rigorous two loop analysis at $\phi_{_{CP}}=\theta_{_2}$
and $\theta_{_3}=\theta_{_t}=0$, (b)dash line represents one loop
results plus threshold radiative corrections at
$\phi_{_{CP}}=\theta_{_2}$ and $\theta_{_3}=\theta_{_t}=0$,
(c)dash-dot line represents rigorous two loop analysis at
$\phi_{_{CP}}=\theta_{_3}$ and $\theta_{_2}=\theta_{_t}=0$, (d)dot
line represents one loop results plus threshold radiative
corrections at $\phi_{_{CP}}=\theta_{_3}$ and
$\theta_{_2}=\theta_{_t}=0$, (e)short-dash line represents rigorous
two loop analysis at $\phi_{_{CP}}=\theta_{_t}$ and
$\theta_{_2}=\theta_{_3}=0$, (f)dash-dot-dot line represents one
loop results plus threshold radiative corrections at
$\phi_{_{CP}}=\theta_{_t}$ and $\theta_{_2}=\theta_{_3}=0$.}
\label{fig6}
\end{center}
\end{figure}

Taking $\tan\beta=20,\;\mu_{_H}=300\;({\rm GeV})$,
we plot $\Delta M_{_{B_{_s}}}$ versus the
CP phases $\phi_{_{CP}}$ in Fig.\ref{fig3}, where the solid line
stands for $\phi_{_{CP}}=\theta_{_2}$ and
$\theta_{_3}=\theta_{_t}=0$, the dash line stands for
$\phi_{_{CP}}=\theta_{_3}$ as well as $\theta_{_2}=\theta_{_t}=0$,
and the dot line stands for $\phi_{_{CP}}=\theta_{_t}$ and
$\theta_{_2}=\theta_{_3}=0$, respectively. Fig.\ref{fig4} is
similar to Fig.\ref{fig3} except there
$\tan\beta=50,\;\mu_{_H}=100\;({\rm GeV})$. Under our assumptions
on the supersymetric parameter space, the theoretical prediction
on $\Delta M_{_{B_{_s}}}$ respects the experimental bound.

\begin{figure}
\setlength{\unitlength}{1mm}
\begin{center}
\begin{picture}(80,120)(55,60)
\put(30,50){\includegraphics{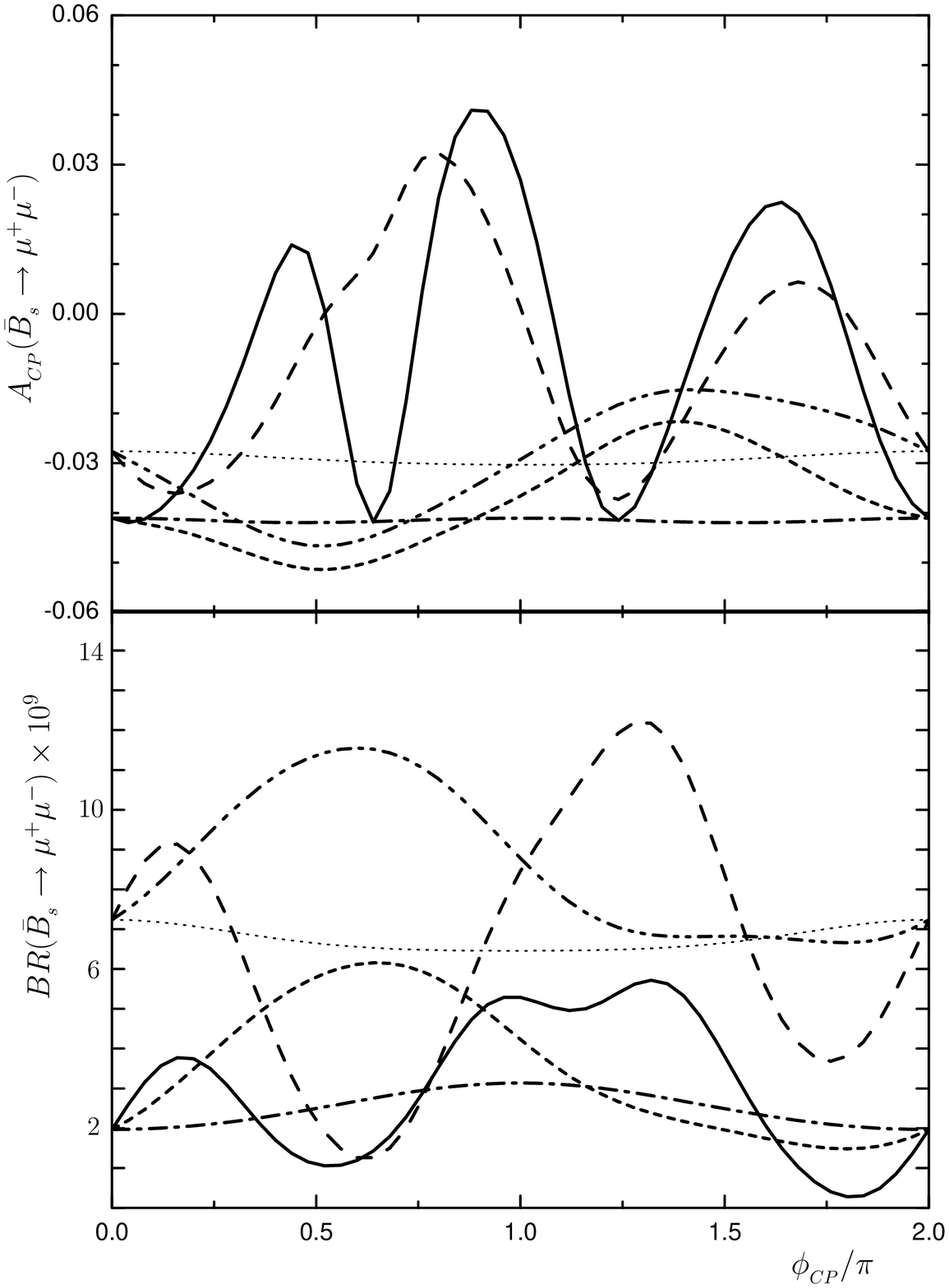}}
\end{picture}
\caption[]{$BR(\bar{B}_{_s}\rightarrow \mu^+\mu^-)$ and
$A_{_{CP}}(\bar{B}_{_s}\rightarrow \mu^+\mu^-)$ when
$m_{_b}^{pole}=4.8{\rm GeV},\;\tan\beta=20$
as well as $\mu_{_H}=300\;({\rm GeV})$. In the
figure (a)solid line stands for rigorous two loop analysis at
$\phi_{_{CP}}=\theta_{_2}$ and $\theta_{_3}=\theta_{_t}=0$,
(b)dash line stands for one loop results plus threshold radiative
corrections at $\phi_{_{CP}}=\theta_{_2}$ and
$\theta_{_3}=\theta_{_t}=0$, (c)dash-dot line stands for rigorous
two loop analysis at $\phi_{_{CP}}=\theta_{_3}$ and
$\theta_{_2}=\theta_{_t}=0$, (d)dot line stands for one loop
results plus threshold radiative corrections at
$\phi_{_{CP}}=\theta_{_3}$ and $\theta_{_2}=\theta_{_t}=0$,
(e)short-dash line stands stands for rigorous two loop analysis at
$\phi_{_{CP}}=\theta_{_t}$ and $\theta_{_2}=\theta_{_3}=0$,
(f)dash-dot-dot line stands for one loop results plus threshold
radiative corrections at $\phi_{_{CP}}=\theta_{_t}$ and
$\theta_{_2}=\theta_{_3}=0$.} \label{fig7}
\end{center}
\end{figure}

For inclusive decay $B\rightarrow X_{_s}\gamma$, the present
experimental observation on the branching ratio sets a constraint
for the parameter space. Furthermore, the CP asymmetry in this
process is highly sensitive to new CP violating phases because the
SM contribution is only $\sim0.5\%$. Taking
$\tan\beta=20,\;\mu_{_H}=300\;({\rm GeV})$, we plot
$BR(B\rightarrow X_{_s}\gamma)$ and $A_{_{CP}}(B\rightarrow
X_{_s}\gamma)$ versus the CP phases $\phi_{_{CP}}$ in
Fig.\ref{fig5}, where the solid and dash lines represent
$\phi_{_{CP}}=\theta_{_2}$ and $\theta_{_3}=\theta_{_t}=0$, the
dash-dot and dot lines represent $\phi_{_{CP}}=\theta_{_3}$ and
$\theta_{_2}=\theta_{_t}=0$, and the short-dash and dash-dot-dot
lines represent $\phi_{_{CP}}=\theta_{_t}$ and
$\theta_{_2}=\theta_{_3}=0$, respectively. From this figure, it is
easy to note that there are very obvious differences between the
exact two loop analysis (solid line for
$\phi_{_{CP}}=\theta_{_2}$, dash-dot line for
$\phi_{_{CP}}=\theta_{_3}$, and short-dash line for
$\phi_{_{CP}}=\theta_{_t}$, respectively) and the theoretical
results which include one-loop contributions and threshold
radiative corrections (dash line for $\phi_{_{CP}}=\theta_{_2}$,
dot line for $\phi_{_{CP}}=\theta_{_3}$, and dash-dot-dot line for
$\phi_{_{CP}}=\theta_{_t}$, respectively). Taking
$\phi_{_{CP}}=\theta_{_3}$ as an example, the CP asymmetry from
rigorous two loop analysis can reach $6\%$, and that from
threshold radiative corrections is smaller than $1\%$, i.e. they
are rather apart from each other, whereas the theoretical
predictions made in the two scenarios on the branching ratios do
not conflict with the present experimental bound.

CP violation is induced by both Standard Model (SM) sector and
SUSY sector. In the SM,  the CP violating parameter $\eta=0.34$ (in the
Wolfenstein parametrization, and corresponds to the central value
permitted by the present data) which indeed induces non-zero CP
violation in concerned processes. Therefore, even the SUSY phase
takes special values as $\phi_{CP}=0,\;\pi,\; 2\pi$, CP asymmetry
may still exist. However, for the process $B\to X_s\gamma$, the CP
violation induced by the SM $\eta$ is much smaller than 1\%, by
contraries, for the rare decay $B_s\rightarrow  l^+ l^-$, the
$\eta-$induced CP asymmetry is greater than 1\%, and the effect is
expected to be observable even as the SUSY phase
$\phi_{CP}=0,\;\pi,\;2\pi$ and does not contribute. In a word,
the process $B\to X_s\gamma$ provides a window for detecting
the new CP sources beside that existing in the CKM matrix.

Similar to Fig.\ref{fig5} except for
$\tan\beta=50,\;\mu_{_H}=100\;({\rm GeV})$,
we plot $BR(B\rightarrow X_{_s}\gamma)$ and
$A_{_{CP}}(B\rightarrow X_{_s}\gamma)$ versus the CP phases
$\phi_{_{CP}}$ in Fig.\ref{fig6}. With the setting for the
parameter space, the branching ratio $BR(B\rightarrow
X_{_s}\gamma)$ varies drastically with the CP phases
$\phi_{_{CP}}=\theta_{_{2,t}}$, but depends on the CP phase
$\phi_{_{CP}}=\theta_{_{3}}$ very gently. At
$\phi_{_{CP}}=\theta_{_{2}}=\pi/4$, the CP asymmetry of rigorous
two loop analysis can reach $5\%$, and that of threshold radiative
corrections is about $3\%$, meanwhile the theoretical predictions
on the branching ratios in both scenarios satisfy the present
experimental bound. When $\phi_{_{CP}}=\theta_{_{3,t}}$, the CP
asymmetry from the strict two loop analysis can reach $4\%$, and
that from threshold radiative corrections is less than $1\%$.

Now, we present our numerical results on the rare decays
$\bar{B}_{_s} \rightarrow l^+l^-,\;(l=\mu,\;\tau)$. The present
experimental upper bound on the branching ratio is
$BR(\bar{B}_{_s}\rightarrow\mu^+\mu^-) \le1.5\times10^{-7}$ at
90\% C.L. \cite{CDF}. Taking $\tan\beta=20, \;\mu_{_H}=300\;({\rm
GeV})$, we plot $BR(\bar{B}_{_s}\rightarrow\mu^+\mu^-)$ and
$A_{_{CP}}(\bar{B}_{_s}\rightarrow\mu^+\mu^-)$ versus the CP
phases $\phi_{_{CP}}$ in Fig.\ref{fig7},  where the solid and dash
lines stand for $\phi_{_{CP}}=\theta_{_2}$ and
$\theta_{_3}=\theta_{_t}=0$, the dash-dot and dot lines stand for
$\phi_{_{CP}}=\theta_{_3}$ and $\theta_{_2}=\theta_{_t}=0$, and
the short-dash and dash-dot-dot lines stand for
$\phi_{_{CP}}=\theta_{_t}$ and $\theta_{_2}=\theta_{_3}=0$,
respectively. From this figure, it is easy to find that there are
evident differences between exact two loop analysis (solid line
for $\phi_{_{CP}}=\theta_{_2}$, dash-dot line for
$\phi_{_{CP}}=\theta_{_3}$, and short-dash line for
$\phi_{_{CP}}=\theta_{_t}$, respectively) and the theoretical
results which  originate from one-loop calculations plus threshold
radiative corrections (dash line for $\phi_{_{CP}}=\theta_{_2}$,
dot line for $\phi_{_{CP}}=\theta_{_3}$, and dash-dot-dot line for
$\phi_{_{CP}}=\theta_{_t}$, respectively). For
$\phi_{_{CP}}=\theta_{_2}\simeq3\pi/2$, the theoretical prediction
on the branching ratio $BR(\bar{B}_{_s}\rightarrow\mu^+\mu^-)$
from one loop results plus threshold radiative corrections (dash
line) can reach $1.2\times10^{-8}$ approximately, the exact two
loop analysis (solid line) modifies the branching ratio to
$5\times10^{-9}$, and the CP asymmetry
$A_{_{CP}}(\bar{B}_{_s}\rightarrow\mu^+\mu^-)$ is about 3\%.
However, it is very difficult to detect this CP asymmetry
$A_{_{CP}}(\bar{B}_{_s}\rightarrow\mu^+\mu^-)$ in near future
experiments with such a small branching ratio. As indicated above,
the SM CP-odd parameter $\eta=0.34$ (in Wolfstein
parametrization)induces a non-vanishing CP asymmetry when
supersymmetric CP phases $\phi_{_{CP}}=0,\;\pi,\;2\pi$.

\begin{figure}
\setlength{\unitlength}{1mm}
\begin{center}
\begin{picture}(80,120)(55,60)
\put(30,50){\includegraphics{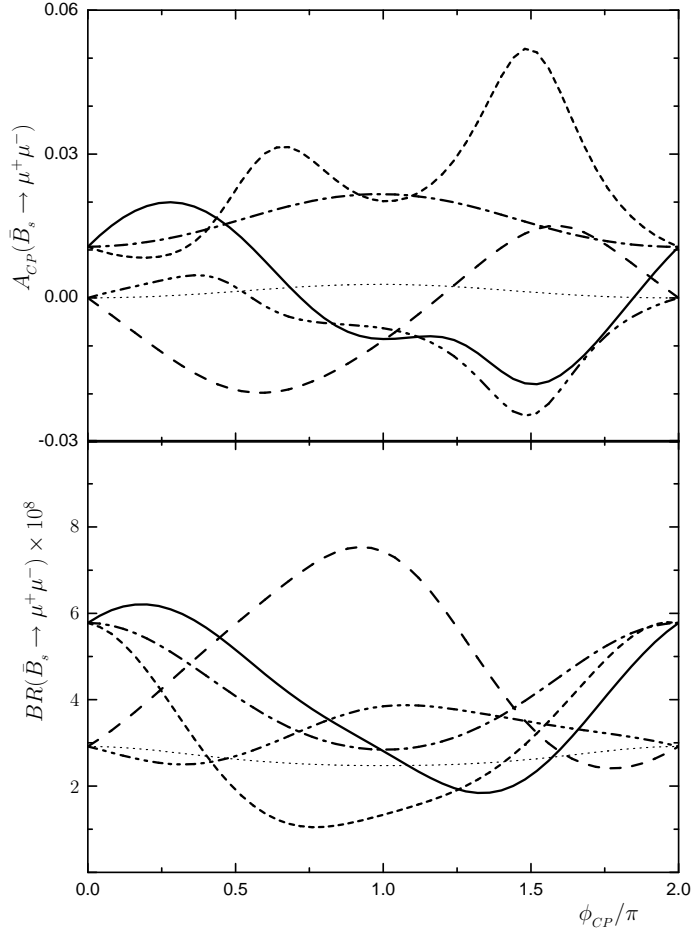}}
\end{picture}
\caption[]{$BR(\bar{B}_{_s}\rightarrow \mu^+\mu^-)$ and
$A_{_{CP}}(\bar{B}_{_s}\rightarrow \mu^+\mu^-)$ when
$m_{_b}^{pole}=4.8{\rm GeV},\;\tan\beta=50$ as well as
$\mu_{_H}=100\;({\rm GeV})$. In the
figure, (a)solid line stands for rigorous two loop analysis at
$\phi_{_{CP}}=\theta_{_2}$ and $\theta_{_3}=\theta_{_t}=0$,
(b)dash line stands for one loop results plus threshold radiative
corrections at $\phi_{_{CP}}=\theta_{_2}$ and
$\theta_{_3}=\theta_{_t}=0$, (c)dash-dot line stands for rigorous
two loop analysis at $\phi_{_{CP}}=\theta_{_3}$ and
$\theta_{_2}=\theta_{_t}=0$, (d)dot line stands for one loop
results plus threshold radiative corrections at
$\phi_{_{CP}}=\theta_{_3}$ and $\theta_{_2}=\theta_{_t}=0$,
(e)short-dash line stands stands for rigorous two loop analysis at
$\phi_{_{CP}}=\theta_{_t}$ and $\theta_{_2}=\theta_{_3}=0$,
(f)dash-dot-dot line stands for one loop results plus threshold
radiative corrections at $\phi_{_{CP}}=\theta_{_t}$ and
$\theta_{_2}=\theta_{_3}=0$.} \label{fig8}
\end{center}
\end{figure}

For larger $\tan\beta$, the situation is  drastically different.
Taking $\tan\beta=50, \;\mu_{_H}=100\;({\rm GeV})$, we plot
$BR(\bar{B}_{_s}\rightarrow\mu^+\mu^-)$ and
$A_{_{CP}}(\bar{B}_{_s}\rightarrow\mu^+\mu^-)$ versus the CP
phases $\phi_{_{CP}}$ in Fig.\ref{fig8}. Clearly, there are
obvious differences between the results of one loop contribution
plus threshold radiative corrections and that of corresponding
rigorous two loop calculations. Additionally, the rigorous two
loop prediction on the branching ratio
$BR(\bar{B}_{_s}\rightarrow\mu^+\mu^-)$  surpasses $10^{-8}$.
Assuming the CP asymmetry to be induced by the CP phase
$\phi_{_{CP}}=\theta_{_t}$, the two loop result for the CP
asymmetry $A_{_{CP}}(\bar{B}_{_s}\rightarrow\mu^+\mu^-)$ is about
5\%, and the corresponding branching ratio is about
$4\times10^{-8}$. Although it is very challenging, this CP
asymmetry could be hopefully measured in forthcoming experiments.

\begin{figure}
\setlength{\unitlength}{1mm}
\begin{center}
\begin{picture}(80,120)(55,60)
\put(30,50){\includegraphics{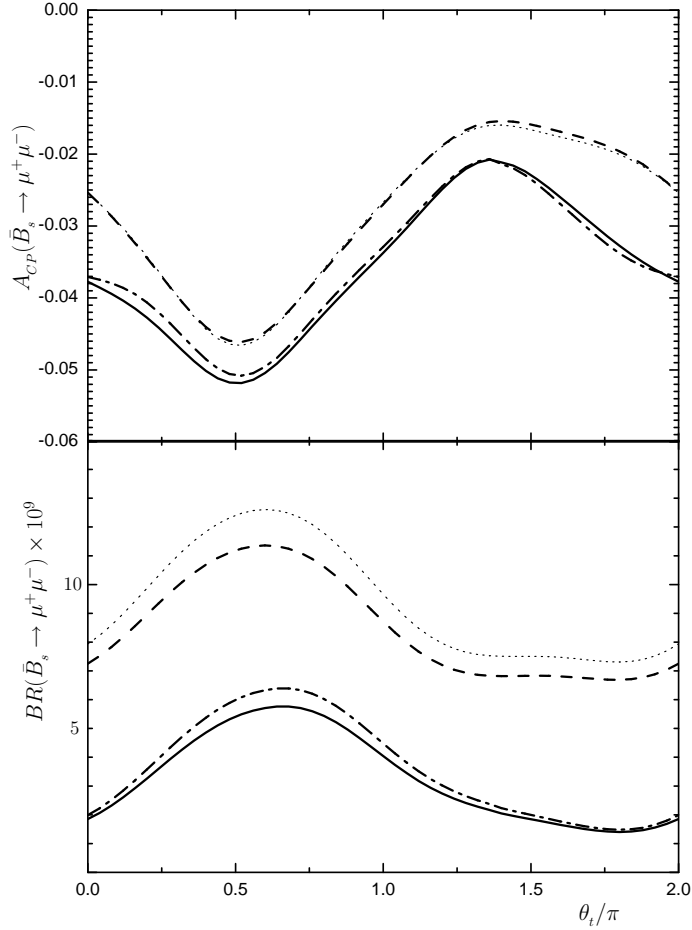}}
\end{picture}
\caption[]{Taking $\tan\beta=20,\;\mu_{_H}=300\;({\rm GeV})$, the branching ratio
$BR(\bar{B}_{_s}\rightarrow\mu^+\mu^-)$ and CP asymmetry
$A_{_{CP}}(\bar{B}_{_s}\rightarrow\mu^+\mu^-)$ vary with the CP
phase $\phi_{_{CP}}=\theta_{_{2}}$, where (a)solid line represents
rigorous two-loop analysis with $m_{_b}^{pole}=4.6\;({\rm GeV})$,
(b)dash line represents  one-loop result plus threshold radiative
correction with $m_{_b}^{pole}=4.6\;({\rm GeV})$, (c)dash-dot line
represents rigorous two-loop analysis with $m_{_b}^{pole}=4.9\;({\rm
GeV})$, (d)dot line represents one-loop result plus threshold
radiative correction with $m_{_b}^{pole}=4.9\;({\rm GeV})$.}
\label{fig9}
\end{center}
\end{figure}

In Fig.\ref{fig7} and Fig.\ref{fig8}, we choose the pole mass of
b-quark as $m_{_b}^{pole}=4.8\;({\rm GeV})$. Since the hadronic
matrix elements depend on  b-quark mass, we let $m_b$ vary within
a certain range which is allowed by the data and see how it
affects the branching ratio
$BR(\bar{B}_{_s}\rightarrow\mu^+\mu^-)$ and the CP asymmetry
$A_{_{CP}}(\bar{B}_{_s}\rightarrow\mu^+\mu^-)$. Taking
$\tan\beta=20,\;\mu_{_H}=300\;({\rm GeV})$, we plot the branching ratio
$BR(\bar{B}_{_s}\rightarrow\mu^+\mu^-)$ and CP asymmetry
$A_{_{CP}}(\bar{B}_{_s}\rightarrow\mu^+\mu^-)$ versus
$\phi_{_{CP}}=\theta_{_{2}}$ in Fig.\ref{fig9}. To investigate the
impaction of the b-quark mass on the measurable quantities, in the
figure, we set it as $m_{_b}^{pole}=4.6$, and $4.9\;({\rm GeV})$,
which correspond to the minimal and maximal values permitted by
the present experiments respectively. Particularly, the branching
ratio $BR(\bar{B}_{_s}\rightarrow\mu^+\mu^-)$ is relatively more
sensitive to the b-quark mass. The present experimental errors
approximately result in $5\%$ theoretical uncertainty. As for the
CP asymmetry $A_{_{CP}}(\bar{B}_{_s}\rightarrow\mu^+\mu^-)$, the
corresponding theoretical uncertainty is only about $2\%$.

\begin{figure}
\setlength{\unitlength}{1mm}
\begin{center}
\begin{picture}(80,120)(55,60)
\put(30,50){\includegraphics{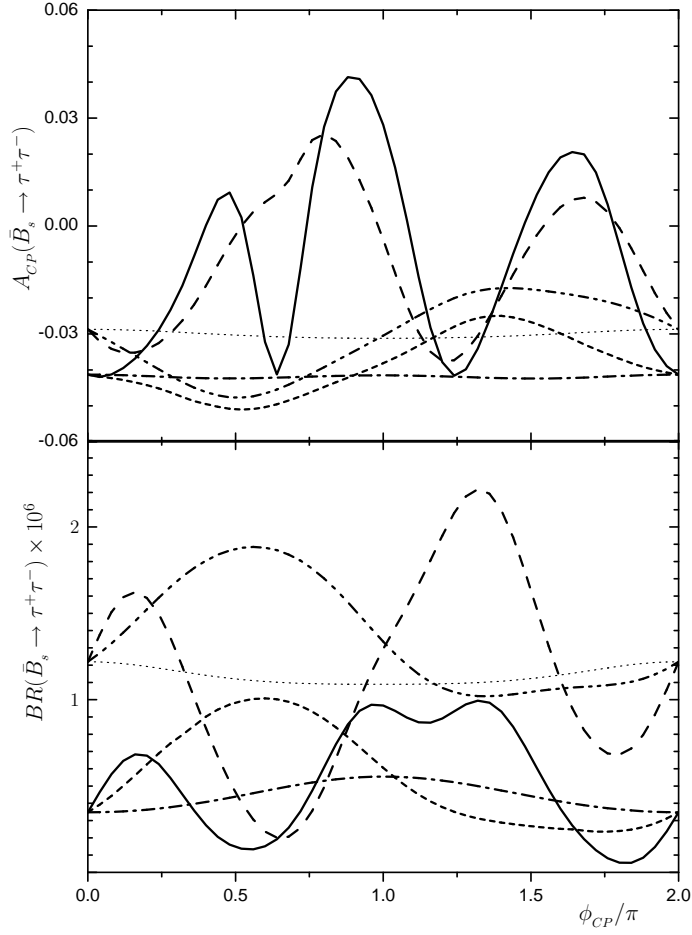}}
\end{picture}
\caption[]{$BR(\bar{B}_{_s}\rightarrow \tau^+\tau^-)$ and
$A_{_{CP}}(\bar{B}_{_s}\rightarrow \tau^+\tau^-)$ when
$m_{_b}^{pole}=4.8{\rm GeV},\;\tan\beta=20$
as well as $\mu_{_H}=300\;({\rm GeV})$. In the
figure, (a)solid line represents rigorous two loop analysis at
$\phi_{_{CP}}=\theta_{_2}$ and $\theta_{_3}=\theta_{_t}=0$,
(b)dash line represents one loop results plus threshold radiative
corrections at $\phi_{_{CP}}=\theta_{_2}$ and
$\theta_{_3}=\theta_{_t}=0$, (c)dash-dot line represents rigorous
two loop analysis at $\phi_{_{CP}}=\theta_{_3}$ and
$\theta_{_2}=\theta_{_t}=0$, (d)dot line represents one loop
results plus threshold radiative corrections at
$\phi_{_{CP}}=\theta_{_3}$ and $\theta_{_2}=\theta_{_t}=0$,
(e)short-dash line represents rigorous two loop analysis at
$\phi_{_{CP}}=\theta_{_t}$ and $\theta_{_2}=\theta_{_3}=0$,
(f)dash-dot-dot line represents one loop results plus threshold
radiative corrections at $\phi_{_{CP}}=\theta_{_t}$ and
$\theta_{_2}=\theta_{_3}=0$.} \label{fig10}
\end{center}
\end{figure}

Comparing with $BR(\bar{B}_{_s}\rightarrow\mu^+\mu^-)$, the
branching ratio $BR(\bar{B}_{_s}\rightarrow\tau^+\tau^-)$ is
enhanced strongly because $\tau$ mass is much heavier than $\mu$
mass. Taking $\tan\beta=20, \;\mu_{_H}=300\;({\rm GeV})$, we plot
$BR(\bar{B}_{_s}\rightarrow\tau^+\tau^-)$ and
$A_{_{CP}}(\bar{B}_{_s}\rightarrow\tau^+\tau^-)$ versus the CP
phases $\phi_{_{CP}}$ in Fig.\ref{fig10}. We find that the exact two
loop predictions on the branching ratio
$BR(\bar{B}_{_s}\rightarrow\tau^+\tau^-)$ can reach $\sim10^{-6}$
approximately, and the CP asymmetry
$A_{_{CP}}(\bar{B}_{_s}\rightarrow\tau^+\tau^-)$ is about $\pm3\%$
correspondingly.

\begin{figure}
\setlength{\unitlength}{1mm}
\begin{center}
\begin{picture}(80,120)(55,60)
\put(30,50){\includegraphics{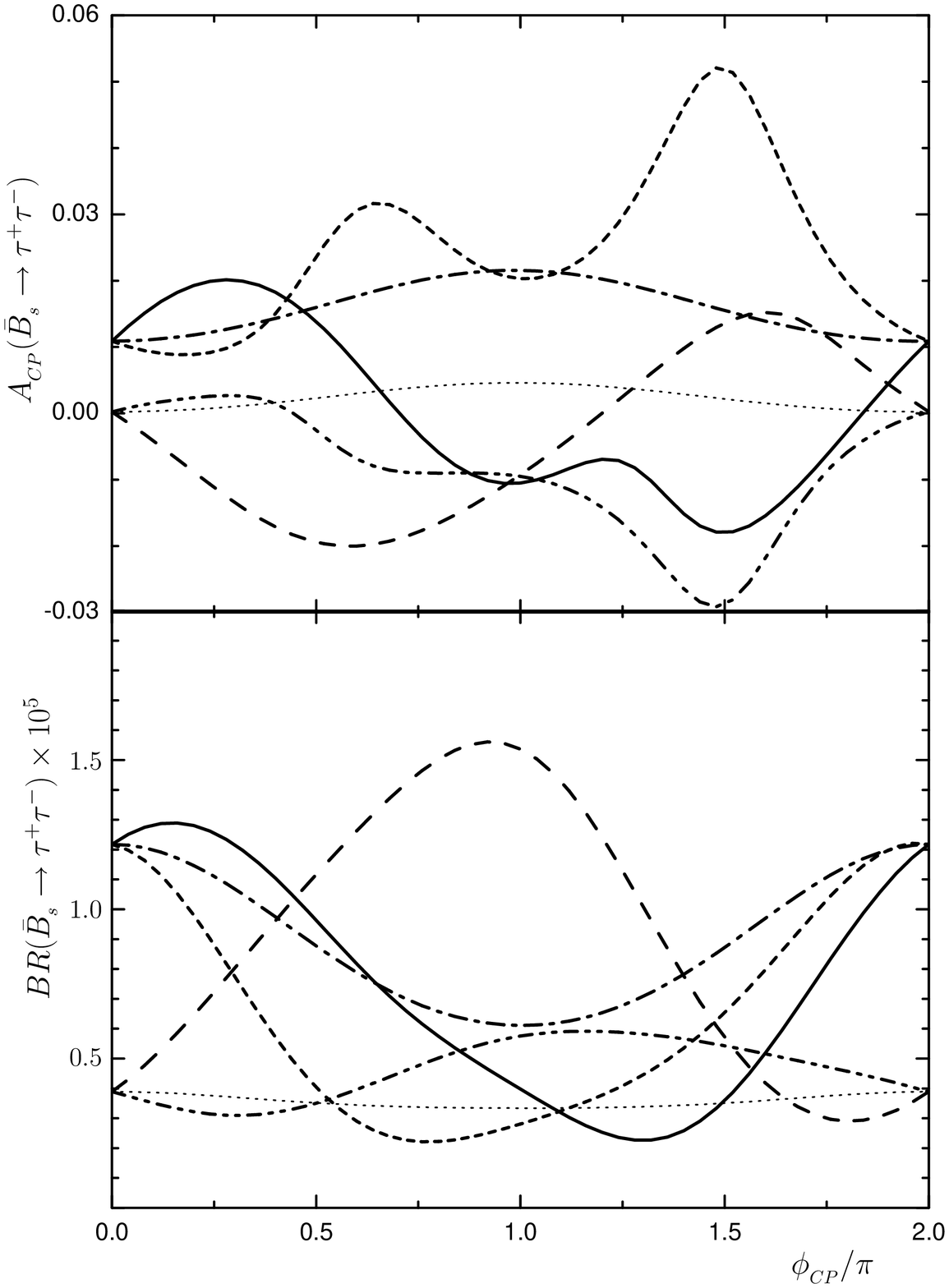}}
\end{picture}
\caption[]{$BR(\bar{B}_{_s}\rightarrow \tau^+\tau^-)$ and
$A_{_{CP}}(\bar{B}_{_s}\rightarrow \tau^+\tau^-)$ when
$m_{_b}^{pole}=4.8{\rm GeV},\;\tan\beta=50$
as well as $\mu_{_H}=100\;({\rm GeV})$ In the
figure, (a)solid line represents rigorous two loop analysis at
$\phi_{_{CP}}=\theta_{_2}$ and $\theta_{_3}=\theta_{_t}=0$,
(b)dash line represents one loop results plus threshold radiative
corrections at $\phi_{_{CP}}=\theta_{_2}$ and
$\theta_{_3}=\theta_{_t}=0$, (c)dash-dot line represents rigorous
two loop analysis at $\phi_{_{CP}}=\theta_{_3}$ and
$\theta_{_2}=\theta_{_t}=0$, (d)dot line represents one loop
results plus threshold radiative corrections at
$\phi_{_{CP}}=\theta_{_3}$ and $\theta_{_2}=\theta_{_t}=0$,
(e)short-dash line represents rigorous two loop analysis at
$\phi_{_{CP}}=\theta_{_t}$ and $\theta_{_2}=\theta_{_3}=0$,
(f)dash-dot-dot line represents one loop results plus threshold
radiative corrections at $\phi_{_{CP}}=\theta_{_t}$ and
$\theta_{_2}=\theta_{_3}=0$.} \label{fig11}
\end{center}
\end{figure}

When $\tan\beta=50$, the branching ratio
$BR(\bar{B}_{_s}\rightarrow\tau^+\tau^-)$ is enhanced further. We
plot $BR(\bar{B}_{_s}\rightarrow\tau^+\tau^-)$ and
$A_{_{CP}}(\bar{B}_{_s}\rightarrow\tau^+\tau^-)$ versus the CP
phases $\phi_{_{CP}}=\theta_{_{2,3,t}}$ in Fig.\ref{fig11}, with
$\tan\beta=50, \;\mu_{_H}=100\;({\rm GeV})$.
Assuming that the CP asymmetry is induced
by the complex trilinear coupling $A_{_{t}}$, the two loop
theoretical prediction on the branching ratio
$BR(\bar{B}_{_s}\rightarrow\tau^+\tau^-)$ is about
$8\times10^{-6}$, whereas
$A_{_{CP}}(\bar{B}_{_s}\rightarrow\tau^+\tau^-) \simeq5\%$ at
$\theta_{_{t}}=3\pi/2$. Certainly, it is difficult to
experimentally measure the rare decay
$\bar{B}_{_s}\rightarrow\tau^+\tau^-$. Similarly, the present experimental
error for b-quark mass causes a theoretical uncertainties of
$\sim5\%$ for $BR(\bar{B}_{_s}\rightarrow\tau^+\tau^-)$, and
$\sim2\%$ for $A_{_{CP}}(\bar{B}_{_s}\rightarrow\tau^+\tau^-)$
respectively.

We now discuss the branching ratio and forward-backward asymmetry
in decays $\bar{B}\rightarrow Kl^+l^-\;(l=\mu,\;\tau)$. The form
factors for  $\bar{B}\rightarrow Kl^+l^-\;(l=\mu,\;\tau)$ decays
are given in Table.\ref{tabf} which correspond to the central
values presented in \cite{Wirbel}.
\begin{table}[h]
\begin{center}
\begin{tabular}{||c|c|c|c|c|c|c|c|c|c|c|c||}
\hline \hline   $f_0(0)$ & $f_+(0)$ & $f_{_T}(0)$ & $c_1^0$ &
$c_1^+$ & $c_1^T$ &
$c_2^0$ & $c_2^+$ & $c_2^T$ & $c_3^0$ & $c_3^+$ & $c_3^T$\\
\hline   $0.319$ & $0.319$ & $0.355$
  & $0.633$ & $1.465$ & $1.478$
 & $-0.095$ & $0.372$ & $0.373$  & $0.591$ &
$0.782$ & $0.700$\\\hline \hline
\end{tabular}
\end{center}
\caption{The parameters for the parametrization Eq.\ref{para},
they correspond to the central values presented in
Ref.\cite{Wirbel}.} \label{tabf}
\end{table}

\begin{figure}
\setlength{\unitlength}{1mm}
\begin{center}
\begin{picture}(80,120)(55,60)
\put(30,50){\includegraphics{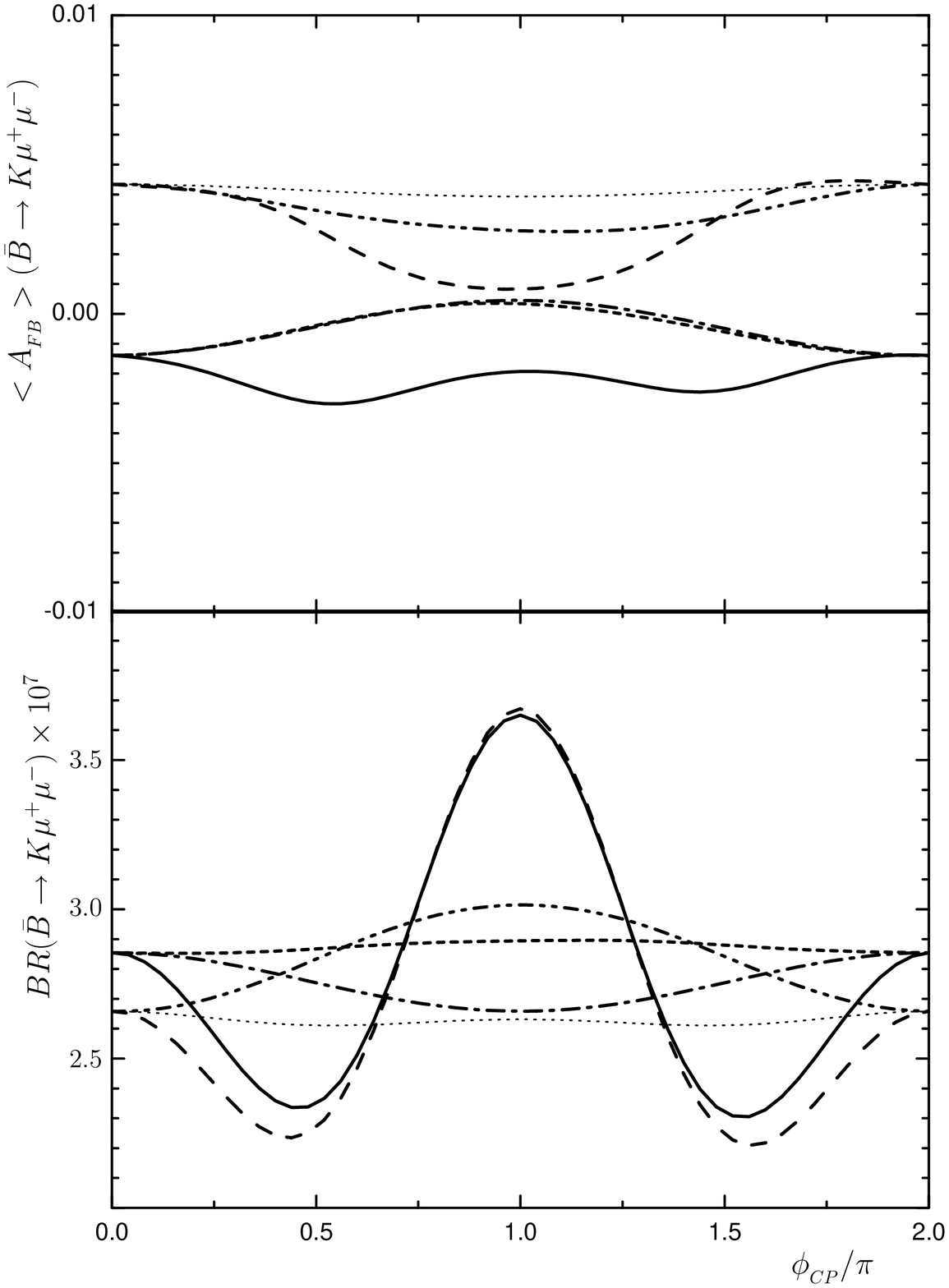}}
\end{picture}
\caption[]{$BR(\bar{B}\rightarrow K\mu^+\mu^-)$ and
$A_{_{CP}}(\bar{B}\rightarrow K\mu^+\mu^-)$ when
$m_{_b}^{pole}=4.8{\rm GeV},\;\tan\beta=50$
as well as $\mu_{_H}=100\;({\rm GeV})$. In the
figure, (a)solid line stands for rigorous two loop analysis at
$\phi_{_{CP}}=\theta_{_2}$ and $\theta_{_3}=\theta_{_t}=0$,
(b)dash line stands for one loop results plus threshold radiative
corrections at $\phi_{_{CP}}=\theta_{_2}$ and
$\theta_{_3}=\theta_{_t}=0$, (c)dash-dot line stands for rigorous
two loop analysis at $\phi_{_{CP}}=\theta_{_3}$ and
$\theta_{_2}=\theta_{_t}=0$, (d)dot line stands for one loop
results plus threshold radiative corrections at
$\phi_{_{CP}}=\theta_{_3}$ and $\theta_{_2}=\theta_{_t}=0$,
(e)short-dash line stands stands for rigorous two loop analysis at
$\phi_{_{CP}}=\theta_{_t}$ and $\theta_{_2}=\theta_{_3}=0$,
(f)dash-dot-dot line stands for one loop results plus threshold
radiative corrections at $\phi_{_{CP}}=\theta_{_t}$ and
$\theta_{_2}=\theta_{_3}=0$.} \label{fig12}
\end{center}
\end{figure}
\begin{figure}
\setlength{\unitlength}{1mm}
\begin{center}
\begin{picture}(80,120)(55,60)
\put(30,50){\includegraphics{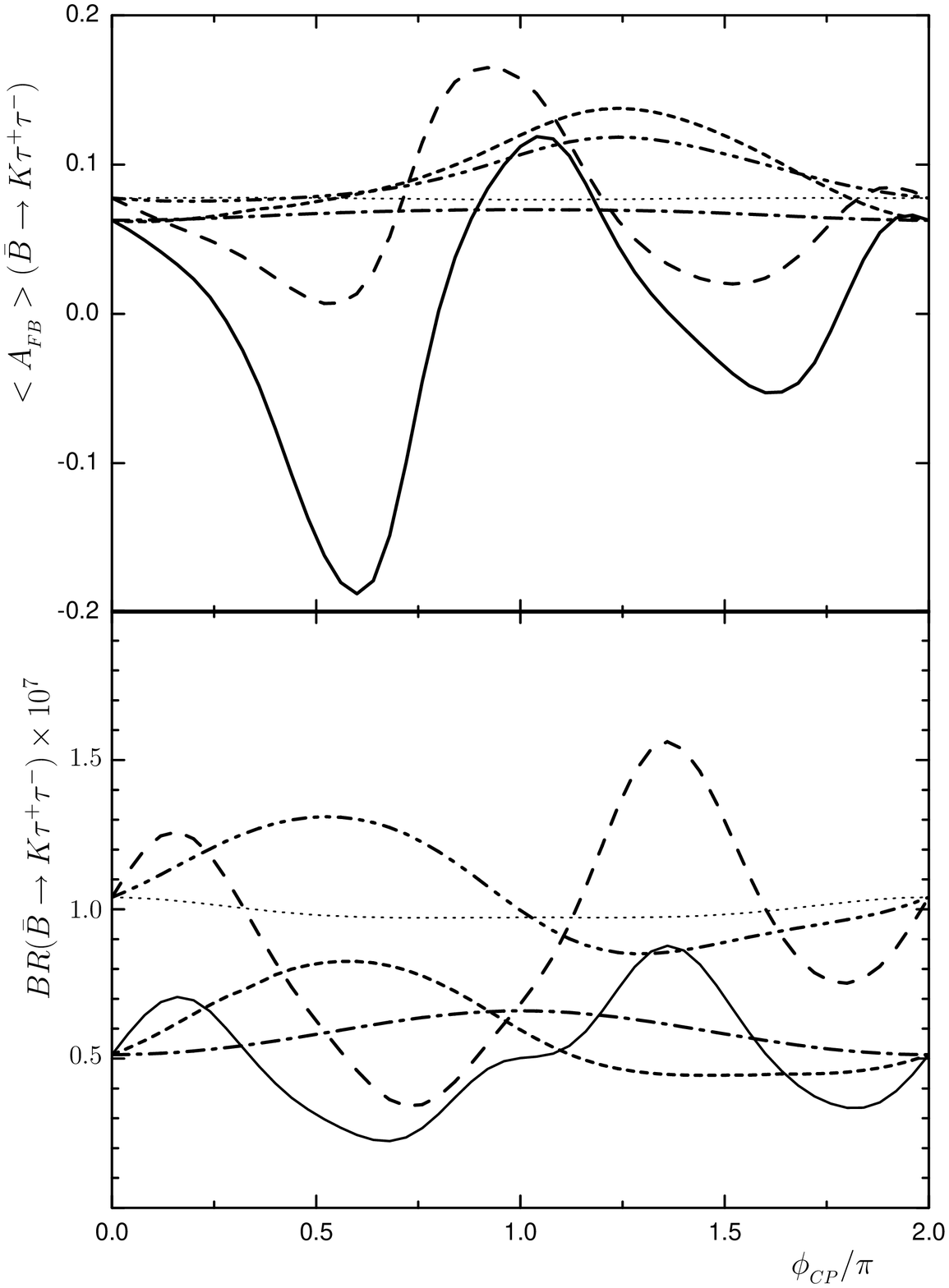}}
\end{picture}
\caption[]{$BR(\bar{B}\rightarrow K\tau^+\tau^-)$ and
$A_{_{CP}}(\bar{B}\rightarrow K\tau^+\tau^-)$ when
$m_{_b}^{pole}=4.8{\rm GeV},\;\tan\beta=20$ as well as
$\mu_{_H}=300\;({\rm GeV})$. In this
figure, (a)solid line stands for rigorous two loop analysis at
$\phi_{_{CP}}=\theta_{_2}$ and $\theta_{_3}=\theta_{_t}=0$,
(b)dash line stands for one loop results plus threshold radiative
corrections at $\phi_{_{CP}}=\theta_{_2}$ and
$\theta_{_3}=\theta_{_t}=0$, (c)dash-dot line stands for rigorous
two loop analysis at $\phi_{_{CP}}=\theta_{_3}$ and
$\theta_{_2}=\theta_{_t}=0$, (d)dot line stands for one loop
results plus threshold radiative corrections at
$\phi_{_{CP}}=\theta_{_3}$ and $\theta_{_2}=\theta_{_t}=0$,
(e)short-dash line stands stands for rigorous two loop analysis at
$\phi_{_{CP}}=\theta_{_t}$ and $\theta_{_2}=\theta_{_3}=0$,
(f)dash-dot-dot line stands for one loop results plus threshold
radiative corrections at $\phi_{_{CP}}=\theta_{_t}$ and
$\theta_{_2}=\theta_{_3}=0$.} \label{fig13}
\end{center}
\end{figure}

Taking $\tan\beta=50, \;\mu_{_H}=100\;({\rm GeV})$, we plot $BR(\bar{B}\rightarrow
K\mu^+\mu^-)$ and $<A_{_{FB}}>(\bar{B}\rightarrow K\mu^+\mu^-)$
versus the various CP phases $\phi_{_{CP}}$ in Fig.\ref{fig12},
where the solid and dash lines stand for
$\phi_{_{CP}}=\theta_{_2}$ and $\theta_{_3}=\theta_{_t}=0$, the
dash-dot and dot lines stand for $\phi_{_{CP}}=\theta_{_3}$ and
$\theta_{_2}=\theta_{_t}=0$, and the short-dash and dash-dot-dot
lines stand for $\phi_{_{CP}}=\theta_{_t}$ and
$\theta_{_2}=\theta_{_3}=0$, respectively. This figure explicitly
indicates that there are also obvious differences between the
strict two loop results (solid line for
$\phi_{_{CP}}=\theta_{_2}$, dash-dot line for
$\phi_{_{CP}}=\theta_{_3}$, and short-dash line for
$\phi_{_{CP}}=\theta_{_t}$, respectively) and the theoretical
predictions which come from one-loop contributions plus threshold
radiative corrections (dash line for $\phi_{_{CP}}=\theta_{_2}$,
dot line for $\phi_{_{CP}}=\theta_{_3}$, and dash-dot-dot line for
$\phi_{_{CP}}=\theta_{_t}$, respectively). It is noted that
although the branching ratio $BR(\bar{B}\rightarrow K\mu^+\mu^-)$
can reach $3\times 10^{-7}$, the average forward-backward
asymmetry is too small to be detected in any forthcoming
experiment.

\begin{figure}
\setlength{\unitlength}{1mm}
\begin{center}
\begin{picture}(80,120)(55,60)
\put(30,50){\includegraphics{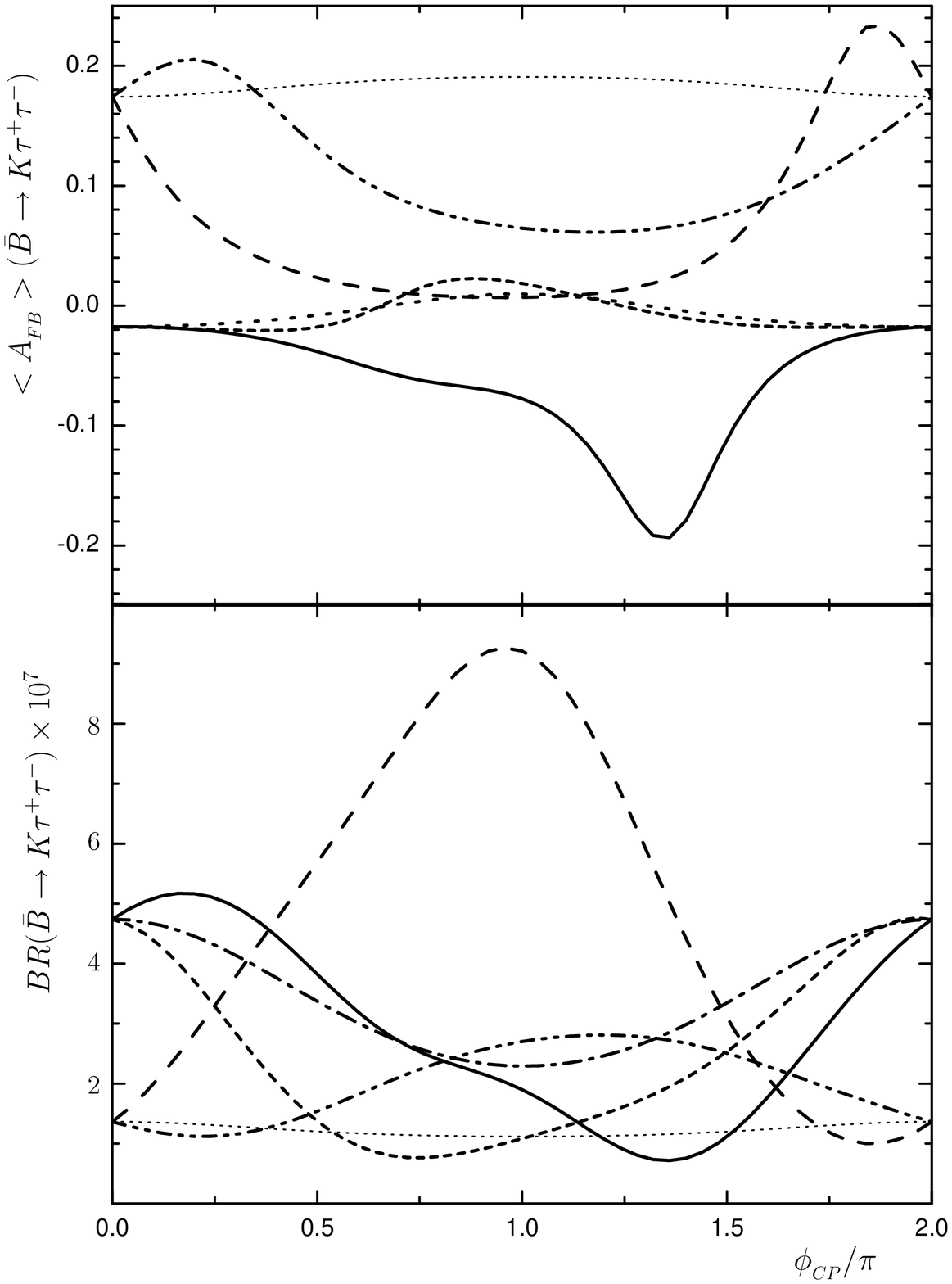}}
\end{picture}
\caption[]{$BR(\bar{B}\rightarrow K\tau^+\tau^-)$ and
$A_{_{CP}}(\bar{B}\rightarrow K\tau^+\tau^-)$ when
$m_{_b}^{pole}=4.8{\rm GeV},\;\tan\beta=50$
as well as $\mu_{_H}=100\;({\rm GeV})$. In this
figure, (a)solid line stands for rigorous two loop analysis at
$\phi_{_{CP}}=\theta_{_2}$ and $\theta_{_3}=\theta_{_t}=0$,
(b)dash line stands for one loop results plus threshold radiative
corrections at $\phi_{_{CP}}=\theta_{_2}$ and
$\theta_{_3}=\theta_{_t}=0$, (c)dash-dot line stands for rigorous
two loop analysis at $\phi_{_{CP}}=\theta_{_3}$ and
$\theta_{_2}=\theta_{_t}=0$, (d)dot line stands for one loop
results plus threshold radiative corrections at
$\phi_{_{CP}}=\theta_{_3}$ and $\theta_{_2}=\theta_{_t}=0$,
(e)short-dash line stands stands for rigorous two loop analysis at
$\phi_{_{CP}}=\theta_{_t}$ and $\theta_{_2}=\theta_{_3}=0$,
(f)dash-dot-dot line stands for one loop results plus threshold
radiative corrections at $\phi_{_{CP}}=\theta_{_t}$ and
$\theta_{_2}=\theta_{_3}=0$.} \label{fig14}
\end{center}
\end{figure}

Probably, the most interesting object to study is the rare decay
$\bar{B}\rightarrow K\tau^+\tau^-$ because the average
forward-backward asymmetry $<A_{_{FB}}>(\bar{B}\rightarrow
K\tau^+\tau^-)$ is much larger than
$<A_{_{FB}}>(\bar{B}\rightarrow K\mu^+\mu^-)$. Taking
$\tan\beta=20,\;\mu_{_H}=300\;({\rm GeV})$, we plot $BR(\bar{B}\rightarrow
K\tau^+\tau^-)$ and $<A_{_{FB}}>(\bar{B}\rightarrow
K\tau^+\tau^-)$ versus the CP phases $\phi_{_{CP}}$ in
Fig.\ref{fig13}. Certainly, there are evident differences between
the theoretical prediction of the exact two loop calculations and
that of one loop result plus threshold radiative corrections. When
$\phi_{_{CP}}=\theta_{_t}=\pi/2$, if considering the one loop
contributions plus threshold radiative corrections, one can have
the branching ratio as $BR(\bar{B}\rightarrow K\tau^+\tau^-)
\simeq10^{-7}$, whereas the strict two-loop calculations modify it
to $8\times10^{-8}$. The average forward-backward asymmetries
$<A_{_{FB}}>(\bar{B}\rightarrow K\tau^+\tau^-)$ are about 10\% for
both cases. For $\tan\beta=50$, the branching ratio is enhanced
further. When $\tan\beta=50,\;\mu_{_H}=100\;({\rm GeV})$ (Fig.\ref{fig14}), the exact
two loop calculations  predict the branching ratio as
$BR(\bar{B}\rightarrow K\tau^+\tau^-)\sim5\times10^{-7}$,
meanwhile the average forward-backward asymmetry
$<A_{_{FB}}>(\bar{B}\rightarrow K\tau^+\tau^-)\sim 20\%$. If we
can accumulate $10^{10}$ $B$ mesons in experiments, one would be
able to detect this asymmetry.

\begin{figure}
\setlength{\unitlength}{1mm}
\begin{center}
\begin{picture}(80,120)(55,60)
\put(30,50){\includegraphics{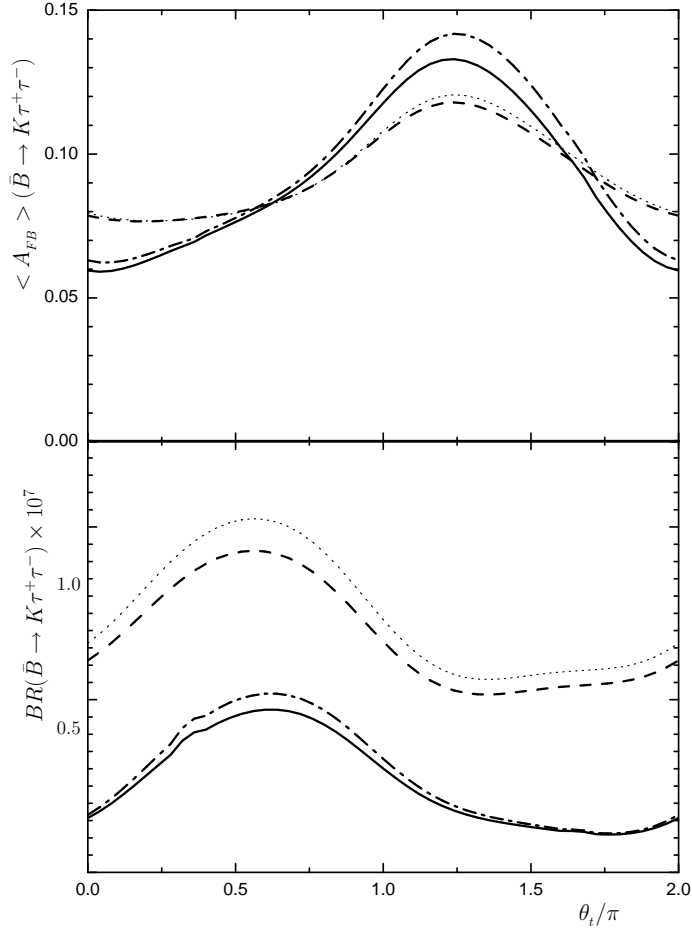}}
\end{picture}
\caption[]{Taking $\tan\beta=20$ as
well as $\mu_{_H}=300\;({\rm GeV})$, the branching ratio
$BR(\bar{B}\rightarrow K\tau^+\tau^-)$ and average
forward-backward asymmetry $A_{_{FB}}(\bar{B}\rightarrow
K\tau^+\tau^-)$ vary with the CP phase
$\phi_{_{CP}}=\theta_{_{2}}$, where (a)solid line represents
rigorous two-loop analysis with $m_{_b}^{pole}=4.6\;({\rm GeV})$;
(b)dash line represents one-loop result plus threshold radiative
correction with $m_{_b}^{pole}=4.6\;({\rm GeV})$; (c)dash-dot line
represents rigorous two-loop analysis with
$m_{_b}^{pole}=4.9\;({\rm GeV})$; (d)dot line represents one-loop
result plus threshold radiative correction with
$m_{_b}^{pole}=4.9\;({\rm GeV})$.} \label{fig15}
\end{center}
\end{figure}

\begin{figure}
\setlength{\unitlength}{1mm}
\begin{center}
\begin{picture}(80,120)(55,60)
\put(30,50){\includegraphics{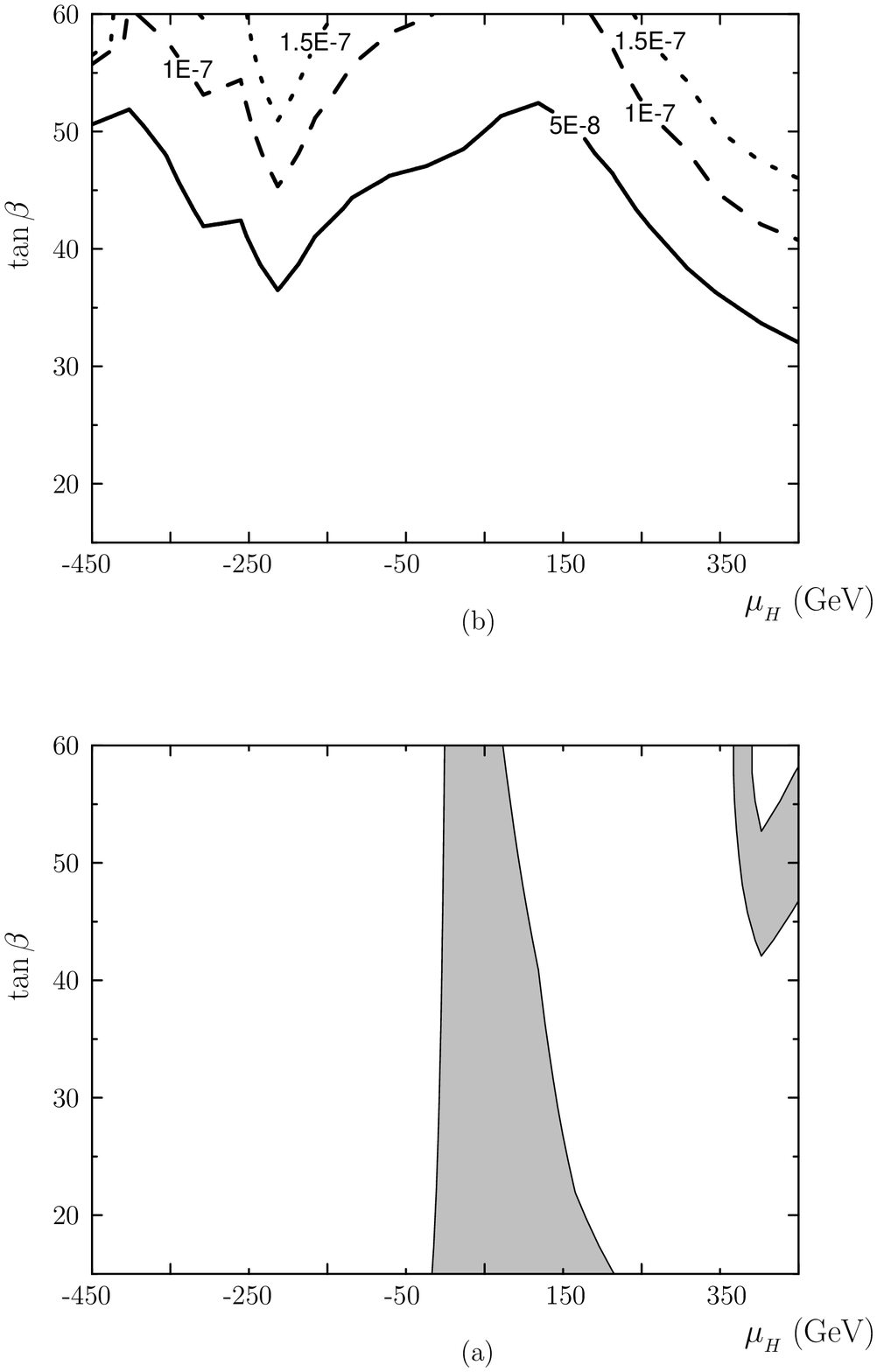}}
\end{picture}
\caption[]{Taking $m_{_b}^{pole}=4.8{\rm
GeV},\;\theta_{_2}=\theta_{_3}=\theta_{_t}=0$, the correlation between $\mu_{_H}$ and
$\tan\beta$. In this figure, (a)the gray regions are permitted by
the experimental bound on the branching ratio $BR(B\rightarrow
X_{_s}\gamma)$, (b)solid line represents
$BR(\bar{B}_{_s}\rightarrow\mu^+\mu^-)=5\times10^{-8}$, dash line
represents $BR(\bar{B}_{_s}\rightarrow\mu^+\mu^-)=10^{-7}$ and dot
line represents
$BR(\bar{B}_{_s}\rightarrow\mu^+\mu^-)=1.5\times10^{-7}$. }
\label{fig16}
\end{center}
\end{figure}

In Fig.\ref{fig13} and Fig.\ref{fig14}, we take the pole mass of
b-quark as $m_{_b}^{pole}=4.8\;({\rm GeV})$. The hadronic matrix
elements depend on concrete values of b-quark mass. Taking
$\tan\beta=20,\;\mu_{_H}=300\;({\rm GeV})$, we plot the branching ratio
$BR(\bar{B}\rightarrow K\tau^+\tau^-)$ and the average
forward-backward asymmetry $<A_{_{FB}}>(\bar{B}\rightarrow
K\tau^+\tau^-)$ versus $\phi_{_{CP}}=\theta_{_{2}}$ in
Fig.\ref{fig15}. As for the b-quark mass, we set it as
$m_{_b}^{pole}=4.6$ and $4.9\;({\rm GeV})$ respectively.
Particularly, the experimental uncertainty for the b-quark mass
leads to a theoretical uncertainty of $\sim$5\% for the branching
ratio $BR(\bar{B}\rightarrow K\tau^+\tau^-)$, and $\sim$2\% for
the average forward-backward asymmetry
$<A_{_{FB}}>(\bar{B}\rightarrow K\tau^+\tau^-)$.

\begin{figure}
\setlength{\unitlength}{1mm}
\begin{center}
\begin{picture}(80,120)(55,60)
\put(30,50){\includegraphics{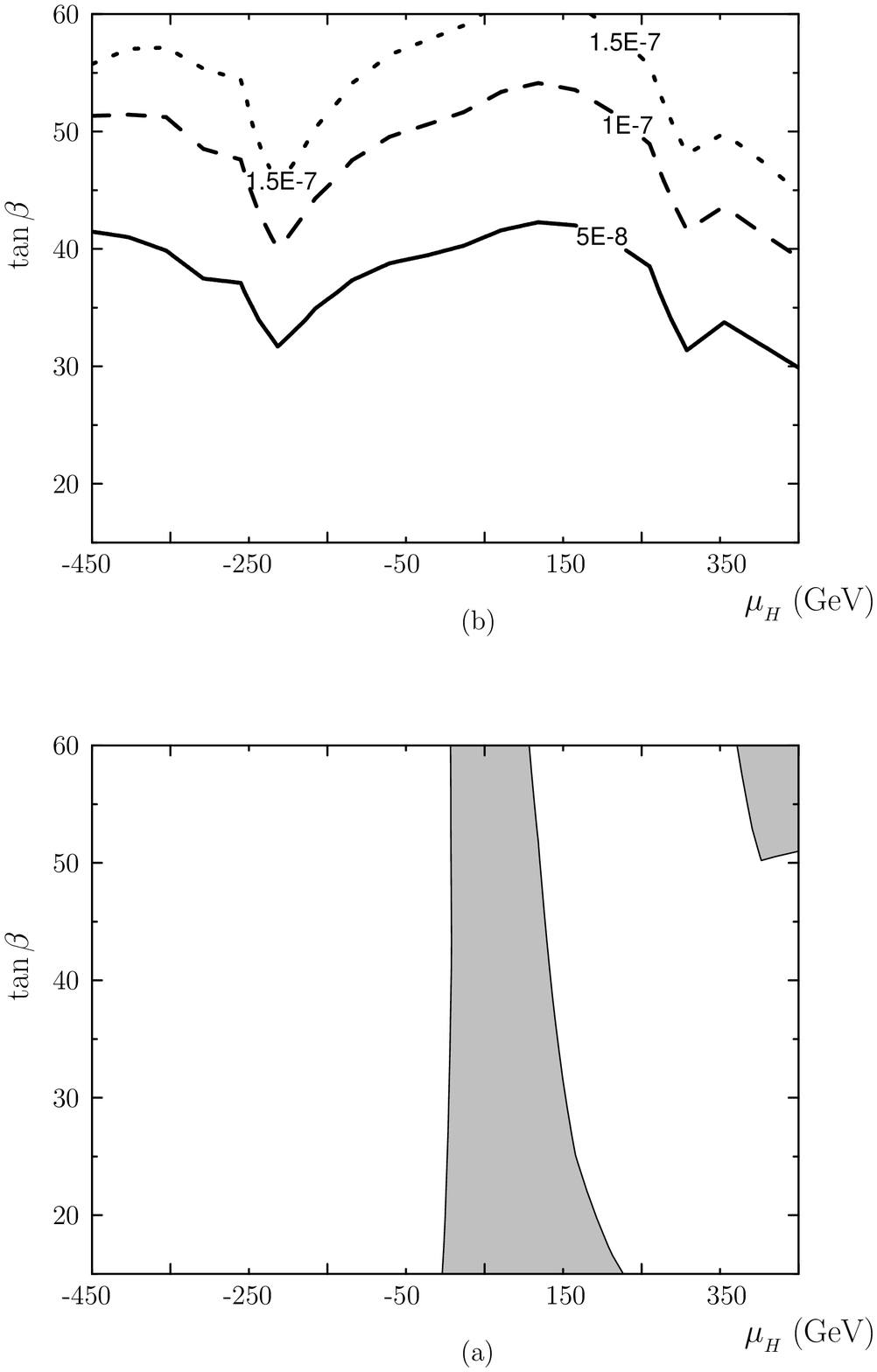}}
\end{picture}
\caption[]{Taking $m_{_b}^{pole}=4.8{\rm
GeV},\;\theta_{_2}=\theta_{_3}=0,\; \theta_{_t}=\pi/2$, the correlation between
$\mu_{_H}$ and $\tan\beta$. In the figure, (a)the gray regions are
permitted by the experimental bound on the branching ratio
$BR(B\rightarrow X_{_s}\gamma)$, (b)solid line represents
$BR(\bar{B}_{_s}\rightarrow\mu^+\mu^-)=5\times10^{-8}$, dash line
represents $BR(\bar{B}_{_s}\rightarrow\mu^+\mu^-)=10^{-7}$ and dot
line represents
$BR(\bar{B}_{_s}\rightarrow\mu^+\mu^-)=1.5\times10^{-7}$. }
\label{fig17}
\end{center}
\end{figure}

The present experimental upper bound of
$BR(\bar{B}_{_s}\rightarrow\mu^+\mu^-)<1.5\times 10^{-7}$ sets a
stringent restriction on the supersymmetric parameter space. Along
with the improvement of experimental precision and the
accumulation of experimental data, this upper bound will be
further modified. If so, we can expect that the new upper bound
may lead to a concrete constraint on the parameter space of our
model. Using the exact two loop results, we plot the correlation
of $\mu_{_H}$ and $\tan\beta$ with the bound of the branching
ratio $BR(B\rightarrow X_{_s}\gamma)$ at
$\theta_{_2}=\theta_{_3}=\theta_{_t}=0$ in Fig.\ref{fig16} (a).
The gray region is permitted by the present experiments.
Corresponding to the two loop results for the branching ratios of
$BR(\bar{B}_{_s}\rightarrow \mu^+\mu^-)$, we plot the correlation
between $\mu_{_H}$ and $\tan\beta$ in Fib. \ref{fig16} (b).
Similar to Fig. \ref{fig16} except for
$\theta_{_t}=\pi/2$, we plot the correlation of $\mu_{_H}$ and
$\tan\beta$ with the bound of the branching ratio $BR(B\rightarrow
X_{_s}\gamma)$ in Fig.\ref{fig17} (a). and possible new upper
bound on the branching ratio $BR(\bar{B}_{_s}\rightarrow
\mu^+\mu^-)$ (Fig. \ref{fig17}(b)). When $\theta_{_t}=0$, the rare
decay $\bar{B}_{_s}\rightarrow\mu^+\mu^-$ will lead to a concrete
constraint on the parameter space if the new experimental upper
bound on the branching ratio
$BR(\bar{B}_{_s}\rightarrow\mu^+\mu^-)$ reaches $10^{-7}$. As for
the case $\theta_{_t}=\pi/2$, the branching ratio
$BR(\bar{B}_{_s}\rightarrow\mu^+\mu^-)$ at $10^{-7}$ level will
raise a stronger constraint on the model discussed here.
Certainly, we plot Fig. \ref{fig16} and Fig. \ref{fig17} under the
hypothesis that the scalar quarks of the third generation are
relatively light. If we push the scalar quark masses of the third
generation to $\ge 1{\rm TeV}$, the situation would change
drastically. Here we do not discuss such cases any further.

\begin{figure}
\setlength{\unitlength}{1mm}
\begin{center}
\begin{picture}(80,120)(55,60)
\put(30,50){\includegraphics{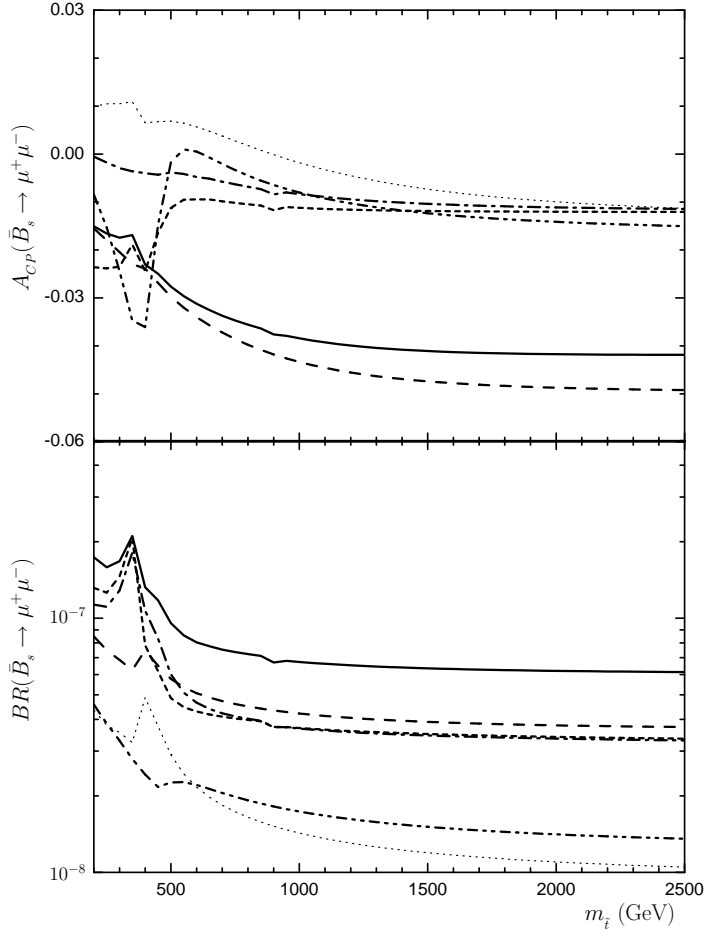}}
\end{picture}
\caption[]{Dependence of the branching ratio
$BR(\bar{B}_{_s}\rightarrow\mu^+\mu^-)$ and CP asymmetry
$A_{_{CP}}(\bar{B}_{_s}\rightarrow\mu^+\mu^-)$ on the right handed
scalar quark mass $m_{_{\tilde t}}$, with $\tan\beta=40$ and
$\mu_{_H}=-50\;({\rm GeV})$. In this figure, (a)solid line stands
for rigorous two loop analysis at $\theta_{_2}=\pi/2$ and
$\theta_{_3}=\theta_{_t}=0$, (b)dash line stands for one loop
results plus threshold radiative corrections at
$\theta_{_2}=\pi/2$ and $\theta_{_3}=\theta_{_t}=0$, (c)dash-dot
line stands for rigorous two loop analysis at $\theta_{_3}=\pi/2$
and $\theta_{_2}=\theta_{_t}=0$, (d)dot line stands for one loop
results plus threshold radiative corrections at
$\theta_{_3}=\pi/2$ and $\theta_{_2}=\theta_{_t}=0$, (e)short-dash
line stands stands for rigorous two loop analysis at
$\theta_{_t}=\pi/2$ and $\theta_{_2}=\theta_{_3}=0$,
(f)dash-dot-dot line stands for one loop results plus threshold
radiative corrections at $\theta_{_t}=-\pi/2$ and
$\theta_{_2}=\theta_{_3}=0$.} \label{fig18}
\end{center}
\end{figure}

\begin{figure}
\setlength{\unitlength}{1mm}
\begin{center}
\begin{picture}(80,120)(55,60)
\put(30,50){\includegraphics{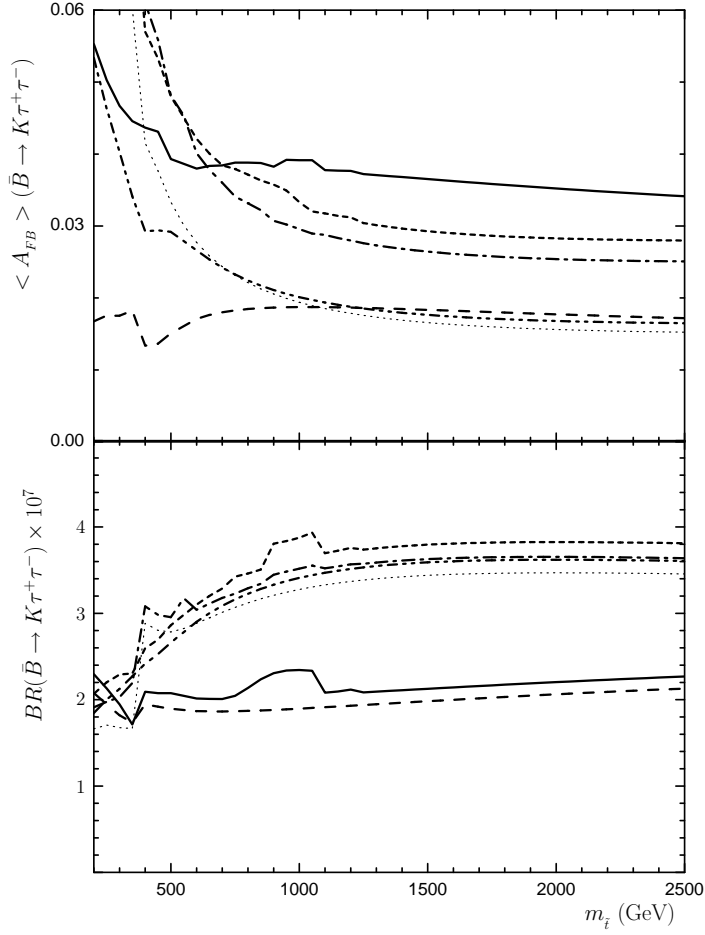}}
\end{picture}
\caption[]{Dependence of the branching ratio
$BR(\bar{B}\rightarrow K\tau^+\tau^-)$ and average
forward-backward asymmetry $A_{_{FB}}(\bar{B}\rightarrow
K\tau^+\tau^-)$ on the right handed scalar quark mass $m_{_{\tilde
t}}$, with $\tan\beta=40$ and $\mu_{_H}=-50\;({\rm GeV})$. In this
figure, (a)solid line stands for rigorous two loop analysis at
$\theta_{_2}=\pi/2$ and $\theta_{_3}=\theta_{_t}=0$, (b)dash line
stands for one loop results plus threshold radiative corrections
at $\theta_{_2}=\pi/2$ and $\theta_{_3}=\theta_{_t}=0$,
(c)dash-dot line stands for rigorous two loop analysis at
$\theta_{_3}=\pi/2$ and $\theta_{_2}=\theta_{_t}=0$, (d)dot line
stands for one loop results plus threshold radiative corrections
at $\theta_{_3}=\pi/2$ and $\theta_{_2}=\theta_{_t}=0$,
(e)short-dash line stands stands for rigorous two loop analysis at
$\theta_{_t}=\pi/2$ and $\theta_{_2}=\theta_{_3}=0$,
(f)dash-dot-dot line stands for one loop results plus threshold
radiative corrections at $\theta_{_t}=-\pi/2$ and
$\theta_{_2}=\theta_{_3}=0$.} \label{fig19}
\end{center}
\end{figure}

Now let us simply discuss the dependance of the branching ratios
and CP asymmetries (or forward-back asymmetries) on squark masses
in the rare processes$\bar{B}_{_s}\rightarrow l^+l^-$ and
$\bar{B}\rightarrow Kl^+l^-$. Taking $\tan\beta=40$ and
$\mu_{_H}=-50\;({\rm GeV})$, we plot the branching ratio
$BR(\bar{B}_{_s}\rightarrow\mu^+\mu^-)$ as well as the CP
asymmetry $A_{_{CP}}(\bar{B}_{_s}\rightarrow\mu^+\mu^-)$ versus
the mass of right handed scalar top $m_{_{\tilde t}}$ in Fig.
\ref{fig18}. Owing to the interference between the contributions
of left handed  and  right handed stop, there is a resonant peak
at $m_{_{\tilde t}}=400\;{\rm GeV}$. When $m_{_{\tilde
t}}>1.5\;{\rm TeV}$, the dependance of the branching ratio
$BR(\bar{B}_{_s}\rightarrow\mu^+\mu^-)$ and the CP asymmetry
$A_{_{CP}}(\bar{B}_{_s}\rightarrow\mu^+\mu^-)$ on $m_{_{\tilde
t}}$ is very gentle. With the same choice of the parameter space,
we plot the branching ratio  $BR(\bar{B}\rightarrow
K\tau^+\tau^-)$ as well as the Forward-Back asymmetry
$A_{_{FB}}(\bar{B}\rightarrow K\tau^+\tau^-)$ versus the mass of
right handed scalar top $m_{_{\tilde t}}$ in Fig. \ref{fig19}. The
resonance at $m_{_{\tilde t}}=400\;{\rm GeV}$ also originates from
the interference between the contributions of left handed and
right handed stop. For the strict two-loop theoretical predictions
on $BR(\bar{B}\rightarrow K\tau^+\tau^-)$, there is a small
resonance around $m_{_{\tilde t}}=1\;{\rm TeV}$ which is due to
the interference between the contributions of right handed stop
and that of right handed scalar s-quark. A similar discussion
about the dependence of the branching ratio $BR(B\rightarrow
X_{_s}\gamma)$ and the CP asymmetry $A_{_{CP}}(B\rightarrow
X_{_s}\gamma)$ on squark masses was given in our previous work
\cite{Feng1}, and here we omit repetitions. It is noted that at
large $\tan\beta$ scenarios, the corrections from sbottom on the
rare process $b\rightarrow sl^+l^-$  only originate from two-loop
box diagrams. This is the reason why the dependence of those
branching ratios, CP asymmetries, and Forward-Back asymmetries on
the sbottom masses is very weak.

\section{Conclusion}

\indent\indent Considering the constraints from the branching
ratio $BR(B\rightarrow X_{_s}\gamma)$ and $\Delta M_{_{B_s}}$, we
discuss the rare processes $B_{_s}\rightarrow\mu^+\mu^-$,
$\bar{B}_{_s}\rightarrow\tau^+\tau^-$, $\bar{B}\rightarrow
K\mu^+\mu^-$ and $\bar{B}\rightarrow K\tau^+\tau^-$ in the CP
violating MSSM at large $\tan\beta$. We find that there are
evident differences between the theoretical predictions of exact
two loop calculations and that of one loop contributions plus
threshold radiative corrections. Additionally, the branching ratio
$BR(\bar{B}_{_s}\rightarrow\mu^+\mu^-)$ of exact two loop
calculations can exceed $10^{-8}$, while the CP asymmetry can
reach 3\%. Although the two loop analysis predicts
$BR(\bar{B}\rightarrow K\mu^+\mu^-)>10^{-7}$, the average
forward-backward asymmetry $<A_{_{FB}}>(\bar{B}\rightarrow
K\mu^+\mu^-)$ is too small to be detected in future experiments.
For the rare decay $\bar{B}_{_s}\rightarrow\tau^+\tau^-$, one has
$BR(\bar{B}_{_s}\rightarrow \tau^+\tau^-)>10^{-6}$, meanwhile
$A_{_{CP}}(\bar{B}_{_s}\rightarrow \tau^+\tau^-)\sim3\%$. Maybe,
the most interesting object to study is the branching ratio
$BR(\bar{B}\rightarrow K\tau^+\tau^-)>10^{-7}$ along with a large
average forward-backward asymmetry $<A_{_{FB}}>(\bar{B}\rightarrow
K\tau^+\tau^-)\sim 20\%$. As a by-product, we also find that the
CP asymmetry of inclusive decay $A_{_{CP}}(B\rightarrow
X_{_s}\gamma)$ can reach 5\%, which is much larger than the SM
prediction also.

\begin{acknowledgments}

The work has been supported by the Academy of Finland under the
contracts no.\ 104915 and 107293, the ABRL Grant No.
R14-2003-012-01001-0 of Korea, and also partly  by the National Natural
Science Foundation of China(NNSFC).
\end{acknowledgments}
\vspace{2.0cm}
\appendix
\section{The one-loop Wilson coefficients at the weak scale\label{1loop}}

The  Wilson coefficients at the weak scale in Eq. (\ref{Ham}) are
given as
\begin{eqnarray}
&&C_7(\mu_{_{\rm W}})={\xi_{_{CKM}}^{ts*}\xi_{_{CKM}}^{tb}\over6}
\bigg\{x_{_t}\Big[\Big({18x_{_t}^2-11x_{_t}-1\over
4(1-x_{_t})^3}+{15x_{_t}^2-16x_{_t} +4\over 2(1-x_{_t})^4}\ln
x_{_t}\Big)
\nonumber\\
&&\hspace{2.0cm}
+\xi_{_G}^b\Big({5x_{_t}-3\over2(1-x_{_t})^2}
-{2-3x_{_t}\over (1-x_{_t})^3}\ln x_{_t}\Big)\Big]
\nonumber\\
&&\hspace{2.0cm}
+{x_{_t}\over\tan^2\beta}\Big[{4x_{_t}^2+x_{_t}x_{_H}+25x_{_H}^2\over
36(x_{_t}-x_{_H})^3} -{(3x_{_t}^2x_{_H}+2x_{_t}x_{_H}^2)(\ln
x_{_t} -\ln x_{_H})\over6(x_{_t}-x_{_H})^4}\Big]
\nonumber\\
&&\hspace{2.0cm}
-\xi_{_H}^bx_{_t}\Big[{1\over
(x_{_t}-x_{_H})}+{(x_{_H}^2+x_{_t}x_{_H})(\ln x_{_t}-\ln
x_{_H})\over (x_{_t}-x_{_H})^3}\Big]
\nonumber\\
&&\hspace{2.0cm}
-|\Gamma_{_{\tilde{t}_i\chi_\alpha}}^L|^2\bigg[\Big({3\over2}x_{_{\chi_\alpha}}
{\partial^2\over\partial^2x_{_{\chi_\alpha}}}+{1\over6}x_{_{\chi_\alpha}}^2
{\partial^3\over\partial^3x_{_{\chi_\alpha}}}\Big)\varrho_{_{1,1}}
(x_{_{\chi_\alpha}},x_{_{\tilde{t}_i}})
\nonumber\\
&&\hspace{2.0cm}
+{\partial^2\over\partial^2x_{_{\tilde{t}_i}}}\varrho_{_{2,1}}
(x_{_{\tilde{t}_i}},x_{_{\chi_\alpha}})
-{1\over3}{\partial^3\over\partial^3x_{_{\tilde{t}_i}}}\varrho_{_{3,1}}
(x_{_{\tilde{t}_i}},x_{_{\chi_\alpha}})\bigg]
\nonumber\\
&&\hspace{2.0cm}
-{\sqrt{2x_{_{\chi_\alpha}}}\over c_{_\beta}}
\Gamma_{_{\tilde{t}_i\chi_\alpha}}^L\Gamma_{_{\tilde{t}_i\kappa_\alpha}}^{R*}
\bigg[{3\over2}{\partial^2\over\partial^2x_{_{\chi_\alpha}}}\varrho_{_{2,1}}
(x_{_{\chi_\alpha}},x_{_{\tilde{t}_i}})
+{\partial^2\over\partial^2x_{_{\tilde{t}_i}}}\varrho_{_{2,1}}
(x_{_{\tilde{t}_i}},x_{_{\chi_\alpha}})
\nonumber\\
&&\hspace{2.0cm}
-2{\partial\over\partial x_{_{\tilde{t}_i}}}\varrho_{_{1,1}}
(x_{_{\tilde{t}_i}},x_{_{\chi_\alpha}})\bigg]\bigg\}
\;,\nonumber\\
&&C_8(\mu_{_{\rm W}})=-{\xi_{_{CKM}}^{ts*}\xi_{_{CKM}}^{tb}\over2}\bigg\{
x_{_t}\Big[{(5x_{_t}-2)\ln x_{_t}\over2(1-x_{_t})^4}
-{3x_{_t}^2-13x_{_t}+4\over4(1-x_{_t})^3}\Big]
\nonumber\\
&&\hspace{2.0cm}
+\xi_{_G}^bx_{_t}\Big[{\ln x_{_t}\over(1-x_{_t})^3}
+{3-x_{_t}\over2(1-x_{_t})^2}\Big]
\nonumber\\
&&\hspace{2.0cm}
+\xi_{_H}^bx_{_t}\Big[{x_{_H}^2(\ln
x_{_t}-\ln x_{_H})\over(x_{_H}-x_{_t})^3}+{3x_{_H}
-x_{_t}\over2(x_{_H}-x_{_t})^2}\Big]
\nonumber\\
&&\hspace{2.0cm}
-{x_{_t}\over\tan^2\beta}\Big[ {x_{_t}x_{_H}^2(\ln
x_{_t}-\ln x_{_H})\over2(x_{_H}-x_{_t})^4}
+{2x_{_H}^2+5x_{_t}x_{_H}-x_{_t}^2\over12(x_{_H}-x_{_t})^3}\Big]
\nonumber\\
&&\hspace{2.0cm}
+{|\Gamma_{_{\tilde{t}_i\chi_\alpha}}^L|^2\over2}\bigg[
{\partial^2\over\partial^2x_{_{\tilde{t}_i}}}\varrho_{_{2,1}}
(x_{_{\tilde{t}_i}},x_{_{\chi_\alpha}})
-{1\over3}{\partial^3\over\partial^3x_{_{\tilde{t}_i}}}\varrho_{_{3,1}}
(x_{_{\tilde{t}_i}},x_{_{\chi_\alpha}})\bigg]
\nonumber\\
&&\hspace{2.0cm}
-{\sqrt{2x_{_{\chi_\alpha}}}\over2c_{_\beta}}
\Gamma_{_{\tilde{t}_i\chi_\alpha}}^L\Gamma_{_{\tilde{t}_i\kappa_\alpha}}^{R*}
\bigg[{\partial^2\over\partial^2x_{_{\tilde{t}_i}}}\varrho_{_{2,1}}
(x_{_{\tilde{t}_i}},x_{_{\chi_\alpha}})
-2{\partial\over\partial x_{_{\tilde{t}_i}}}\varrho_{_{1,1}}
(x_{_{\tilde{t}_i}},x_{_{\chi_\alpha}})\bigg]\bigg\}
\;,\nonumber\\
&&C_9^\gamma(\mu_{_{\rm W}})={\xi_{_{CKM}}^{ts*}\xi_{_{CKM}}^{tb}\over9}
\bigg\{\Big[{-19x_{_t}^3+25x_{_t}^2\over2(1-x_{_t})^3}+{3x_{_t}^4-30x_{_t}^3
+54x_{_t}^2-32x_{_t}+8\over(1-x_{_t})^4}\ln x_{_t}\Big]
\nonumber\\
&&\hspace{2.0cm}
+{x_{_t}\over\tan^2\beta}\Big[-{19x_{_t}^2+109x_{_t}x_{_H}-98x_{_H}^2\over18(x_{_t}
-x_{_H})^3}+{(3x_{_t}^3-9x_{_t}^2x_{_H}-4x_{_H}^3)(\ln x_{_t}-\ln
x_{_H})\over6(x_{_t}-x_{_H})^4}\Big)\Big]
\nonumber\\
&&\hspace{2.0cm}
-|\Gamma_{_{\tilde{t}_i\chi_\alpha}}^{L}|^2\bigg[
\Big(3{\partial\over\partial x_{_{\chi_\alpha}}}
-{15\over4}x_{_{\chi_\alpha}}{\partial^2\over\partial^2x_{_{\chi_\alpha}}}
-{7\over4}x_{_{\chi_\alpha}}^2{\partial^3\over\partial^3x_{_{\chi_\alpha}}}\Big)
\varrho_{_{1,1}}(x_{_{\chi_\alpha}},x_{_{\tilde{t}_i}})
\nonumber\\
&&\hspace{2.0cm}
-{1\over6}{\partial^3\over\partial x_{_{\tilde{t}_i}}^3}
\varrho_{_{3,1}}(x_{_{\tilde{t}_i}},x_{_{\chi_\alpha}})\bigg]
\bigg\}
\;,\nonumber\\
&&C_9^Z(\mu_{_{\rm W}})=-{1-4s_{_{\rm W}}^2\over
8s_{_{\rm W}}^2}\xi_{_{CKM}}^{ts*}\xi_{_{CKM}}^{tb}
\bigg\{\Big[{x_{_t}(6-x_{_t})\over1-x_{_t}}
+{(3x_{_t}^2+2x_{_t})\ln x_{_t}\over(1-x_{_t})^2}\Big]
\nonumber\\
&&\hspace{2.0cm}
+\Big[{(3-2s_{_{\rm W}}^2)\over
6\tan^2\beta}x_{_t}{\partial\over\partial x_{_t}}
\varrho_{_{2,1}}(x_{_t},x_{_H})+{(3-4s_{_{\rm W}}^2)\over
3\tan^2\beta}x_{_t}^2{\partial\over\partial x_{_t}}
\varrho_{_{1,1}}(x_{_t},x_{_H})\Big]
\nonumber\\
&&\hspace{2.0cm}
+\bigg[\Gamma_{_{\tilde{t}_i\chi_\alpha}}^{L}
\Gamma_{_{\tilde{t}_i\chi_\beta}}^{L*}
\Big(\xi_{_{\chi}}^4\Big)_{_{\alpha\beta}}
T_1(x_{_{\tilde{t}_i}},x_{_{\chi_\alpha}},x_{_{\chi_\beta}})
-\Gamma_{_{\tilde{t}_i\chi_\alpha}}^{L}
\Gamma_{_{\tilde{t}_j\chi_\alpha}}^{L*}\Big(\xi_{_{t}}\Big)_{_{ij}}
T_1(x_{_{\tilde{t}_i}},x_{_{\chi_\alpha}},x_{_{\chi_\beta}})
\nonumber\\
&&\hspace{2.0cm}
-2\sqrt{x_{_{\chi_\alpha}}x_{_{\chi_\beta}}}
\Gamma_{_{\tilde{t}_i\chi_\alpha}}^{L}
\Gamma_{_{\tilde{t}_i\chi_\beta}}^{L*}
\Big(\xi_{_{\chi}}^1\Big)_{_{\alpha\beta}}
T_0(x_{_{\tilde{t}_i}},x_{_{\chi_\alpha}},x_{_{\chi_\beta}})
\nonumber\\
&&\hspace{2.0cm}
+{3-2s_{_{\rm W}}^2\over12}|\Gamma_{_{\tilde{t}_i\chi_\alpha}}^{L}|^2
{\partial\over\partial x_{_{\tilde{t}_i}}}\varrho_{_{2,1}}(
x_{_{\tilde{t}_i}},x_{_{\chi_\alpha}})\bigg\}
\;,\nonumber\\
&&C_{10}^Z(\mu_{_{\rm W}})={1\over 1-4s_{_{\rm W}}^2}C_9^Z(\mu_{_{\rm
W}})\;,\nonumber\\
&&C_{_{S}}^{H}(\mu_{_{\rm W}})
=-{\sqrt{x_{_l}x_{_b}}\over4s_{_{\rm W}}^2s_{_\beta}c_{_\beta}}
\xi_{_{CKM}}^{ts*}\xi_{_{CKM}}^{tb}\sum\limits_{k=1}^3
{{\cal Z}_{_H}^{2k}\over x_{_{H_k^0}}}\bigg\{\xi_{_G}^bs_{_\beta}\Big(c_{_\beta}{\cal
Z}_{_H}^{2k}-s_{_\beta}{\cal Z}_{_H}^{3k}\Big)E(x_{_t})
\nonumber\\
&&\hspace{2.0cm}
+\xi_{_H}^bs_{_\beta}\tan\beta\Big(s_{_\beta}{\cal Z}_{_H}^{2k}+c_{_\beta}{\cal
Z}_{_H}^{3k}+i{\cal Z}_{_H}^{1k}\Big)Q_+(x_{_t},x_{_{H}})
+\xi_{_{G}}^bx_{_t}^2\bigg[\Big({\cal Z}_{_H}^{3k}
\nonumber\\
&&\hspace{2.0cm}
+i3c_{_\beta}{\cal Z}_{_H}^{1k}\Big)R_1(x_{_t})+\Big(c_{_\beta}^2
(1+2s_{_\beta}^2){\cal Z}_{_H}^{3k}\Big)
-s_{_\beta}c_{_\beta}(1-2s_{_\beta}^2){\cal Z}_{_H}^{2k}+ic_{_\beta}{\cal
Z}_{_H}^{1k}\Big)R_2(x_{_t})\bigg]
\nonumber\\
&&\hspace{2.0cm}
-\xi_{_{H}}^bx_{_t}\bigg[\Big({\cal
Z}_{_H}^{3k}+ic_{_\beta}{\cal
Z}_{_H}^{1k}\Big){\partial\over\partial x_{_t}}\varrho_{_{2,1}}(x_{_t},x_{_{H}})
+\Big({\cal Z}_{_H}^{3k}-ic_{_\beta}{\cal Z}_{_H}^{1k}\Big)x_{_t}
{\partial\over\partial x_{_t}}\varrho_{_{1,1}}(x_{_t},x_{_{H}})
\nonumber\\
&&\hspace{2.0cm}
-s_{_\beta}\Big(c_{_\beta}(1+2s_{_\beta}^2){\cal
Z}_{_H}^{2k}+s_{_\beta}(1+2c_{_\beta}^2){\cal
Z}_{_H}^{3k}\Big)x_{_t}{\partial\over\partial x_{_{H}}}
\varrho_{_{1,1}}(x_{_H},x_{_t})\bigg]
\nonumber\\
&&\hspace{2.0cm}
+\bigg[{c_{_\beta}^2-s_{_\beta}^2\over c_{_\beta}}
\Big(s_{_\beta}[s_{_\beta}-(1+2{s_{_{\rm
W}}^2\over c_{_{\rm W}}^2})c_{_\beta}^2]{\cal
Z}_{_H}^{2k}+c_{_\beta}[(1+2{s_{_{\rm W}}^2\over c_{_{\rm W}}
^2})s_{_\beta}^2-c_{_\beta}^2]{\cal Z}_{_H}^{3k}\Big)
\nonumber\\
&&\hspace{2.0cm}
+{i\over c_{_\beta}}{\cal Z}_{_H}^{1k}\bigg]x_{_t}P_+(x_{_t},x_{_H})
+2\tan\beta\bigg[\Gamma_{_{\tilde{t}_i\chi_\alpha}}^{L}
\Gamma_{_{\tilde{t}_i\chi_\beta}}^{R*}
\Big(\kappa_{_{H^{^k}}}^1\Big)_{_{\alpha\beta}}T_1(
x_{_{\tilde{t}_i}},x_{_{\chi_\alpha}},x_{_{\chi_\beta}})
\nonumber\\
&&\hspace{2.0cm}
+\sqrt{x_{_{\chi_\alpha}}x_{_{\chi_\beta}}}
\Gamma_{_{\tilde{t}_i\chi_\alpha}}^{L}
\Gamma_{_{\tilde{t}_i\chi_\beta}}^{R*}
\Big(\kappa_{_{H^{^k}}}^2\Big)_{_{\alpha\beta}}T_0(
x_{_{\tilde{t}_i}},x_{_{\chi_\alpha}},x_{_{\chi_\beta}})
\nonumber\\
&&\hspace{2.0cm}
-{\sqrt{2x_{_{\chi_\alpha}}}\over c_{_{\rm w}}^2}
\Gamma_{_{\tilde{t}_i\chi_\alpha}}^{L}
\Gamma_{_{\tilde{t}_j\chi_\alpha}}^{R*}
\Big(\zeta_{_{tH}}^k\Big)_{_{ij}}T_0
(x_{_{\chi_\alpha}},x_{_{\tilde{t}_i}},x_{_{\tilde{t}_j}})
\bigg\}
\;,\nonumber\\
&&C_{_{P}}^{H}(\mu_{_{\rm W}})=i{s_{_\beta}{\cal Z }_{_H}^{1k}\over
{\cal Z }_{_H}^{2k}}C_{_{S}}^{H}(\mu_{_{\rm W}})\;,\nonumber\\
&&C_{_{S}}^{count}(\mu_{_{\rm
W}})=-{G_{_F}\sqrt{x_{_b}x_{_l}}\over4\sqrt{2}c_{_\beta}^2}
\xi_{_{CKM}}^{ts*}\xi_{_{CKM}}^{tb}\chi_{_B}\sum\limits_{k=1}^3
{1\over x_{_{H_k^0}}}{\cal Z}_{_H}^{2k}\Big({\cal Z}_{_H}^{2k}+\Delta{\cal Z
}_{_H}^{3k}
\nonumber\\
&&\hspace{2.5cm}
+i[s_{_\beta}-c_{_\beta}\Delta]{\cal Z}_{_H}^{1k}\Big)
\bigg[\sqrt{2}x_{_t}(1-x_{_H})P_+(x_{_H},x_{_t})
\nonumber\\
&&\hspace{2.5cm}
+{\sqrt{x_{_{\chi_\alpha}}}\over c_{_\beta}}\Gamma_{_{\tilde{t}_i\chi_\alpha}}^{R*}
\Gamma_{_{\tilde{t}_i\chi_\alpha}}^{L}
\varrho_{_{1,1}}(x_{_{\tilde{t}_i}} ,x_{_{\chi_\alpha}})
\bigg]\;,\nonumber\\
&&C_{_{P}}^{count}(\mu_{_{\rm
W}})=i{G_{_F}s_{_\beta}\sqrt{x_{_b}x_{_l}}\over4\sqrt{2}c_{_\beta}^2}
\xi_{_{CKM}}^{ts*}\xi_{_{CKM}}^{tb}\chi_{_B}\sum\limits_{k=1}^3
{1\over x_{_{H_k^0}}}{\cal Z}_{_H}^{1k}
\Big({\cal Z}_{_H}^{2k}+\Delta{\cal Z}_{_H}^{3k}
\nonumber\\
&&\hspace{2.5cm}
+i[s_{_\beta}-c_{_\beta}\Delta]{\cal Z}_{_H}^{1k}\Big)
\bigg[\sqrt{2}x_{_t}(1-x_{_H})P_+(x_{_H},x_{_t})
\nonumber\\
&&\hspace{2.5cm}
+{\sqrt{x_{_{\chi_\alpha}}}\over c_{_\beta}}\Gamma_{_{\tilde{t}_i\chi_\alpha}}^{R*}
\Gamma_{_{\tilde{t}_i\chi_\alpha}}^{L}
\varrho_{_{1,1}}(x_{_{\tilde{t}_i}} ,x_{_{\chi_\alpha}})
\bigg]\;,\nonumber\\
&&C_9^{box}(\mu_{_{\rm W}})=-{\xi_{_{CKM}}^{ts*}\xi_{_{CKM}}^{tb}
\over16s_{_{\rm w}}^2}\bigg\{\Big[-x_{_t}x_{_l}R_2(x_{_t})
+x_{_t}x_{_l}{\partial\over\partial x_{_H}}\varrho_{_{1,1}}(x_{_H},x_{_t})
\nonumber\\
&&\hspace{2.0cm}
- 2x_{_t}x_{_l}P_+(x_{_t},x_{_H})
-4x_{_t}R_1(x_{_t})\Big]\nonumber\\
&&\hspace{2.0cm}
+4\Gamma_{_{\tilde{t}_i\chi_\beta}}^{L}
\Gamma_{_{\tilde{t}_i\chi_\alpha}}^{L*}\Big({\cal Z}_+\Big)^\dagger_{\alpha1}
\Big({\cal Z}_+\Big)_{1\beta}D_1(x_{_{\tilde{t}_i}},x_{_{\tilde{\nu}_I}},
x_{_{\chi_\alpha}},x_{_{\chi_\beta}})
\nonumber\\
&&\hspace{2.0cm}
-{4x_{_l}\over c_{_\beta}^2}\Gamma_{_{\tilde{t}_i\chi_\beta}}^{L}
\Gamma_{_{\tilde{t}_\beta\chi_\alpha}}^{L*}\Big({\cal Z}_-\Big)_{2\alpha}
\Big({\cal Z}_-\Big)^\dagger_{\beta2}\sqrt{x_{_{\chi_\alpha}}x_{_{\chi_\beta}}}D_0
(x_{_{\tilde{t}_i}},x_{_{\tilde{\nu}_I}},
x_{_{\chi_\alpha}},x_{_{\chi_\beta}})\bigg\}
\;,\nonumber\\
&&C_{10}^{box}(\mu_{_{\rm W}})=-{\xi_{_{CKM}}^{ts*}\xi_{_{CKM}}^{tb}
\over16s_{_{\rm w}}^2}\bigg\{\Big[-x_{_t}x_{_l}R_2(x_{_t})
+x_{_t}x_{_l}{\partial\over\partial x_{_H}}\varrho_{_{1,1}}(x_{_H},x_{_t})
\nonumber\\
&&\hspace{2.0cm}
- 2x_{_t}x_{_l}P_+(x_{_t},x_{_H})
-4x_{_t}R_1(x_{_t})\Big]\nonumber\\
&&\hspace{2.0cm}
-4\Gamma_{_{\tilde{t}_i\chi_\beta}}^{L}
\Gamma_{_{\tilde{t}_i\chi_\alpha}}^{L*}\Big({\cal Z}_+\Big)^\dagger_{\alpha1}
\Big({\cal Z}_+\Big)_{1\beta}D_1(x_{_{\tilde{t}_i}},x_{_{\tilde{\nu}_I}},
x_{_{\chi_\alpha}},x_{_{\chi_\beta}})
\nonumber\\
&&\hspace{2.0cm}
+{4x_{_l}\over c_{_\beta}^2}\Gamma_{_{\tilde{t}_i\chi_\beta}}^{L}
\Gamma_{_{\tilde{t}_\beta\chi_\alpha}}^{L*}\Big({\cal Z}_-\Big)_{2\alpha}
\Big({\cal Z}_-\Big)^\dagger_{\beta2}\sqrt{x_{_{\chi_\alpha}}x_{_{\chi_\beta}}}D_0
(x_{_{\tilde{t}_i}},x_{_{\tilde{\nu}_I}},
x_{_{\chi_\alpha}},x_{_{\chi_\beta}})\bigg\}
\;,\nonumber\\
&&C_{9}^{\prime box}(\mu_{_{\rm W}})=-{\xi_{_{CKM}}^{ts*}\xi_{_{CKM}}^{tb}
\over16s_{_{\rm w}}^2}x_{_l}\sqrt{x_{_b}x_{_s}}\tan^4\beta\bigg\{\xi_{_H}^b
\Big[{\partial\over\partial x_{_H}}\varrho_{_{1,1}}(x_{_H},x_{_t})
-{1\over x_{_H}}\Big]
\nonumber\\&&\hspace{2.0cm}
+{2\chi_{_B}\over s_{_\beta}^4}\Gamma_{_{\tilde{t}_i\chi_\alpha}}^{R}
\Gamma_{_{\tilde{t}_i\chi_\beta}}^{R*}\Big({\cal Z}_-\Big)_{2\alpha}
\Big({\cal Z}_-\Big)^\dagger_{\beta2}
D_1(x_{_{\tilde{t}_i}},x_{_{\tilde{\nu}_I}},
x_{_{\chi_\alpha}},x_{_{\chi_\beta}})\bigg\}\;,\nonumber\\
&&C_{10}^{\prime box}(\mu_{_{\rm W}})=C_{9}^{\prime box}(\mu_{_{\rm W}})
\;,\nonumber\\
&&C_{_{S}}^{box}(\mu_{_{\rm W}})=-{\xi_{_{CKM}}^{ts*}\xi_{_{CKM}}^{tb}
\over 4s_{_{\rm w}}^2}\sqrt{x_{_l}x_{_b}}\bigg\{\Big[\xi_{_G}^b
E(x_{_t})+\xi_{_G}^b\tan^2\beta E_+(x_{_H},x_{_t})\Big]
\nonumber\\
&&\hspace{2.0cm}
-{\chi_{_B}\over c_{_\beta}^2}
\Gamma_{_{\tilde{t}_i\chi_\beta}}^{L}
\Gamma_{_{\tilde{t}_i\chi_\alpha}}^{R*}\Big({\cal Z}_-\Big)_{2\alpha}
\Big({\cal Z}_+\Big)_{1\beta}D_1(x_{_{\tilde{t}_i}}
,x_{_{\tilde{\nu}_I}},x_{_{\chi_\alpha}},x_{_{\chi_\beta}})
\;,\nonumber\\
&&C_{_{P}}^{box}(\mu_{_{\rm W}})=C_{_{S}}^{box}(\mu_{_{\rm
W}})\;.\nonumber\\
\label{coefficient}
\end{eqnarray}
The one-loop integrands are defined as
\begin{eqnarray}
&&E(x)={x\ln x\over(1-x)^2}\;,\nonumber\\
&&R_1(x)={1\over1-x}+{\ln x\over(1-x)^2}\;,\nonumber\\
&&R_2(x)=-\Big({1\over1-x}+{x\ln x\over(1-x)^2}\Big)\;,\nonumber\\
&&R_3(x)=-\Big({1\over1-x}+{x^2\ln x\over(1-x)^2}\Big)\;,\nonumber\\
&&E_+(x,y)={y\over x-y}\Big[{\ln x\over 1-x}-{\ln y\over 1-y}\Big]\;,\nonumber\\
&&P_+(x,y)={1\over x-y}\Big[{x\ln x\over 1-x}-{y\ln y\over 1-y}\Big]\;,\nonumber\\
&&Q_+(x,y)=-{x\over x-y}\Big[{\ln x\over (x-1)^2}-{y\ln y\over y-1}\Big]
\;,\nonumber\\
&&\varrho_{_{m,n}}(x,y)={x^m\ln^nx-y^m\ln^ny\over x-y}
\;,\nonumber\\
&&T_0(x_1,x_2,x_3)=\sum\limits_{i=1}^3{x_i\ln x_i\over\prod\limits_{j\neq i}
(x_j-x_i)}
\;,\nonumber\\
&&T_1(x_1,x_2,x_3)=\sum\limits_{i=1}^3{x_i^2\ln x_i\over\prod\limits_{j\neq i}
(x_j-x_i)}\;,\nonumber\\
&&D_0(x_1,x_2,x_3,x_4)=\sum\limits_{i=1}^4{x_i\ln x_i\over\prod\limits_{j\neq i}
(x_j-x_i)}
\;,\nonumber\\
&&D_1(x_1,x_2,x_3,x_4)=\sum\limits_{i=1}^4{x_i^2\ln x_i\over\prod\limits_{j\neq i}
(x_j-x_i)}\;.
\label{1lf}
\end{eqnarray}

\section{The two-loop corrections from the $Z,\;H$ penguin-diagrams\label{2t}}
\indent\indent
\begin{eqnarray}
&&{\cal P}_{_Z}^{(1)}=
-{2\over3}(c_{_{\rm w}}^2-s_{_{\rm w}}^2)({\cal Z}_{_{\tilde s}})_{_{2,i}}
\Big(\xi_{_{H}}\Big)_{_{ij}}
\bigg\{\Big[({\cal Z}_{_{\tilde t}}^\dagger)_{_{j,2}}
(x_{_{\rm w}}x_{_t})^{1/2}{\partial\over\partial x_{_H}}\Psi_{_{1b}}
\nonumber\\
&&\hspace{1.3cm}
-({\cal Z}_{_{\tilde t}}^\dagger)_{_{j,1}}
e^{-i\theta_{_3}}(x_{_{\rm w}}x_{_{\tilde g}})^{1/2}
{\partial\over\partial x_{_H}}\Psi_{_{1a}}\Big](x_{_{\tilde{t}_j}};
x_{_t},x_{_H};x_{_{\tilde g}},x_{_{\tilde{s}_i}})\bigg\}
\;,\nonumber\\
&&{\cal P}_{_Z}^{(2)}=-{2\over3}({\cal Z}_{_{\tilde s}})_{_{2,i}}
\Big(\xi_{_{H}}\Big)_{_{ij}}
\bigg\{\Big[-\Big({1\over4}+{1\over3}s_{_{\rm w}}^2\Big)({\cal Z}_{_{\tilde t}}^\dagger)_{_{j,2}}
(x_{_{\rm w}}x_{_t})^{1/2}{\partial\over\partial x_{_t}}\Psi_{_{1b}}
\nonumber\\
&&\hspace{1.3cm}
+\Big({1\over4}-{1\over3}s_{_{\rm w}}^2\Big)
({\cal Z}_{_{\tilde t}}^\dagger)_{_{j,1}}
e^{-i\theta_{_3}}(x_{_{\rm w}}x_{_{\tilde g}})^{1/2}
{\partial\over\partial x_{_t}}\Psi_{_{1a}}
\nonumber\\
&&\hspace{1.3cm}
+{2\over3}s_{_{\rm w}}^2({\cal Z}_{_{\tilde t}}^\dagger)_{_{j,1}}
e^{i\theta_{_3}}(x_{_{\rm w}}x_{_{\tilde g}})^{1/2}x_{_t}
{\partial\over\partial x_{_t}}\Psi_{_{0}}\Big](x_{_{\tilde{t}_j}};
x_{_t},x_{_H};x_{_{\tilde g}},x_{_{\tilde{s}_i}})
\bigg\}\;,\nonumber\\
&&{\cal P}_{_Z}^{(3)}=-{2\over3}({\cal Z}_{_{\tilde s}})_{_{2,i}}
\Big(\xi_{_{s}}\Big)_{_{ik}}\Big(\xi_{_{H}}\Big)_{_{kj}}
\Big[({\cal Z}_{_{\tilde t}}^\dagger)_{_{j,2}}
(x_{_{\rm w}}x_{_t})^{1/2}
-e^{-i\theta_{_3}}(x_{_{\rm w}}x_{_{\tilde g}})^{1/2}
({\cal Z}_{_{\tilde t}}^\dagger)_{_{j,1}}\Big]
\nonumber\\
&&\hspace{1.3cm}\times
{\partial\over\partial x_{_{\tilde{s}_i}}}
\Psi_{_{1b}}(x_{_{\tilde{t}_j}};x_{_t},x_{_H};x_{_{\tilde g}},x_{_{\tilde{s}_i}})
\;,\nonumber\\
&&{\cal P}_{_Z}^{(4)}=
-{1\over6}({\cal Z}_{_{\tilde s}})_{_{2,i}}
\Big(\xi_{_{H}}\Big)_{_{ik}}\Big(\xi_{_{t}}\Big)_{_{kj}}
{1\over x_{_{\tilde{t}_j}}-x_{_{\tilde{t}_k}}}
\bigg\{\Big[({\cal Z}_{_{\tilde t}}^\dagger)_{_{j,2}}
(x_{_{\rm w}}x_{_t})^{1/2}\Big(\Psi_{_{1b}}-\Psi_{_{1a}}\Big)
\nonumber\\
&&\hspace{1.3cm}
-({\cal Z}_{_{\tilde t}}^\dagger)_{_{j,1}}
e^{-i\theta_{_3}}(x_{_{\rm w}}x_{_{\tilde g}})^{1/2}
\Big(\Psi_{_{1a}}-\Psi_{_{1b}}\Big)\Big](x_{_{\tilde{t}_j}};
x_{_{\tilde g}},x_{_{\tilde{s}_i}};x_{_t},x_{_H})
\nonumber\\
&&\hspace{1.3cm}
-(x_{_{\tilde{t}_j}}\rightarrow x_{_{\tilde{t}_k}})\bigg\}
\;,\nonumber\\
&&{\cal P}_{_Z}^{(5)}=
{1\over3\sqrt{2}s_{_\beta}}({\cal Z}_{_{\tilde s}})_{_{2,i}}
({\cal Z}_{_{\tilde s}}^\dagger)_{_{i,1}}({\cal Z}_{_{\tilde t}}^\dagger)_{_{j,1}}
({\cal Z}_-^\dagger)_{_{\beta,2}}{1\over x_{_{\chi_\alpha}}-x_{_{\chi_\beta}}}
\bigg\{\Big[({\cal Z}_{_{\tilde t}})_{_{1,j}}
({\cal Z}_-)_{_{1,\alpha}}\Big(\xi_{_{\chi}}^1\Big)_{_{\alpha\beta}}
\Big(\Psi_{_{2b}}
\nonumber\\
&&\hspace{1.3cm}
-\Psi_{_{2d}}-\varrho_{_{2,1}}(x_{_{\chi_\alpha}},x_{_{\tilde{t}_j}})\Big)
-{\sqrt{2}m_{_{\chi_\beta}}x_{_t}\over m_{_{\rm w}}s_{_\beta}}
({\cal Z}_{_{\tilde t}})_{_{1,j}}({\cal Z}_+^\dagger)_{_{\alpha,2}}
\Big(\xi_{_{\chi}}^4\Big)_{_{\alpha\beta}}\Psi_{_{1b}}
\nonumber\\
&&\hspace{1.3cm}
+{m_{_{\chi_\alpha}}x_{_t}\over\sqrt{2}m_{_{\rm w}}s_{_\beta}}
({\cal Z}_{_{\tilde t}})_{_{1,j}}({\cal Z}_+^\dagger)_{_{\alpha,2}}
\Big(\xi_{_{\chi}}^1\Big)_{_{\alpha\beta}}\Psi_{_{1b}}
-e^{-i\theta_{_3}}(x_{_{\tilde g}}x_{_t})^{1/2}
({\cal Z}_{_{\tilde t}})_{_{2,j}}({\cal Z}_-)_{_{1,\alpha}}
\Big(\xi_{_{\chi}}^1\Big)_{_{\alpha\beta}}\Psi_{_{1a}}
\nonumber\\
&&\hspace{1.3cm}
+2e^{-i\theta_{_3}}(x_{_{\tilde g}}x_{_t}x_{_{\chi_\alpha}}x_{_{\chi_\beta}})^{1/2}
({\cal Z}_{_{\tilde t}})_{_{2,j}}({\cal Z}_-)_{_{1,\alpha}}
\Big(\xi_{_{\chi}}^4\Big)_{_{\alpha\beta}}\Psi_{_{0}}
\nonumber\\
&&\hspace{1.3cm}
+e^{-i\theta_{_3}}{\sqrt{2}m_{_t}(x_{_{\tilde g}}
x_{_{\chi_\beta}})^{1/2}\over m_{_{\rm w}}s_{_\beta}}
({\cal Z}_{_{\tilde t}})_{_{2,j}}({\cal Z}_+^\dagger)_{_{\alpha,2}}
\Big(\xi_{_{\chi}}^4\Big)_{_{\alpha\beta}}
\Big(\Psi_{_{1a}}-\Psi_{_{1b}}\Big)
\nonumber\\
&&\hspace{1.3cm}
-e^{-i\theta_{_3}}{m_{_t}(x_{_{\tilde g}}
x_{_{\chi_\alpha}})^{1/2}\over\sqrt{2}m_{_{\rm w}}s_{_\beta}}
({\cal Z}_{_{\tilde t}})_{_{2,j}}({\cal Z}_+^\dagger)_{_{\alpha,2}}
\Big(\xi_{_{\chi}}^1\Big)_{_{\alpha\beta}}
\Big(\Psi_{_{1a}}-\Psi_{_{1b}}\Big)\Big]
(x_{_t};x_{_{\chi_\alpha}},x_{_{\tilde{t}_j}};
x_{_{\tilde g}},x_{_{\tilde{s}_i}})
\nonumber\\
&&\hspace{1.3cm}
-2(x_{_{\chi_\alpha}}x_{_{\chi_\beta}})^{1/2}
({\cal Z}_{_{\tilde t}})_{_{1,j}}({\cal Z}_-)_{_{1,\alpha}}
\Big(\xi_{_{\chi}}^4\Big)_{_{\alpha\beta}}
\Big(\Psi_{_{1b}}-\Psi_{_{1a}}\Big)(x_{_t};x_{_{\tilde g}},
x_{_{\tilde{s}_i}};x_{_{\chi_\alpha}},x_{_{\tilde{t}_j}})
\nonumber\\
&&\hspace{1.3cm}
-(x_{_{\chi_\alpha}}\rightarrow x_{_{\chi_\beta}})\bigg\}
\;,\nonumber\\
&&{\cal P}_{_Z}^{(6)}=
-{1\over3\sqrt{2}s_{_\beta}}({\cal Z}_{_{\tilde s}})_{_{2,i}}
({\cal Z}_{_{\tilde s}}^\dagger)_{_{i,1}}\Big(\xi_{_{t}}\Big)_{_{jk}}
({\cal Z}_{_{\tilde t}}^\dagger)_{_{k,1}}({\cal Z}_-^\dagger)_{_{\alpha,2}}
{1\over x_{_{\tilde{t}_j}}-x_{_{\tilde{t}_k}}}
\bigg\{\Big[({\cal Z}_{_{\tilde t}})_{_{2,j}}({\cal Z}_-)_{_{1,\alpha}}
\Big(\Psi_{_{2b}}
\nonumber\\
&&\hspace{1.3cm}
-\Psi_{_{2d}}-\varrho_{_{2,1}}(x_{_{\chi_\alpha}},x_{_{\tilde{t}_j}})\Big)
-{m_{_t}\over\sqrt{2}m_{_{\rm w}}s_{_\beta}}
({\cal Z}_{_{\tilde t}})_{_{2,j}}({\cal Z}_+^\dagger)_{_{\alpha,2}}
(x_{_t}x_{_{\chi_\alpha}})^{1/2}\Psi_{_{1b}}
\nonumber\\
&&\hspace{1.3cm}
-e^{-i\theta_{_3}}({\cal Z}_{_{\tilde t}})_{_{2,j}}({\cal Z}_-)_{_{1,\alpha}}
(x_{_t}x_{_{\tilde g}})^{1/2}\Psi_{_{1a}}
\nonumber\\
&&\hspace{1.3cm}
+e^{-i\theta_{_3}}{m_{_t}\over\sqrt{2}m_{_{\rm w}}s_{_\beta}}
({\cal Z}_{_{\tilde t}})_{_{2,j}}({\cal Z}_+^\dagger)_{_{\alpha,2}}
(x_{_{\tilde g}}x_{_{\chi_\alpha}})^{1/2}
\Big(\Psi_{_{1a}}-\Psi_{_{1b}}\Big)\Big](x_{_t};x_{_{\chi_\alpha}},
x_{_{\tilde{t}_j}};x_{_{\tilde g}},x_{_{\tilde{s}_i}})
\nonumber\\
&&\hspace{1.3cm}
-(x_{_{\tilde{t}_j}}\rightarrow x_{_{\tilde{t}_k}})\bigg\}
\;,\nonumber\\
&&{\cal P}_{_Z}^{(7)}=
-{\over3\sqrt{2}s_{_\beta}}({\cal Z}_-^\dagger)_{_{\alpha,2}}
({\cal Z}_{_{\tilde s}})_{_{2,i}}\Big(\xi_{_{s}}\Big)_{_{ik}}
({\cal Z}_{_{\tilde s}}^\dagger)_{_{k,1}}
({\cal Z}_{_{\tilde t}}^\dagger)_{_{j,1}}
{1\over x_{_{\tilde{s}_i}}-x_{_{\tilde{s}_k}}}
\nonumber\\
&&\hspace{1.3cm}\times
\bigg\{\Big[-({\cal Z}_{_{\tilde t}})_{_{1,j}}({\cal Z}_-)_{_{1,\alpha}}
\Big(\varrho_{_{2,1}}(x_{_{\tilde g}},x_{_{\tilde{s}_i}})
+\Psi_{_{2d}}-\Psi_{_{2b}}\Big)
\nonumber\\
&&\hspace{1.3cm}
-{m_{_t}\over\sqrt{2}m_{_{\rm w}}s_{_\beta}}({\cal Z}_{_{\tilde t}})_{_{1,j}}
({\cal Z}_+^\dagger)_{_{\alpha,2}}(x_{_t}x_{_{\chi_\alpha}})^{1/2}
\Psi_{_{1a}}
+e^{-i\theta_{_3}}({\cal Z}_{_{\tilde t}})_{_{2,j}}({\cal Z}_-)_{_{1,\alpha}}
(x_{_{\tilde g}}x_{_t})^{1/2}\Psi_{_{1b}}
\nonumber\\
&&\hspace{1.3cm}
+e^{-i\theta_{_3}}{m_{_t}\over\sqrt{2}
m_{_{\rm w}}s_{_\beta}}({\cal Z}_{_{\tilde t}})_{_{2,j}}
({\cal Z}_+^\dagger)_{_{\alpha,2}}(x_{_{\tilde g}}x_{_{\chi_\alpha}})^{1/2}
\Big(\Psi_{_{1b}}-\Psi_{_{1a}}\Big)\Big](x_{_t};x_{_{\tilde g}},
x_{_{\tilde{s}_i}};x_{_{\tilde{t}_j}},x_{_{\chi_\alpha}})
\nonumber\\
&&\hspace{1.3cm}
-(x_{_{\tilde{s}_i}}\rightarrow x_{_{\tilde{s}_k}})\bigg\}
\;,\nonumber\\
&&{\cal P}_{_Z}^{(8)}=
-{\sqrt{2}\over3s_{_\beta}}({\cal Z}_{_{\tilde s}})_{_{2,i}}
({\cal Z}_{_{\tilde s}}^\dagger)_{_{i,1}}({\cal Z}_{_{\tilde t}}^\dagger)_{_{j,1}}
({\cal Z}_-^\dagger)_{_{\alpha,2}}
\bigg\{\bigg[\Big({1\over2}-{2\over3}s_{_{\rm w}}^2\Big)
\bigg(({\cal Z}_{_{\tilde t}})_{_{1,j}}({\cal Z}_-)_{_{1,\alpha}}
x_{_t}{\partial\over\partial x_{_t}}\Psi_{_{1b}}
\nonumber\\
&&\hspace{1.3cm}
-e^{-i\theta_{_3}}({\cal Z}_{_{\tilde t}})_{_{2,j}}({\cal Z}_-)_{_{1,\alpha}}
(x_{_{\tilde g}}x_{_t})^{1/2}{\partial\over\partial x_{_t}}
\Big(\Psi_{_{1a}}-\Psi_{_{1b}}\Big)
\nonumber\\
&&\hspace{1.3cm}
-e^{-i\theta_{_3}}{m_{_t}\over\sqrt{2}m_{_{\rm w}}s_{_\beta}}
({\cal Z}_{_{\tilde t}})_{_{2,j}}({\cal Z}_+^\dagger)_{_{\alpha,2}}
(x_{_{\chi_\alpha}}x_{_{\tilde g}})^{1/2}{\partial\over\partial x_{_t}}
\Big(\Psi_{_{1a}}-\Psi_{_{1b}}\Big)\bigg)
\nonumber\\
&&\hspace{1.3cm}
-{4\over3}s_{_{\rm w}}^2e^{-i\theta_{_3}}\bigg(
({\cal Z}_{_{\tilde t}})_{_{2,j}}({\cal Z}_-)_{_{1,\alpha}}
(x_{_t}x_{_{\tilde g}})^{1/2}{\partial\over\partial x_{_t}}
\Big(\Psi_{_{1a}}-\Psi_{_{1b}}\Big)
\nonumber\\
&&\hspace{1.3cm}
+{m_{_t}\over\sqrt{2}m_{_{\rm w}}s_{_\beta}}
({\cal Z}_{_{\tilde t}})_{_{2,j}}({\cal Z}_+^\dagger)_{_{\alpha,2}}
x_{_t}(x_{_{\chi_\alpha}}x_{_{\tilde g}})^{1/2}{\partial\over\partial x_{_t}}
\Psi_{_{0}}\bigg)\bigg](x_{_t};x_{_{\tilde{t}_j}},x_{_{\chi_\alpha}};x_{_{\tilde g}},
x_{_{\tilde{s}_i}})
\nonumber\\
&&\hspace{1.3cm}
+\bigg[\Big({1\over2}-{2\over3}s_{_{\rm w}}^2\Big)
{m_{_t}\over\sqrt{2}m_{_{\rm w}}s_{_\beta}}\bigg(
({\cal Z}_{_{\tilde t}})_{_{1,j}}({\cal Z}_+^\dagger)_{_{\alpha,2}}
(x_{_{\chi_\alpha}}x_{_t})^{1/2}{\partial\over\partial x_{_t}}
\Big(\Psi_{_{1b}}-\Psi_{_{1a}}\Big)
\nonumber\\
&&\hspace{1.3cm}
-e^{-i\theta_{_3}}({\cal Z}_{_{\tilde t}})_{_{2,j}}({\cal Z}_+^\dagger)_{_{\alpha,2}}
(x_{_{\chi_\alpha}}x_{_{\tilde g}})^{1/2}{\partial\over\partial x_{_t}}
\Big(\Psi_{_{1a}}-\Psi_{_{1b}}\Big)\bigg)
\nonumber\\
&&\hspace{1.3cm}
+{4\over3}s_{_{\rm w}}^2\bigg(({\cal Z}_{_{\tilde t}})_{_{1,j}}
({\cal Z}_-)_{_{1,\alpha}}{\partial\over\partial x_{_t}}\Big[\Psi_{_{2c}}
-2\Psi_{_{2b}}+\Psi_{_{2a}}\Big]
\nonumber\\
&&\hspace{1.3cm}
+{m_{_t}\over\sqrt{2}m_{_{\rm w}}s_{_\beta}}
({\cal Z}_{_{\tilde t}})_{_{2,j}}({\cal Z}_+^\dagger)_{_{\alpha,2}}
(x_{_{\chi_\alpha}}x_{_t})^{1/2}{\partial\over\partial x_{_t}}
\Big(\Psi_{_{1b}}-\Psi_{_{1a}}\Big)\bigg)
\bigg](x_{_t};x_{_{\tilde g}},x_{_{\tilde{s}_i}};x_{_{\tilde{t}_j}},x_{_{\chi_\alpha}})
\bigg\}.
\label{ap3-1}
\end{eqnarray}

\begin{eqnarray}
&&{\cal P}_{_G}^{(2)}=-{2\over3}({\cal Z}_{_{\tilde s}})_{_{1,i}}
\Big(\xi_{_{H}}\Big)_{_{ij}}({\cal Z}_{_{\tilde t}}^\dagger)_{_{j,2}}
e^{i\theta_{_3}}(x_{_t}x_{_{\tilde g}})^{1/2}
\Psi_{_{0}}(x_{_{\tilde{t}_j}};x_{_t},x_{_H};x_{_{\tilde g}},x_{_{\tilde{s}_i}})
\;,\nonumber\\
&&{\cal P}_{_G}^{(4)}=
-{1\over3}({\cal Z}_{_{\tilde s}})_{_{1,i}}
\Big(\xi_{_{H}}\Big)_{_{ik}}\Big(\eta_{_{H}}\Big)_{_{kj}}
{x_{_{\rm w}}\over x_{_{\tilde{t}_j}}-x_{_{\tilde{t}_k}}}
\bigg\{\Big[({\cal Z}_{_{\tilde t}}^\dagger)_{_{j,1}}
\Psi_{_{1b}}
\nonumber\\
&&\hspace{1.3cm}
-e^{i\theta_{_3}}(x_{_t}x_{_{\tilde g}})^{1/2}
({\cal Z}_{_{\tilde t}}^\dagger)_{_{j,2}}
\Psi_{_{0}}\Big](x_{_{\tilde{t}_j}};x_{_t},x_{_H};
x_{_{\tilde g}},x_{_{\tilde{s}_i}})
-(x_{_{\tilde{t}_j}}\rightarrow x_{_{\tilde{t}_k}})\bigg\}
\;,\nonumber\\
&&{\cal P}_{_G}^{(5)}={2\over3s_{_\beta}}
({\cal Z}_{_{\tilde s}})_{_{1,i}}({\cal Z}_{_{\tilde s}}^\dagger)_{_{i,1}}
({\cal Z}_{_{\tilde t}}^\dagger)_{_{j,1}}({\cal Z}_-^\dagger)_{_{\beta,2}}
{1\over x_{_{\chi_\alpha}}-x_{_{\chi_\beta}}}
\nonumber\\
&&\hspace{1.3cm}\times
\bigg\{\bigg[-{m_{_t}\over\sqrt{2}m_{_{\rm w}}s_{_\beta}}
({\cal Z}_{_{\tilde t}})_{_{2,j}}({\cal Z}_+^\dagger)_{_{\alpha,2}}
\Big(\xi_{_{\chi}}^2\Big)_{_{\alpha\beta}}\Big(\Psi_{_{2b}}-\Psi_{_{2d}}
-\varrho_{_{2,1}}(x_{_{\chi_\alpha}},x_{_{\tilde{t}_j}})\Big)
\nonumber\\
&&\hspace{1.3cm}
-({\cal Z}_{_{\tilde t}})_{_{2,j}}({\cal Z}_-)_{_{1,\alpha}}
\bigg(\Big(\xi_{_{\chi}}^2\Big)_{_{\alpha\beta}}
(x_{_t}x_{_{\chi_\alpha}})^{1/2}
-\Big(\xi_{_{\chi}}^3\Big)_{_{\alpha\beta}}
(x_{_t}x_{_{\chi_\beta}})^{1/2}\bigg)\Psi_{_{1b}}
\nonumber\\
&&\hspace{1.3cm}
+e^{i\theta_{_3}}({\cal Z}_{_{\tilde t}})_{_{1,j}}
({\cal Z}_-)_{_{1,\alpha}}\Big(\xi_{_{\chi}}^2\Big)_{_{\alpha\beta}}
\bigg((x_{_{\tilde g}}x_{_{\chi_\alpha}})^{1/2}
-\Big(\xi_{_{\chi}}^3\Big)_{_{\alpha\beta}}
(x_{_{\tilde g}}x_{_{\chi_\beta}})^{1/2}\bigg)
\Big(\Psi_{_{1a}}-\Psi_{_{1b}}\Big)
\nonumber\\
&&\hspace{1.3cm}
+e^{i\theta_{_3}}{m_{_t}\over\sqrt{2}
m_{_{\rm w}}s_{_\beta}}({\cal Z}_{_{\tilde t}})_{_{1,j}}
({\cal Z}_+^\dagger)_{_{\alpha,2}}\Big(\xi_{_{\chi}}^2\Big)_{_{\alpha\beta}}
(x_{_t}x_{_{\tilde g}})^{1/2}\Psi_{_{1a}}
\nonumber\\
&&\hspace{1.3cm}
-e^{i\theta_{_3}}{m_{_t}\over\sqrt{2}
m_{_{\rm w}}s_{_\beta}}({\cal Z}_{_{\tilde t}})_{_{1,j}}
({\cal Z}_+^\dagger)_{_{\alpha,2}}\Big(\xi_{_{\chi}}^3\Big)_{_{\alpha\beta}}
(x_{_t}x_{_{\tilde g}}x_{_{\chi_\alpha}}x_{_{\chi_\beta}})^{1/2}
\Psi_{_{0}}\bigg](x_{_t};x_{_{\chi_\alpha}},x_{_{\tilde{t}_j}};
x_{_{\tilde g}},x_{_{\tilde{s}_i}})
\nonumber\\
&&\hspace{1.3cm}
+{m_{_t}\over\sqrt{2}m_{_{\rm w}}s_{_\beta}}
({\cal Z}_{_{\tilde t}})_{_{2,j}}({\cal Z}_+^\dagger)_{_{\alpha,2}}
\Big(\xi_{_{\chi}}^3\Big)_{_{\alpha\beta}}
(x_{_{\chi_\alpha}}x_{_{\chi_\beta}})^{1/2}
\Big(\Psi_{_{1b}}-\Psi_{_{1a}}\Big)
(x_{_t};x_{_{\tilde g}},x_{_{\tilde{s}_i}};
x_{_{\chi_\alpha}},x_{_{\tilde{t}_j}})
\nonumber\\
&&\hspace{1.3cm}
-(x_{_{\chi_\alpha}}\rightarrow x_{_{\chi_\beta}})\bigg\}
\;,\nonumber\\
&&{\cal P}_{_G}^{(6)}=
-{\sqrt{2}\over3s_{_\beta}}({\cal Z}_{_{\tilde s}})_{_{1,i}}
({\cal Z}_{_{\tilde s}}^\dagger)_{_{i,1}}
({\cal Z}_{_{\tilde t}}^\dagger)_{_{k,1}}({\cal Z}_-^\dagger)_{_{\alpha,2}}
\Big(\eta_{_H}\Big)_{_{jk}}{1\over x_{_{\tilde{t}_j}}-x_{_{\tilde{t}_k}}}
\nonumber\\
&&\hspace{1.3cm}\times
\bigg\{\bigg[({\cal Z}_{_{\tilde t}})_{_{2,j}}({\cal Z}_-)_{_{1,\alpha}}
x_{_t}^{1/2}\Psi_{_{1b}}
-e^{i\theta_{_3}}({\cal Z}_{_{\tilde t}})_{_{1,j}}
({\cal Z}_-)_{_{1,\alpha}}x_{_{\tilde g}}^{1/2}
\Big(\Psi_{_{1a}}-\Psi_{_{1b}}\Big)
\nonumber\\
&&\hspace{1.3cm}
-e^{i\theta_{_3}}{m_{_t}\over\sqrt{2}m_{_{\rm w}}s_{_\beta}}({\cal Z}_{_{\tilde t}})_{_{1,j}}
({\cal Z}_+^\dagger)_{_{\alpha,2}}(x_{_{\tilde g}}x_{_t}x_{_{\chi_\alpha}})^{1/2}
\Psi_{_{0}}\bigg](x_{_t};x_{_{\chi_\alpha}},
x_{_{\tilde{t}_j}};x_{_{\tilde g}},x_{_{\tilde{s}_i}})
\nonumber\\
&&\hspace{1.3cm}
-{m_{_t}\over\sqrt{2}m_{_{\rm w}}s_{_\beta}}({\cal Z}_{_{\tilde t}})_{_{2,j}}
({\cal Z}_+^\dagger)_{_{\alpha,2}}x_{_{\chi_\alpha}}^{1/2}
\Big(\Psi_{_{1b}}-\Psi_{_{1a}}\Big)(x_{_t};x_{_{\tilde g}},
x_{_{\tilde{s}_i}};x_{_{\chi_\alpha}},x_{_{\tilde{t}_j}})
\nonumber\\
&&\hspace{1.3cm}
-(x_{_{\tilde{t}_j}}\rightarrow x_{_{\tilde{t}_k}})\bigg\}
\;,\nonumber\\
&&{\cal P}_{_G}^{(8)}=
{2\over3s_{_\beta}}({\cal Z}_{_{\tilde s}})_{_{1,i}}
({\cal Z}_{_{\tilde s}}^\dagger)_{_{i,1}}({\cal Z}_{_{\tilde t}})_{_{1,j}}
({\cal Z}_{_{\tilde t}}^\dagger)_{_{j,1}}({\cal Z}_-^\dagger)_{_{\alpha,2}}
\bigg\{({\cal Z}_-)_{_{1,\alpha}}\Psi_{_{1b}}
\nonumber\\
&&\hspace{1.3cm}
+{m_{_t}\over\sqrt{2}m_{_{\rm w}}s_{_\beta}}
({\cal Z}_+^\dagger)_{_{\alpha,2}}(x_{_{\chi_\alpha}}x_{_{\tilde g}})^{1/2}
\Psi_{_{0}}\bigg\}(x_{_t};x_{_{\tilde{t}_j}},x_{_{\chi_\alpha}};x_{_{\tilde g}},
x_{_{\tilde{s}_i}})\;.
\label{app3-2}
\end{eqnarray}

\begin{eqnarray}
&&{\cal P}_{_{H^{^\rho}}}^{(1)}={1\over3}s_{_\beta}
({\cal Z}_{_H})_{_{3,\rho}}({\cal Z}_{_{\tilde s}})_{_{1,i}}
\Big(\xi_{_{H}}\Big)_{_{ij}}
\bigg\{({\cal Z}_{_{\tilde t}}^\dagger)_{_{j,1}}x_{_{\rm w}}
{\partial\over\partial x_{_H}}\Psi_{_{1b}}
\nonumber\\
&&\hspace{1.3cm}
-e^{i\theta_{_3}}({\cal Z}_{_{\tilde t}}^\dagger)_{_{j,2}}
(x_{_t}x_{_{\tilde g}})^{1/2}x_{_{\rm w}}{\partial\over\partial x_{_H}}
\Psi_{_{0}}\bigg\}(x_{_{\tilde{t}_j}};x_{_t},
x_{_H};x_{_{\tilde{s}_i}},x_{_{\tilde g}})
\;,\nonumber\\
&&{\cal P}_{_{H^{^\rho}}}^{(2)}=
{1\over3}\Big(s_{_\beta}({\cal Z}_{_H})_{_{2,\rho}}
+i({\cal Z}_{_H})_{_{1,\rho}}\Big)({\cal Z}_{_{\tilde t}})_{_{1,j}}
({\cal Z}_{_{\tilde s}})_{_{1,i}}({\cal Z}_{_{\tilde s}}^\dagger)_{_{i,1}}
{1\over x_{_H}-x_{_{\rm w}}}
\bigg\{\bigg[({\cal Z}_{_{\tilde t}}^\dagger)_{_{j,1}}
\Big(\Psi_{_{2b}}-2\Psi_{_{2c}}
\nonumber\\
&&\hspace{1.3cm}
-\varrho_{_{2,1}}(x_{_H},x_{_t})\Big)
-e^{i\theta_{_3}}({\cal Z}_{_{\tilde t}}^\dagger)_{_{j,2}}(x_{_{\tilde g}}x_{_t})^{1/2}
\Big(\Psi_{_{1a}}-2\Psi_{_{1b}}\Big)\bigg](x_{_{\tilde{t}_j}};x_{_H},x_{_t};
x_{_{\tilde g}},x_{_{\tilde{s}_i}})
\nonumber\\
&&\hspace{1.3cm}
-(x_{_H}\rightarrow x_{_{\rm w}})\bigg\}
\;,\nonumber\\
&&{\cal P}_{_{H^{^\rho}}}^{(3)}={1\over3}\Big[s_{_\beta}^3
({\cal Z}_{_H})_{_{2,\rho}}-i({\cal Z}_{_H})_{_{1,\rho}}\Big]s_{_\beta}
({\cal Z}_{_H})_{_{3,\rho}}
({\cal Z}_{_{\tilde s}})_{_{1,i}}\Big(\xi_{_{H}}\Big)_{_{ij}}
{x_{_{\rm w}}\over x_{_H}-x_{_{\rm w}}}\bigg\{\bigg[
({\cal Z}_{_{\tilde t}}^\dagger)_{_{j,1}}\Psi_{_{1b}}
\nonumber\\
&&\hspace{1.3cm}
-e^{i\theta_{_3}}({\cal Z}_{_{\tilde t}}^\dagger)_{_{j,2}}
(x_{_t}x_{_{\tilde g}})^{1/2}\Psi_{_{0}}\bigg](x_{_{\tilde{t}_j}};x_{_H},x_{_t};
x_{_{\tilde g}},x_{_{\tilde{s}_i}})-(x_{_H}\rightarrow x_{_{\rm w}})\bigg\}
\;,\nonumber\\
&&{\cal P}_{_{H^{^\rho}}}^{(4)}=-{1\over3}({\cal Z}_{_H})_{_{3,\rho}}
({\cal Z}_{_{\tilde s}})_{_{1,i}}\Big(\xi_{_{H}}\Big)_{_{ij}}
\bigg\{2({\cal Z}_{_{\tilde t}}^\dagger)_{_{j,1}}
x_{_t}{\partial\over\partial x_{_t}}\Psi_{_{1b}}
\nonumber\\
&&\hspace{1.0cm}
-e^{i\theta_{_3}}({\cal Z}_{_{\tilde t}}^\dagger)_{_{j,2}}
(x_{_{\tilde g}}x_{_t})^{1/2}{\partial\over\partial x_{_t}}
\Big(\Psi_{_{1a}}+x_{_t}\Psi_{_{0}}\Big)\bigg\}(x_{_{\tilde{t}_j}};x_{_H},x_{_t};
x_{_{\tilde g}},x_{_{\tilde{s}_i}})
\;,\nonumber\\
&&{\cal P}_{_{H^{^\rho}}}^{(5)}=-{2\over3}({\cal Z}_{_{\tilde s}})_{_{1,i}}
\Big(\zeta_{_{sH}}^\rho\Big)_{_{ik}}\Big(\xi_{_{H}}\Big)_{_{kj}}
{x_{_{\rm w}}\over x_{_{\tilde{s}_i}}-x_{_{\tilde{s}_k}}}
\bigg\{\bigg[({\cal Z}_{_{\tilde t}}^\dagger)_{_{j,1}}\Psi_{_{1b}}
\nonumber\\
&&\hspace{1.3cm}
-e^{i\theta_{_3}}({\cal Z}_{_{\tilde t}}^\dagger)_{_{j,2}}(x_{_{\tilde g}}x_{_t})^{1/2}
\Psi_{_{0}}\bigg](x_{_{\tilde{t}_j}};x_{_H},x_{_t};
x_{_{\tilde g}},x_{_{\tilde{s}_i}})-(x_{_{\tilde{s}_i}}\rightarrow x_{_{\tilde{s}_k}})\bigg\}
\;,\nonumber\\
&&{\cal P}_{_{H^{^\rho}}}^{(6)}={2\over3}({\cal Z}_{_{\tilde s}})_{_{1,i}}
\Big(\xi_{_{H}}\Big)_{_{ik}}\Big(\zeta_{_{tH}}^\rho\Big)_{_{kj}}
{x_{_{\rm w}}\over x_{_{\tilde{t}_j}}-x_{_{\tilde{t}_k}}}
\bigg\{\bigg[({\cal Z}_{_{\tilde t}}^\dagger)_{_{j,1}}\Psi_{_{1b}}
\nonumber\\
&&\hspace{1.3cm}
-e^{i\theta_{_3}}({\cal Z}_{_{\tilde t}}^\dagger)_{_{j,2}}(x_{_t}x_{_{\tilde g}})^{1/2}
\Psi_{_{0}}\bigg](x_{_{\tilde{t}_j}};x_{_H},x_{_t};x_{_{\tilde g}},x_{_{\tilde{s}_i}})
-(x_{_{\tilde{t}_j}}\rightarrow x_{_{\tilde{t}_k}})\bigg\}\;,\nonumber\\
&&{\cal P}_{_{H^{^\rho}}}^{(7)}=
-{2\over3s_{_\beta}}({\cal Z}_{_{\tilde s}})_{_{1,i}}
({\cal Z}_{_{\tilde s}}^\dagger)_{_{i,1}}({\cal Z}_{_{\tilde t}}^\dagger)_{_{j,1}}
({\cal Z}_-^\dagger)_{_{\beta,2}}{1\over x_{_{\chi_\alpha}}-x_{_{\chi_\beta}}}
\nonumber\\
&&\hspace{1.3cm}\times
\bigg\{\bigg[-{m_{_t}\over\sqrt{2}m_{_{\rm w}}s_{_\beta}}
({\cal Z}_{_{\tilde t}})_{_{2,j}}({\cal Z}_+^\dagger)_{_{\alpha,2}}
\Big(\kappa_{_{H^{^\rho}}}^1\Big)_{_{\alpha\beta}}\Big(\Psi_{_{2b}}-\Psi_{_{2d}}
-\varrho_{_{2,1}}(x_{_{\chi_\alpha}},x_{_{\tilde{t}_j}})\Big)
\nonumber\\
&&\hspace{1.3cm}
-({\cal Z}_{_{\tilde t}})_{_{2,j}}({\cal Z}_-)_{_{1,\alpha}}
\bigg(\Big(\kappa_{_{H^{^\rho}}}^2\Big)_{_{\alpha\beta}}
(x_{_t}x_{_{\chi_\beta}})^{1/2}+\Big(\kappa_{_{H^{^\rho}}}^1\Big)_{_{\alpha\beta}}
(x_{_t}x_{_{\chi_\alpha}})^{1/2}\bigg)\Psi_{_{1b}}
\nonumber\\
&&\hspace{1.3cm}
+e^{i\theta_{_3}}({\cal Z}_{_{\tilde t}})_{_{1,j}}({\cal Z}_-)_{_{1,\alpha}}
\bigg(\Big(\kappa_{_{H^{^\rho}}}^2\Big)_{_{\alpha\beta}}
(x_{_{\tilde g}}x_{_{\chi_\beta}})^{1/2}
+\Big(\kappa_{_{H^{^\rho}}}^1\Big)_{_{\alpha\beta}}
(x_{_{\tilde g}}x_{_{\chi_\alpha}})^{1/2}\bigg)
\Big(\Psi_{_{1a}}-\Psi_{_{1b}}\Big)
\nonumber\\
&&\hspace{1.3cm}
+e^{i\theta_{_3}}{m_{_t}\over\sqrt{2}m_{_{\rm w}}s_{_\beta}}
({\cal Z}_{_{\tilde t}})_{_{1,j}}({\cal Z}_+^\dagger)_{_{\alpha,2}}
\Big(\kappa_{_{H^{^\rho}}}^1\Big)_{_{\alpha\beta}}
(x_{_t}x_{_{\tilde g}})^{1/2}\Psi_{_{1a}}
\nonumber\\
&&\hspace{1.3cm}
+e^{i\theta_{_3}}{m_{_t}\over\sqrt{2}m_{_{\rm w}}s_{_\beta}}
({\cal Z}_{_{\tilde t}})_{_{1,j}}({\cal Z}_+^\dagger)_{_{\alpha,2}}
\Big(\kappa_{_{H^{^\rho}}}^2\Big)_{_{\alpha\beta}}
(x_{_t}x_{_{\tilde g}}x_{_{\chi_\alpha}}x_{_{\chi_\beta}})^{1/2}
\Psi_{_{0}}\bigg](x_{_t};x_{_{\chi_\alpha}},
x_{_{\tilde{t}_j}};x_{_{\tilde g}},x_{_{\tilde{s}_i}})
\nonumber\\
&&\hspace{1.3cm}
-{m_{_t}\over\sqrt{2}m_{_{\rm w}}s_{_\beta}}
({\cal Z}_{_{\tilde t}})_{_{2,j}}({\cal Z}_+^\dagger)_{_{\alpha,2}}
\Big(\kappa_{_{H^{^\rho}}}^2\Big)_{_{\alpha\beta}}
(x_{_{\chi_\alpha}}x_{_{\chi_\beta}})^{1/2}
\Big(\Psi_{_{1b}}-\Psi_{_{1a}}\Big)(x_{_t};x_{_{\tilde g}},
x_{_{\tilde{s}_i}};x_{_{\chi_\alpha}},x_{_{\tilde{t}_j}})
\nonumber\\
&&\hspace{1.3cm}
-(x_{_{\chi_\alpha}}\rightarrow x_{_{\chi_\beta}})\bigg\}
\;,\nonumber\\
&&{\cal P}_{_{H^{^\rho}}}^{(8)}=
{2\over3s_{_\beta}}({\cal Z}_{_{\tilde s}})_{_{1,i}}
({\cal Z}_{_{\tilde s}}^\dagger)_{_{i,1}}({\cal Z}_{_{\tilde t}}^\dagger)_{_{k,1}}
({\cal Z}_-^\dagger)_{_{\alpha,2}}\Big(\zeta_{_{tH}}^\rho\Big)_{_{jk}}
{1\over x_{_{\tilde{t}_j}}-x_{_{\tilde{t}_k}}}
\nonumber\\
&&\hspace{1.3cm}\times
\bigg\{\bigg[({\cal Z}_{_{\tilde t}})_{_{2,j}}
({\cal Z}_-)_{_{1,\alpha}}(x_{_{\rm w}}x_{_t})^{1/2}\Psi_{_{1b}}
-e^{i\theta_{_3}}({\cal Z}_{_{\tilde t}})_{_{1,j}}
({\cal Z}_-)_{_{1,\alpha}}(x_{_{\rm w}}x_{_{\tilde g}})^{1/2}
\Big(\Psi_{_{1a}}-\Psi_{_{1b}}\Big)
\nonumber\\
&&\hspace{1.3cm}
-e^{i\theta_{_3}}{m_{_t}
\over\sqrt{2}m_{_{\rm w}}s_{_\beta}}({\cal Z}_{_{\tilde t}})_{_{1,j}}
({\cal Z}_+^\dagger)_{_{\alpha,2}}
(x_{_{\rm w}}x_{_{\tilde g}}x_{_t}x_{_{\chi_\alpha}})^{1/2}
\Psi_{_{0}}\bigg](x_{_t};x_{_{\chi_\alpha}},
x_{_{\tilde{t}_j}};x_{_{\tilde g}},x_{_{\tilde{s}_i}})
\nonumber\\
&&\hspace{1.3cm}
-{m_{_t}\over\sqrt{2}m_{_{\rm w}}s_{_\beta}}
({\cal Z}_{_{\tilde t}})_{_{2,j}}({\cal Z}_+^\dagger)_{_{\alpha,2}}
(x_{_{\rm w}}x_{_{\chi_\alpha}})^{1/2}\Big[\Big(\Psi_{_{1b}}-\Psi_{_{1a}}\Big)
(x_{_t};x_{_{\tilde g}},x_{_{\tilde{s}_i}};x_{_{\chi_\alpha}},
x_{_{\tilde{t}_j}})
\nonumber\\
&&\hspace{1.3cm}
-(x_{_{\tilde{t}_j}}\rightarrow x_{_{\tilde{t}_k}})\bigg\}\;,\nonumber\\
&&{\cal P}_{_{H^{^\rho}}}^{(10)}=
-{1\over3s_{_\beta}}({\cal Z}_{_{\tilde s}})_{_{1,i}}
({\cal Z}_{_{\tilde s}}^\dagger)_{_{i,1}}({\cal Z}_{_H})_{_{3,\rho}}
({\cal Z}_{_{\tilde t}}^\dagger)_{_{j,1}}({\cal Z}_-^\dagger)_{_{\alpha,2}}
\bigg\{\bigg[-({\cal Z}_{_{\tilde t}})_{_{2,j}}
({\cal Z}_-)_{_{1,\alpha}}\Big(1+2x_{_t}{\partial\over\partial x_{_t}}\Big)
\Psi_{_{1b}}
\nonumber\\
&&\hspace{1.3cm}
+2e^{i\theta_{_3}}({\cal Z}_{_{\tilde t}})_{_{1,j}}
({\cal Z}_-)_{_{1,\alpha}}(x_{_t}x_{_{\tilde g}})^{1/2}
{\partial\over\partial x_{_t}}\Big(\Psi_{_{1a}}-\Psi_{_{1b}}\Big)
\nonumber\\
&&\hspace{1.3cm}
+e^{i\theta_{_3}}{m_{_t}\over \sqrt{2}m_{_{\rm w}}s_{_\beta}}
({\cal Z}_{_{\tilde t}})_{_{2,j}}
({\cal Z}_+^\dagger)_{_{\alpha,2}}(x_{_{\tilde g}}x_{_{\chi_\alpha}})^{1/2}
\Big(1+2x_{_t}{\partial\over\partial x_{_t}}\Big)
\Psi_{_{0}}\bigg](x_{_t};x_{_{\chi_\alpha}},
x_{_{\tilde{t}_j}};x_{_{\tilde g}},x_{_{\tilde{s}_i}})
\nonumber\\
&&\hspace{1.3cm}
-{\sqrt{2}m_{_t}\over m_{_{\rm w}}s_{_\beta}}({\cal Z}_{_{\tilde t}})_{_{2,j}}
({\cal Z}_+^\dagger)_{_{\alpha,2}}(x_{_t}x_{_{\chi_\alpha}})^{1/2}
{\partial\over\partial x_{_t}}\Big(\Psi_{_{1b}}-\Psi_{_{1a}}\Big)
(x_{_t};x_{_{\tilde g}},x_{_{\tilde{s}_i}};x_{_{\chi_\alpha}},
x_{_{\tilde{t}_j}})\bigg\}\;.
\label{ap3-3}
\end{eqnarray}

\section{The two-loop corrections from the box diagrams\label{2b}}
\indent\indent

\begin{eqnarray}
&&{\cal B}_{V}=-{m_{_b}\over\sqrt{2}m_{_{\rm w}}s_{_\beta}}
({\cal Z}_{_{\tilde s}})_{_{1,j}}({\cal Z}_{_{\tilde b}})^\dagger_{_{i,1}}
({\cal Z}_+)^\dagger_{\alpha1}({\cal Z}_+)_{1\beta}
\Big(\Gamma_{_{\tilde{b}_i\chi_\alpha}}^{^R}\Big)
\Big(\Gamma_{_{\tilde{s}_j\chi_\beta}}^{^R}\Big)^*
(x_{_t}x_{_{\chi_\beta}})^{1/2}
\nonumber\\
&&\hspace{1.4cm}\times
\sum\limits_{\rho=\{\chi_\alpha,\chi_\beta\}}
{1\over\prod\limits_{\sigma\neq\rho}
(x_{_\rho}-x_{_\sigma})}\sum\limits_{\varrho=\{\tilde{b}_i,\tilde{s}_j\}}
{1\over\prod\limits_{\varsigma\neq\varrho}(x_{_\varrho}-x_{_\varsigma})}
\Psi_{_{1b}}(x_{_t};x_{_\rho},x_{_{\tilde{\nu}_I}};
x_{_\varrho},x_{_{\tilde g}})\;,\nonumber\\
&&{\cal B}_{A}=-{\cal B}_{V}\;.
\label{va}
\end{eqnarray}

\begin{eqnarray}
&&{\cal B}_{V}^{\prime(1)}={\cal B}_{V}\;,\nonumber\\
&&{\cal B}_{A}^{\prime(1)}=-{\cal B}_{V}^{\prime(1)}\;,\nonumber\\
&&{\cal B}_{V}^{\prime(2)}=-{m_{_{l^I}}m_{_t}\over m_{_{\rm w}}^2s_{_\beta}^2}
({\cal Z}_-)^\dagger_{\alpha2}({\cal Z}_+)_{1\beta}
\Big(\Gamma_{_{\tilde{t}_j\chi_\beta}}^{^L}\Big)
\Big(\Gamma_{_{\tilde{t}_i\chi_\alpha}}^{^R}\Big)^*
\sum\limits_{\rho=\{\chi_{\alpha,\beta},\tilde{t}_{i,j},\tilde{\nu}_k\}}
{1\over\prod\limits_{\sigma\neq\rho}(x_{_\rho}-x_{_\sigma})}
\nonumber\\
&&\hspace{1.4cm}\times
\Bigg\{\Big[({\cal Z}_{_{\tilde t}})_{_{1,i}}({\cal Z}_{_{\tilde t}})_{_{j,1}}^\dagger
+({\cal Z}_{_{\tilde t}})_{_{2,i}}({\cal Z}_{_{\tilde t}})_{_{j,2}}^\dagger\Big]
\Bigg[-{1\over4}\Big(2x_{_t}-x_{_\rho}+2x_{_{\tilde g}}\Big)x_{_\rho}^2\ln x_{_\rho}
\nonumber\\
&&\hspace{1.4cm}
+\Theta_{_2}(x_{_t},x_{_\rho},x_{_{\tilde g}})\Bigg]
-\Big[({\cal Z}_{_{\tilde t}})_{_{1,i}}({\cal Z}_{_{\tilde t}})_{_{j,2}}^\dagger
e^{i\theta_{_3}}
+({\cal Z}_{_{\tilde t}})_{_{2,i}}({\cal Z}_{_{\tilde t}})_{_{j,2}}^\dagger
e^{-i\theta_{_3}}\Big]
\nonumber\\
&&\hspace{1.4cm}\times
(x_{_t}x_{_{\tilde g}})^{1/2}
\Bigg[{1\over2}x_{_\rho}^2\ln x_{_\rho}
+\Theta_{_{1a}}(x_{_t},x_{_\rho},x_{_{\tilde g}})\Bigg]\Bigg\}\;,\nonumber\\
&&{\cal B}_{A}^{\prime(2)}={\cal B}_{V}^{\prime(2)}\;.
\label{pva}
\end{eqnarray}

\begin{eqnarray}
&&{\cal B}_{S}^{(1)}=-\Big(x_{_{\tilde g}}x_{_{l^I}}\Big)^{1/2}
e^{i\theta_{_3}}({\cal Z}_{_{\tilde b}})_{_{1,i}}
({\cal Z}_{_{\tilde b}})^\dagger_{_{i,1}}({\cal Z}_{_{\tilde s}})_{_{1,k}}
\Big({\cal A}_{_{st}}\Big)^\dagger_{kj}({\cal Z}_{_{\tilde t}})^\dagger_{_{j,1}}
\sum\limits_{\rho=\{\nu,W\}}{1\over\prod\limits_{\sigma\neq\rho}
(x_{_\rho}-x_{_\sigma})}
\nonumber\\
&&\hspace{1.4cm}\times
\sum\limits_{\varrho=\{\tilde{b}_i,\tilde{s}_k\}}
{1\over\prod\limits_{\varsigma\neq\varrho}
(x_{_\varrho}-x_{_\varsigma})}\Bigg\{
\Big(\Psi_{_{1a}}-2\Psi_{_{1b}}\Big)
(x_{_{\tilde{t}_j}};x_{_\rho},x_{_H};x_{_{\tilde g}},x_{_\varrho})
\Bigg\}\;,\nonumber\\
&&{\cal B}_{P}^{(1)}={\cal B}_{S}^{(1)}\;,\nonumber\\
&&{\cal B}_{S}^{(2)}=-\sum\limits_{\rho=W,H}{1\over\prod\limits_{\sigma\neq\rho}
(x_{_\rho}-x_{_\sigma})}\Bigg\{\Bigg[\Big(x_{_{\tilde g}}x_{_{l^I}}\Big)^{1/2}
e^{i\theta_{_3}}({\cal Z}_{_{\tilde t}})_{_{1,i}}\Big({\cal A}_{_{bt}}\Big)_{ij}
({\cal Z}_{_{\tilde b}})^\dagger_{_{j,2}}
\nonumber\\
&&\hspace{1.2cm}
-\Big(x_{_t}x_{_{l^I}}\Big)^{1/2}({\cal Z}_{_{\tilde t}})_{_{2,i}}
\Big({\cal A}_{_{bt}}\Big)_{ij}({\cal Z}_{_{\tilde b}})^\dagger_{_{j,2}}\Bigg]
\Psi_{_{0}}(x_{_{\tilde{t}_i}};x_{_\rho},x_{_t};x_{_{\tilde g}},x_{_{\tilde{b}_j}})
\Bigg\}\;,\nonumber\\
&&{\cal B}_{P}^{(2)}=-{\cal B}_{S}^{(2)}\;,\nonumber\\
&&{\cal B}_{S}^{(3)}=-{m_{_b}m_{_e}t_{_\beta}\over m_{_{\rm w}}^2}\Bigg\{
({\cal Z}_{_{\tilde t}})_{_{1,i}}({\cal Z}_{_{\tilde t}})^\dagger_{_{i,1}}
({\cal Z}_{_{\tilde s}})^\dagger_{_{j,1}}({\cal Z}_{_{\tilde s}})_{_{1,j}}
\Bigg[-\varrho_{_{1,1}}(1,x_{_t})-\varrho_{_{1,1}}(x_{_H},x_{_t})
\nonumber\\
&&\hspace{1.4cm}
+\sum\limits_{\rho=\nu,W,H}{1\over\prod\limits_{\sigma\neq\rho}
(x_{_\rho}-x_{_\sigma})}\Bigg(-3\varrho_{_{2,1}}(x_{_\rho},x_{_t})
+\Big[\Psi_{_{2b}}-2\Psi_{_{2c}}\Big](x_{_{\tilde{t}_i}};x_{_\rho},x_{_t};
x_{_{\tilde g}},x_{_{\tilde{s}_j}})\Bigg)\Bigg]
\nonumber\\
&&\hspace{1.4cm}
-\Big(x_{_{\tilde g}}x_{_{t}}\Big)^{1/2}e^{i\theta_{_3}}
({\cal Z}_{_{\tilde t}})_{_{1,i}}({\cal Z}_{_{\tilde t}})^\dagger_{_{i,2}}
({\cal Z}_{_{\tilde s}})^\dagger_{_{j,1}}({\cal Z}_{_{\tilde s}})_{_{2,j}}
\sum\limits_{\rho=\nu,W,H}{1\over\prod\limits_{\sigma\neq\rho}
(x_{_\rho}-x_{_\sigma})}\Bigg[\Psi_{_{1a}}
\nonumber\\
&&\hspace{1.4cm}
-2\Psi_{_{1b}}\Bigg](x_{_{\tilde{t}_i}};x_{_\rho},x_{_t};
x_{_{\tilde g}},x_{_{\tilde{s}_j}})\Bigg\}
\;,\nonumber\\
&&{\cal B}_{P}^{(3)}=-{\cal B}_{S}^{(3)}\;,\nonumber\\
&&{\cal B}_{S}^{(4)}=-\Big(x_{_{\tilde g}}x_{_{l^I}}\Big)^{1/2}
e^{i\theta_{_3}}({\cal Z}_{_{\tilde s}})_{_{1,k}}
({\cal Z}_{_{\tilde s}})^\dagger_{_{k,1}}({\cal Z}_{_{\tilde t}})_{_{1,j}}
\Big({\cal A}_{_{bt}}\Big)_{ji}({\cal Z}_{_{\tilde b}})^\dagger_{_{i,2}}
\sum\limits_{\rho=\nu,W}{1\over\prod\limits_{\sigma\neq\rho}
(x_{_\rho}-x_{_\sigma})}
\nonumber\\
&&\hspace{1.4cm}\times
\sum\limits_{\varrho=\tilde{b}_i,\tilde{s}_k}
{1\over\prod\limits_{\varsigma\neq\varrho}(x_{_\varrho}-x_{_\varsigma})}
\Bigg\{\Big(\Psi_{_{1a}}-2\Psi_{_{1b}}\Big)
(x_{_{\tilde{t}_j}};x_{_\rho},x_{_H};x_{_{\tilde g}},x_{_\varrho})
\Bigg\}
\;,\nonumber\\
&&{\cal B}_{P}^{(4)}=-{\cal B}_{S}^{(4)}\;,\nonumber\\
&&{\cal B}_{S}^{(5)}=-{m_{_b}m_{_{l^I}}t_{_\beta}\over m_{_{\rm w}}^2}
\Bigg\{({\cal Z}_{_{\tilde t}})_{_{1,i}}({\cal Z}_{_{\tilde t}})^\dagger_{_{i,1}}
{\partial\over\partial x_{_t}}\Bigg[\sum\limits_{\rho=t,W,H}
{1\over\prod\limits_{\sigma\neq\rho}(x_{_\rho}-x_{_\sigma})}
\Bigg({1\over4}x_{_\rho}^2\ln x_{_\rho}
\nonumber\\
&&\hspace{1.4cm}
+\Theta_{_{1b}}(x_{_{\tilde{t}_i}},x_{_\rho},x_{_{\tilde g}})\Bigg)\Bigg]
-(x_{_t}x_{_{\tilde g}})^{1/2}e^{-i\theta_{_3}}({\cal Z}_{_{\tilde t}})_{_{2,i}}
({\cal Z}_{_{\tilde t}})^\dagger_{_{i,1}}
\nonumber\\
&&\hspace{1.4cm}\times
{\partial\over\partial x_{_t}}\Bigg[\sum\limits_{\rho=t,W,H}
{1\over\prod\limits_{\sigma\neq\rho}(x_{_\rho}-x_{_\sigma})}
\Bigg({1\over2}x_{_\rho}\ln x_{_\rho}
+\Theta_{_{0}}(x_{_{\tilde{t}_i}},
x_{_\rho},x_{_{\tilde g}})\Bigg)\Bigg]
\nonumber\\
&&\hspace{1.4cm}
-(x_{_t}x_{_{\tilde g}})^{1/2}e^{i\theta_{_3}}({\cal Z}_{_{\tilde t}})_{_{1,i}}
({\cal Z}_{_{\tilde t}})^\dagger_{_{i,2}}
{\partial\over\partial x_{_t}}\Bigg[\sum\limits_{\rho=t,W,H}
{1\over\prod\limits_{\sigma\neq\rho}(x_{_\rho}-x_{_\sigma})}
\Bigg({1\over2}x_{_\rho}\ln x_{_\rho}
\nonumber\\
&&\hspace{1.4cm}
+\Theta_{_{0}}(x_{_{\tilde{t}_i}},x_{_\rho},x_{_{\tilde g}})\Bigg)\Bigg]
+x_{_t}({\cal Z}_{_{\tilde t}})_{_{2,i}}({\cal Z}_{_{\tilde t}})^\dagger_{_{i,2}}
{\partial\over\partial x_{_t}}\Bigg[\sum\limits_{\rho=\nu,t,W,H}
{1\over\prod\limits_{\sigma\neq\rho}(x_{_\rho}-x_{_\sigma})}
\Bigg({1\over4}x_{_\rho}^2\ln x_{_\rho}
\nonumber\\
&&\hspace{1.4cm}
+\Theta_{_{1b}}(x_{_{\tilde{t}_i}},
x_{_\rho},x_{_{\tilde g}})\Bigg)\Bigg]\Bigg\}
\;,\nonumber\\
&&{\cal B}_{P}^{(5)}=-{\cal B}_{S}^{(5)}\;,\nonumber\\
&&{\cal B}_{S}^{(6)}=({\cal Z}_-)^\dagger_{\alpha2}({\cal Z}_+)_{1\beta}
\Big(\Gamma_{_{\tilde{t}_i\chi_\beta}}^{^L}\Big)
\sum\limits_{\rho=\{\chi_\alpha,\chi_\beta,\tilde{t}_i\}}
{1\over\prod\limits_{\sigma\neq\rho}(x_{_\rho}-x_{_\sigma})}
\Bigg\{-{\sqrt{2}m_{_{l^I}}\over m_{_{\rm w}}s_{_\beta}}
\Big(\Gamma_{_{\tilde{b}_j\chi_\alpha}}^{^L}\Big)
\nonumber\\
&&\hspace{1.4cm}\times
({\cal Z}_{_{\tilde t}})^\dagger_{_{i,1}}({\cal Z}_{_{\tilde b}})^\dagger_{_{j,2}}
\Bigg[{1\over2}\varrho_{_{2,1}}(x_{_\rho},x_{_{\tilde{\nu}_I}})
+\Bigg(\Psi_{_{2b}}-\Psi_{_{2d}}\Bigg)(x_{_t};x_{_\rho},x_{_{\tilde{\nu}_I}};
x_{_{\tilde g}},x_{_{\tilde{b}_j}})\Bigg]
\nonumber\\
&&\hspace{1.4cm}
+{\sqrt{2}m_{_{l^I}}\over m_{_{\rm w}}s_{_\beta}}
\Big(\Gamma_{_{\tilde{b}_j\chi_\alpha}}^{^L}\Big)
({\cal Z}_{_{\tilde t}})^\dagger_{_{i,2}}({\cal Z}_{_{\tilde b}})^\dagger_{_{j,2}}
(x_{_t}x_{_{\tilde g}})^{1/2}e^{i\theta_{_3}}\Psi_{_{1a}}
(x_{_t};x_{_\rho},x_{_{\tilde{\nu}_I}};
x_{_{\tilde g}},x_{_{\tilde{b}_j}})
\nonumber\\
&&\hspace{1.4cm}
+{m_{_{l^I}}m_{_t}\over m_{_{\rm w}}^2s_{_\beta}^2}
\Big(\Gamma_{_{\tilde{b}_j\chi_\alpha}}^{^R}\Big)
({\cal Z}_{_{\tilde t}})^\dagger_{_{i,1}}({\cal Z}_{_{\tilde b}})^\dagger_{_{j,2}}
(x_{_t}x_{_{\chi_\alpha}})^{1/2}\Psi_{_{1b}}
(x_{_t};x_{_\rho},x_{_{\tilde{\nu}_I}};
x_{_{\tilde g}},x_{_{\tilde{b}_j}})
\nonumber\\
&&\hspace{1.4cm}
-{m_{_{l^I}}m_{_t}\over m_{_{\rm w}}^2s_{_\beta}^2}
\Big(\Gamma_{_{\tilde{b}_j\chi_\alpha}}^{^R}\Big)
({\cal Z}_{_{\tilde t}})^\dagger_{_{i,2}}({\cal Z}_{_{\tilde b}})^\dagger_{_{j,2}}
(x_{_t}x_{_{\tilde g}})^{1/2}e^{i\theta_{_3}}\Bigg(\Psi_{_{1a}}
-\Psi_{_{1b}}\Bigg)(x_{_t};x_{_\rho},x_{_{\tilde{\nu}_I}};
x_{_{\tilde g}},x_{_{\tilde{b}_j}})\Bigg\}
\;,\nonumber\\
&&{\cal B}_{P}^{(6)}=-{\cal B}_{S}^{(6)}\;,\nonumber\\
&&{\cal B}_{S}^{(7)}={m_{_{l^I}}m_{_t}\over m_{_{\rm w}}^2s_{_\beta}^2}
({\cal Z}_-)^\dagger_{\alpha2}({\cal Z}_+)_{1\beta}
\Big(\Gamma_{_{\tilde{s}_j\chi_\beta}}^{^L}\Big)^*
\Big(\Gamma_{_{\tilde{t}_i\chi_\alpha}}^{^R}\Big)^*
\sum\limits_{\rho=\{\chi_\alpha,\chi_\beta,\tilde{t}_i\}}
{1\over\prod\limits_{\sigma\neq\rho}(x_{_\rho}-x_{_\sigma})}
\nonumber\\
&&\hspace{1.4cm}\times
\Bigg\{\Bigg[({\cal Z}_{_{\tilde s}})_{_{1,j}}({\cal Z}_{_{\tilde t}})_{_{2,i}}
(x_{_t}x_{_{\chi_\beta}})^{1/2}\Psi_{_{1b}}
(x_{_t};x_{_\rho},x_{_{\tilde{\nu}_I}};
x_{_{\tilde g}},x_{_{\tilde{b}_j}})
\nonumber\\
&&\hspace{1.4cm}
-({\cal Z}_{_{\tilde s}})_{_{1,j}}({\cal Z}_{_{\tilde t}})_{_{1,i}}
(x_{_{\chi_\beta}}x_{_{\tilde g}})^{1/2}e^{i\theta_{_3}}
\Bigg(\Psi_{_{1a}}-\Psi_{_{1b}}\Bigg)(x_{_t};x_{_\rho},x_{_{\tilde{\nu}_I}};
x_{_{\tilde g}},x_{_{\tilde{b}_j}})\Bigg]\Bigg\}
\;,\nonumber\\
&&{\cal B}_{P}^{(7)}=-{\cal B}_{S}^{(7)}\;,\nonumber\\
&&{\cal B}_{S}^{(8)}={2\sqrt{2}m_{_{l^I}}\over m_{_{\rm w}}s_{_\beta}}
({\cal Z}_-)^\dagger_{\alpha2}({\cal Z}_+)_{1\beta}
\Big(\Gamma_{_{\tilde{b}_i\chi_\alpha}}^{^L}\Big)
\Big(\Gamma_{_{\tilde{s}_j\chi_\beta}}^{^L}\Big)^*
({\cal Z}_{_{\tilde s}})_{_{2,j}}({\cal Z}_{_{\tilde b}})^\dagger_{_{i,1}}
(x_{_{\chi_\beta}}x_{_{\tilde g}})^{1/2}e^{-i\theta_{_3}}
\nonumber\\
&&\hspace{1.4cm}\times
\sum\limits_{\rho=\{\chi_\alpha,\chi_\beta\}}
{1\over\prod\limits_{\sigma\neq\rho}
(x_{_\rho}-x_{_\sigma})}\sum\limits_{\varrho=\{\tilde{b}_i,\tilde{s}_j\}}
{1\over\prod\limits_{\varsigma\neq\varrho}(x_{_\varrho}-x_{_\varsigma})}
\Bigg\{\Big(\Psi_{_{1a}}-\Psi_{_{1b}}\Big)(x_{_t};x_{_\rho},x_{_{\tilde{\nu}_I}};
x_{_\varrho},x_{_{\tilde g}})\Bigg\}
\;,\nonumber\\
&&{\cal B}_{P}^{(8)}=-{\cal B}_{S}^{(8)}\;,\nonumber\\
&&{\cal B}_{S}^{(9)}=\Big(\Gamma_{_{3\tilde{s}_j\chi_\beta}}^{^L}\Big)
({\cal Z}_{_{\tilde s}})_{_{2,j}}({\cal Z}_{_{\tilde b}})^\dagger_{_{i,1}}
(x_{_{\chi_\beta}}x_{_{\tilde g}})^{1/2}e^{-i\theta_{_3}}
\sum\limits_{\rho=\{\chi_\alpha,\chi_\beta\}}
{1\over\prod\limits_{\sigma\neq\rho}(x_{_\rho}-x_{_\sigma})}
\nonumber\\
&&\hspace{1.4cm}\times
\sum\limits_{\varrho=\{\tilde{b}_i,\tilde{s}_j\}}
{1\over\prod\limits_{\varsigma\neq\varrho}(x_{_\varrho}-x_{_\varsigma})}
\Bigg\{\Bigg[{2\sqrt{2}m_{_{l^I}}\over m_{_{\rm w}}s_{_\beta}}
({\cal Z}_+)^\dagger_{\alpha1}({\cal Z}_-)_{2\beta}
\Big(\Gamma_{_{\tilde{b}_i\chi_\alpha}}^{^L}\Big)
\Big(\Psi_{_{1a}}-\Psi_{_{1b}}\Big)
\nonumber\\
&&\hspace{1.4cm}
-{2m_{_{l^I}}m_{_b}\over m_{_{\rm w}}^2s_{_\beta}c_{_\beta}}
({\cal Z}_+)^\dagger_{\alpha1}({\cal Z}_-)_{2\beta}
\Big(\Gamma_{_{\tilde{b}_i\chi_\alpha}}^{^R}\Big)
\Psi_{_{1a}}\nonumber\\
&&\hspace{1.4cm}
-{2m_{_{l^I}}m_{_b}\over m_{_{\rm w}}^2s_{_\beta}c_{_\beta}}
({\cal Z}_-)^\dagger_{\alpha2}({\cal Z}_+)_{1\beta}
\Big(\Gamma_{_{\tilde{b}_i\chi_\alpha}}^{^R}\Big)
(x_{_t}x_{_{\chi_\alpha}})^{1/2}
\Psi_{_{0}}\Bigg](x_{_t};x_{_\rho},x_{_{\tilde{\nu}_I}};
x_{_\varrho},x_{_{\tilde g}})\Bigg\}
\;,\nonumber\\
&&{\cal B}_{P}^{(9)}={\cal B}_{S}^{(9)}\;.
\label{sp1}
\end{eqnarray}

\begin{eqnarray}
&&{\cal B}_{S}^{\prime(1)}=-\sum\limits_{\rho=\nu,W,H}{1\over\prod\limits_{\sigma\neq\rho}
(x_{_\rho}-x_{_\sigma})}\Bigg\{\Bigg[\Big(x_{_t}x_{_{l^I}}\Big)^{1/2}
({\cal Z}_{_{\tilde s}})_{_{2,j}}\Big({\cal A}_{_{st}}\Big)^\dagger_{ji}
({\cal Z}_{_{\tilde t}})^\dagger_{_{i,2}}\Psi_{_{1b}}
\nonumber\\
&&\hspace{1.2cm}
+\Big(x_{_{\tilde g}}x_{_{l^I}}\Big)^{1/2}e^{-i\theta_{_3}}
({\cal Z}_{_{\tilde s}})_{_{2,j}}\Big({\cal A}_{_{st}}\Big)^\dagger_{ji}
({\cal Z}_{_{\tilde t}})^\dagger_{_{i,1}}\Psi_{_{1a}}
\Bigg](x_{_{\tilde{t}_i}};x_{_\rho},x_{_t};x_{_{\tilde g}},x_{_{\tilde{s}_j}})
\Bigg\}\;,\nonumber\\
&&{\cal B}_{P}^{\prime(1)}=-{\cal B}_{S}^{\prime(1)}\;,\nonumber\\
&&{\cal B}_{S}^{\prime(2)}=-\Big(x_{_{\tilde g}}x_{_{l^I}}\Big)^{1/2}
e^{-i\theta_{_3}}({\cal Z}_{_{\tilde b}})_{_{1,i}}
({\cal Z}_{_{\tilde b}})^\dagger_{_{i,1}}({\cal Z}_{_{\tilde s}})_{_{2,k}}
\Big({\cal A}_{_{st}}\Big)^\dagger_{kj}({\cal Z}_{_{\tilde t}})^\dagger_{_{j,1}}
\sum\limits_{\rho=\{\nu,W\}}{1\over\prod\limits_{\sigma\neq\rho}
(x_{_\rho}-x_{_\sigma})}
\nonumber\\
&&\hspace{1.4cm}\times
\sum\limits_{\varrho=\{\tilde{b}_i,\tilde{s}_k\}}
{1\over\prod\limits_{\varsigma\neq\varrho}
(x_{_\varrho}-x_{_\varsigma})}\Bigg\{
\Big(\Psi_{_{1a}}-2\Psi_{_{1b}}\Big)
(x_{_{\tilde{t}_j}};x_{_\rho},x_{_H};x_{_{\tilde g}},x_{_\varrho})
\Bigg\}\;,\nonumber\\
&&{\cal B}_{P}^{\prime(2)}={\cal B}_{S}^{\prime(2)}\;,\nonumber\\
&&{\cal B}_{S}^{\prime(3)}=-\Big(x_{_{\tilde g}}x_{_{l^I}}\Big)^{1/2}
e^{-i\theta_{_3}}({\cal Z}_{_{\tilde s}})_{_{2,k}}
({\cal Z}_{_{\tilde s}})^\dagger_{_{k,1}}({\cal Z}_{_{\tilde t}})_{_{1,j}}
\Big({\cal A}_{_{bt}}\Big)_{ji}({\cal Z}_{_{\tilde b}})^\dagger_{_{i,1}}
\sum\limits_{\rho=\{\nu,W\}}{1\over\prod\limits_{\sigma\neq\rho}
(x_{_\rho}-x_{_\sigma})}
\nonumber\\
&&\hspace{1.4cm}\times
\sum\limits_{\varrho=\{\tilde{b}_i,\tilde{s}_k\}}
{1\over\prod\limits_{\varsigma\neq\varrho}
(x_{_\varrho}-x_{_\varsigma})}\Bigg\{\Big(\Psi_{_{1a}}-2\Psi_{_{1b}}\Big)
(x_{_{\tilde{t}_j}};x_{_\rho},x_{_H};x_{_{\tilde g}},x_{_\varrho})
\Bigg\}\;,\nonumber\\
&&{\cal B}_{P}^{\prime(3)}=-{\cal B}_{S}^{\prime(3)}\;,\nonumber\\
&&{\cal B}_{S}^{\prime(4)}=\sum\limits_{\rho=\{\chi_\alpha,\chi_\beta,\tilde{t}_i\}}
{1\over\prod\limits_{\sigma\neq\rho}(x_{_\rho}-x_{_\sigma})}
\Bigg\{-{\sqrt{2}m_{_{l^I}}\over m_{_{\rm w}}s_{_\beta}}
({\cal Z}_+)^\dagger_{\alpha1}({\cal Z}_-)_{2\beta}
\Big(\Gamma_{_{\tilde{s}_j\chi_\beta}}^{^L}\Big)^*
\Big(\Gamma_{_{\tilde{t}_i\chi_\alpha}}^{^L}\Big)^*
\nonumber\\
&&\hspace{1.4cm}\times
\Bigg[({\cal Z}_{_{\tilde s}})_{_{2,j}}({\cal Z}_{_{\tilde t}})_{_{1,i}}
\Bigg({1\over2}\varrho_{_{2,1}}(x_{_\rho},x_{_{\tilde{\nu}_k}})
+\Big[\Psi_{_{2b}}-\Psi_{_{2d}}\Big](x_{_t};x_{_\rho},x_{_{\tilde{\nu}_k}};
x_{_{\tilde g}},x_{_{\tilde{b}_j}})\Bigg)
\nonumber\\
&&\hspace{1.4cm}
-({\cal Z}_{_{\tilde s}})_{_{2,j}}({\cal Z}_{_{\tilde t}})_{_{2,i}}
(x_{_t}x_{_{\tilde g}},)^{1/2}e^{-i\theta_{_3}}\Psi_{_{1a}}
(x_{_t};x_{_\rho},x_{_{\tilde{\nu}_k}};
x_{_{\tilde g}},x_{_{\tilde{b}_j}})\Bigg]\Bigg\}
\;,\nonumber\\
&&{\cal B}_{P}^{\prime(4)}={\cal B}_{S}^{\prime(4)}\;,\nonumber\\
&&{\cal B}_{S}^{\prime(5)}=
{2\sqrt{2}m_{_{l^I}}\over m_{_{\rm w}}s_{_\beta}}
({\cal Z}_-)^\dagger_{\alpha2}({\cal Z}_+)_{1\beta}
\Big(\Gamma_{_{\tilde{b}_i\chi_\alpha}}^{^L}\Big)
\Big(\Gamma_{_{\tilde{s}_j\chi_\beta}}^{^L}\Big)^*
({\cal Z}_{_{\tilde s}})_{_{2,j}}({\cal Z}_{_{\tilde b}})^\dagger_{_{i,1}}
(x_{_{\chi_\beta}}x_{_{\tilde g}})^{1/2}e^{-i\theta_{_3}}
\nonumber\\
&&\hspace{1.4cm}\times
\sum\limits_{\rho=\{\chi_\alpha,\chi_\beta\}}
{1\over\prod\limits_{\sigma\neq\rho}
(x_{_\rho}-x_{_\sigma})}\sum\limits_{\varrho=\{\tilde{b}_i,\tilde{s}_j\}}
{1\over\prod\limits_{\varsigma\neq\varrho}(x_{_\varrho}-x_{_\varsigma})}
\Bigg\{\Big(\Psi_{_{1a}}-\Psi_{_{1b}}\Big)(x_{_t};x_{_\rho},x_{_{\tilde{\nu}_k}};
x_{_\varrho},x_{_{\tilde g}})\Bigg\}
\;,\nonumber\\
&&{\cal B}_{P}^{\prime(5)}=-{\cal B}_{S}^{\prime(5)}\;,\nonumber\\
&&{\cal B}_{S}^{\prime(6)}=
\Big(\Gamma_{_{\tilde{s}_j\chi_\beta}}^{^L}\Big)^*
({\cal Z}_{_{\tilde s}})_{_{2,j}}({\cal Z}_{_{\tilde b}})^\dagger_{_{i,1}}
(x_{_{\chi_\beta}}x_{_{\tilde g}})^{1/2}e^{-i\theta_{_3}}
\sum\limits_{\rho=\{\chi_\alpha,\chi_\beta\}}
{1\over\prod\limits_{\sigma\neq\rho}(x_{_\rho}-x_{_\sigma})}
\nonumber\\
&&\hspace{1.4cm}\times
\sum\limits_{\varrho=\{\tilde{b}_i,\tilde{s}_j\}}
{1\over\prod\limits_{\varsigma\neq\varrho}(x_{_\varrho}-x_{_\varsigma})}
\Bigg\{\Bigg[{2\sqrt{2}m_{_{l^I}}\over m_{_{\rm w}}s_{_\beta}}
({\cal Z}_+)^\dagger_{\alpha1}({\cal Z}_-)_{2\beta}
\Big(\Gamma_{_{\tilde{b}_i\chi_\alpha}}^{^L}\Big)
\Big(\Psi_{_{1a}}-\Psi_{_{1b}}\Big)
\nonumber\\
&&\hspace{1.4cm}
-{2m_{_{l^I}}m_{_b}\over m_{_{\rm w}}^2s_{_\beta}c_{_\beta}}
({\cal Z}_+)^\dagger_{\alpha1}({\cal Z}_-)_{2\beta}
\Big(\Gamma_{_{\tilde{b}_i\chi_\alpha}}^{^R}\Big)\Psi_{_{1a}}
\nonumber\\
&&\hspace{1.4cm}
-{2m_{_{l^I}}m_{_b}\over m_{_{\rm w}}^2s_{_\beta}c_{_\beta}}
({\cal Z}_-)^\dagger_{\alpha2}({\cal Z}_+)_{1\beta}
\Big(\Gamma_{_{\tilde{b}_i\chi_\alpha}}^{^R}\Big)
(x_{_t}x_{_{\chi_\alpha}})^{1/2}\Psi_{_{0}}\Bigg]
(x_{_t};x_{_\rho},x_{_{\tilde{\nu}_k}};x_{_\varrho},x_{_{\tilde g}})\Bigg\}
\;,\nonumber\\
&&{\cal B}_{P}^{\prime(6)}={\cal B}_{S}^{\prime(6)}\;.
\label{psp}
\end{eqnarray}

Here, some two-loop functions are defined as
\begin{eqnarray}
&&\Theta_{_{0}}(x_{_0},x_{_1},x_{_2})={1\over2}\Bigg[2x_{_1}\ln x_{_1}
-x_{_1}\ln^2x_{_1}-\Phi(x_{_0},x_{_1},x_{_2})\Bigg]\;,\nonumber\\
&&\Theta_{_{1a}}(x_{_0},x_{_1},x_{_2})=-{1\over2}\Bigg[x_{_1}^2\ln^2x_{_1}
+x_{_1}\Phi(x_{_0},x_{_1},x_{_2})\Bigg]\;,\nonumber\\
&&\Theta_{_{1b}}(x_{_0},x_{_1},x_{_2})={1\over4}\Bigg[-4(x_{_0}-x_{_2})
x_{_1}\ln x_{_1}+x_{_1}^2\ln^2x_{_1}+(x_{_0}+x_{_1}+x_{_2})
\Phi(x_{_0},x_{_1},x_{_2})\Bigg]\;,\nonumber\\
&&\Theta_{_{2}}(x_{_0},x_{_1},x_{_2})={1\over4}\Bigg[2x_{_1}^3\ln x_{_1}
+(2x_{_0}-x_{_1}+2x_{_2})x_{_1}^2\ln^2x_{_1}+x_{_1}(x_{_0}-x_{_1}+x_{_2})
\Phi(x_{_0},x_{_1},x_{_2})\Bigg]\;.\nonumber\\
\label{ap4-eq2}
\end{eqnarray}
For the functions $\Psi_{_{2b,2c,2d,1a,1b}}$ as well as
$\Phi(x_0,x_1,x_2)$ can be found in our forthcoming work
\cite{Feng2}

In those expressions, we have defined the short notations
\begin{eqnarray}
&&\Gamma_{_{\tilde{t}_i\chi_\alpha}}^{L}=\Big({\cal Z}_{_{\tilde t}}\Big)_{1i}
\Big({\cal Z}_+\Big)^\dagger_{\alpha1}-{m_{_{t}}\over \sqrt{2}m_{_{\rm w}}s_{_\beta}}
\Big({\cal Z}_{_{\tilde t}}\Big)_{2i}\Big({\cal Z}_+\Big)^\dagger_{\alpha2}
\;,\nonumber\\
&&\Gamma_{_{\tilde{t}_i\chi_\alpha}}^{R}=\Big({\cal Z}_{_{\tilde t}}\Big)_{1i}
\Big({\cal Z}_-\Big)_{2\alpha}\;,\nonumber\\
&&\Gamma_{_{\tilde{D}^I_i\chi_\alpha}}^{L}=\Big({\cal Z}_{_{\tilde{D}^I}}\Big)_{1i}
\Big({\cal Z}_-\Big)^\dagger_{\alpha1}-{m_{_{d^I}}\over \sqrt{2}m_{_{\rm w}}s_{_\beta}}
\Big({\cal Z}_{_{\tilde{D}^I}}\Big)_{2i}\Big({\cal Z}_-\Big)^\dagger_{\alpha2}
\;,\nonumber\\
&&\Gamma_{_{\tilde{D}^I_i\chi_\alpha}}^{R}=\Big({\cal Z}_{_{\tilde{D}^I}}\Big)_{1i}
\Big({\cal Z}_+\Big)_{2\alpha}\;,\nonumber\\
&&\Big(\xi_{_{H}}\Big)_{_{ij}}=\Big[{m_{_t}|\mu|\over m_{_{\rm w}}^2}e^{-i\theta_{_\mu}}
({\cal Z}_{_{\tilde t}})_{_{2,j}}({\cal Z}_{_{\tilde s}}^\dagger)_{_{i,1}}
+s_{_\beta}{\sqrt{2}s_{_{\rm w}}A_{_s}\over em_{_{\rm w}}}
({\cal Z}_{_{\tilde t}})_{_{1,j}}({\cal Z}_{_{\tilde s}}^\dagger)_{_{i,1}}\Big]
\;,\nonumber\\
&&\Big(\xi_{_{q}}\Big)_{_{ij}}=\left\{\begin{array}{l}({\cal Z}_{_{\tilde s}}^\dagger)_{_{i,1}}
({\cal Z}_{_{\tilde s}})_{_{1,j}}-{2\over3}s_{_{\rm w}}^2\delta_{_{ij}}\;,\;q=s\\\\
({\cal Z}_{_{\tilde t}}^\dagger)_{_{i,1}}
({\cal Z}_{_{\tilde t}})_{_{1,j}}-{4\over3}s_{_{\rm w}}^2\delta_{_{ij}}\;,\;q=t
\end{array}\right.\;,\nonumber\\
&&\Big(\eta_{_{H}}\Big)_{_{ij}}=\Big[{m_{_t}|\mu|\over m_{_{\rm w}}^2t_{_\beta}}
\Big(e^{i\theta_{_\mu}}({\cal Z}_{_{\tilde t}}^\dagger)_{_{i,1}}({\cal Z}_{_{\tilde t}})_{_{2,j}}
-e^{-i\theta_{_\mu}}({\cal Z}_{_{\tilde t}}^\dagger)_{_{i,2}}({\cal Z}_{_{\tilde t}})_{_{1,j}}\Big)
\nonumber\\
&&\hspace{1.6cm}
+s_{_\beta}{\sqrt{2}s_{_{\rm w}}\over em_{_{\rm w}}}
\Big(A_{_t}({\cal Z}_{_{\tilde t}}^\dagger)_{_{i,2}}
({\cal Z}_{_{\tilde t}})_{_{1,j}}-A_{_t}^*({\cal Z}_{_{\tilde t}}^\dagger)_{_{i,1}}
({\cal Z}_{_{\tilde t}})_{_{2,j}}\Big)\Big]
\;,\nonumber\\
&&\Big(\xi_{_{\chi}}^i\Big)_{_{\alpha\beta}}=\left\{\begin{array}{l}
2\delta_{_{\alpha\beta}}(c_{_{\rm w}}^2-s_{_{\rm w}}^2)
+({\cal Z}_+)_{_{1,\alpha}}({\cal Z}_+^\dagger)_{_{\beta,1}}\;,i=1\\\\
c_{_\beta}({\cal Z}_+)_{_{1,\alpha}}({\cal Z}_-)_{_{2,\beta}}
-s_{_\beta}({\cal Z}_+)_{_{2,\alpha}}({\cal Z}_-)_{_{1,\beta}}\;,i=2\\\\
c_{_\beta}({\cal Z}_-^\dagger)_{_{\alpha,2}}({\cal Z}_+^\dagger)_{_{\beta,1}}
-s_{_\beta}({\cal Z}_-^\dagger)_{_{\alpha,1}}({\cal Z}_+^\dagger)_{_{\beta,2}}
\;,i=3\\\\
2\delta_{_{\alpha\beta}}(c_{_{\rm w}}^2-s_{_{\rm w}}^2)
+({\cal Z}_-^\dagger)_{_{\alpha,1}}({\cal Z}_-)_{_{1,\beta}}\;,i=4\\\\
\end{array}\right.\;,\nonumber\\
&&\Big(\zeta_{_{sH}}^\rho\Big)_{_{ij}}={s_{_\beta}\over6c_{_{\rm w}}^2}
\Big[(1+2c_{_{\rm w}}^2)({\cal Z}_{_{\tilde s}}^\dagger)_{_{i,1}}
({\cal Z}_{_{\tilde s}})_{_{1,j}}+2s_{_{\rm w}}^2
({\cal Z}_{_{\tilde s}}^\dagger)_{_{i,1}}({\cal Z}_{_{\tilde s}})_{_{1,j}}
\Big]({\cal Z}_{_H})_{_{3,\rho}}
\nonumber\\
&&\hspace{1.6cm}
+{s_{_{\rm w}}\over\sqrt{2}em_{_{\rm w}}}
\Big[A_{_{\tilde s}}^*({\cal Z}_{_{\tilde s}}^\dagger)_{_{i,1}}
({\cal Z}_{_{\tilde s}})_{_{2,j}}
+A_{_{\tilde s}}({\cal Z}_{_{\tilde s}}^\dagger)_{_{i,2}}
({\cal Z}_{_{\tilde s}})_{_{1,j}}\Big]({\cal Z}_{_H})_{_{2,\rho}}
\nonumber\\
&&\hspace{1.6cm}
-i{s_{_\beta}s_{_{\rm w}}\over\sqrt{2}em_{_{\rm w}}}
\Big[A_{_{\tilde s}}^*({\cal Z}_{_{\tilde s}}^\dagger)_{_{i,1}}
({\cal Z}_{_{\tilde s}})_{_{2,j}}-A_{_{\tilde s}}({\cal Z}_{_{\tilde s}}^\dagger)_{_{i,2}}
({\cal Z}_{_{\tilde s}})_{_{1,j}}\Big]({\cal Z}_{_H})_{_{1,\rho}}
\;,\nonumber\\
&&\Big(\zeta_{_{tH}}^\rho\Big)_{_{ij}}={m_{_t}|\mu|\over2m_{_{\rm w}}^2s_{_\beta}}
\Big[e^{i\theta_{_\mu}}({\cal Z}_{_{\tilde t}})_{_{2,j}}
({\cal Z}_{_{\tilde t}}^\dagger)_{_{i,1}}
-e^{-i\theta_{_\mu}}({\cal Z}_{_{\tilde t}})_{_{1,j}}
({\cal Z}_{_{\tilde t}}^\dagger)_{_{i,2}}\Big]({\cal Z}_{_H})_{_{2,\rho}}
\nonumber\\
&&\hspace{1.6cm}
+\Big[\Big({(4c_{_{\rm w}}^2-1)s_{_\beta}\over6c_{_{\rm w}}^2}
-{m_{_t}^2\over m_{_{\rm w}}^2s_{_\beta}}\Big)({\cal Z}_{_{\tilde t}})_{_{1,j}}
({\cal Z}_{_{\tilde t}}^\dagger)_{_{i,1}}
+\Big({2s_{_{\rm w}}^2s_{_\beta}\over3c_{_{\rm w}}^2}
-{m_{_t}^2\over m_{_{\rm w}}^2s_{_\beta}}\Big)({\cal Z}_{_{\tilde t}})_{_{2,j}}
({\cal Z}_{_{\tilde t}}^\dagger)_{_{i,2}}
\nonumber\\
&&\hspace{1.6cm}
-{s_{_{\rm w}}\over\sqrt{2}em_{_{\rm w}}}\Big(A_{_{\tilde t}}^*
({\cal Z}_{_{\tilde t}})_{_{2,j}}({\cal Z}_{_{\tilde t}}^\dagger)_{_{i,1}}
+A_{_{\tilde t}}({\cal Z}_{_{\tilde t}})_{_{1,j}}({\cal Z}_{_{\tilde t}})_{_{i,2}}
\Big)\Big]({\cal Z}_{_H})_{_{3,\rho}}
\nonumber\\
&&\hspace{1.6cm}
+i{m_{_t}|\mu|\over2m_{_{\rm w}}^2}\Big[e^{i\theta_{_\mu}}({\cal Z}_{_{\tilde t}})_{_{2,j}}
({\cal Z}_{_{\tilde t}}^\dagger)_{_{i,1}}
-e^{-i\theta_{_\mu}}({\cal Z}_{_{\tilde t}})_{_{1,j}}
({\cal Z}_{_{\tilde t}}^\dagger)_{_{i,2}}\Big]
({\cal Z}_{_H})_{_{1,\rho}}\;,\nonumber\\
&&\Big(\kappa_{_{H^{^\rho}}}^i\Big)_{_{\alpha\beta}}=\left\{\begin{array}{l}
({\cal Z}_+)_{_{1,\alpha}}({\cal Z}_-)_{_{2,\beta}}({\cal Z}_{_H})_{_{2,\rho}}
+({\cal Z}_+)_{_{2,\alpha}}({\cal Z}_-)_{_{1,\beta}}
({\cal Z}_{_H})_{_{3,\rho}}\\\\
+is_{_\beta}({\cal Z}_+)_{_{1,\alpha}}({\cal Z}_-)_{_{2,\beta}}
({\cal Z}_{_H})_{_{1,\rho}}\;,\;\;i=1\\\\
({\cal Z}_-^\dagger)_{_{\alpha,2}}
({\cal Z}_+^\dagger)_{_{\beta,1}}({\cal Z}_{_H})_{_{2,\rho}}
+({\cal Z}_-^\dagger)_{_{\alpha,1}}({\cal Z}_+^\dagger)_{_{\beta,2}}
({\cal Z}_{_H})_{_{3,\rho}}\\\\
-is_{_\beta}({\cal Z}_-^\dagger)_{_{\alpha,2}}({\cal Z}_+^\dagger)_{_{\beta,1}}
({\cal Z}_{_H})_{_{1,\rho}}\;,\;\;i=2
\end{array}\right.\;.
\label{ap4-eq3}
\end{eqnarray}

\end{document}